\documentclass{aa}
\usepackage{txfonts} 
\usepackage{graphicx} 
\usepackage{longtable}
 
\usepackage{url} 
\usepackage{natbib} 
\usepackage{txfonts}
\begin{document}


   \title{Barlenses in the CALIFA survey: combining the photometric and stellar population analysis }

   \author{E. Laurikainen
          \inst{1}
          \and
          H. Salo\inst{1}
          \and
          J. Laine\inst{1,2,3}
          \and 
          J. Janz\inst{1,4} 
          }

   \institute{Astronomy Research Unit, University of Oulu, FI-90014 Finland\\
              \email{eija.laurikainen@oulu.fi}
         \and
             Hamburg Sternwarte, Universit\"at Hamburg, Gojenbergsweg 112, 21029, Hamburg, Germany
         \and Instituto de Astronom\'ia, Universidad Nacional Aut\'onoma de M\'exico, A. P. 70-264 04510, M\'exico D.F., M\'exico   
         \and 
            Finnish Centre of Astronomy with ESO (FINCA), University of Turku, V\"ais\"al\"antie 20, FI-21500 Piikki\"o, Finland\\
             }
   

   \date{Received; 30.3.2018 accepted; }

 
  \abstract
   {}
{It is theoretically predicted that the Boxy/Peanut bar components
  have a barlens appearance (a round central component embedded in the
  narrow bar) at low galaxy inclinations.  Here we investigate
  barlenses in the Calar Alto Legacy Integral Field Area (CALIFA)
  survey galaxies, studying their morphologies, stellar
  populations and metallicities.  We show that, when present, barlenses
  account for a significant portion of light of photometric bulges
  (i.e. the excess light on top of the disks), which highlights the
  importance of bars in accumulating the central galaxy mass
  concentrations in the cosmic timescale.}
{We make multi-component decompositions for a sample of 46 barlens
  galaxies drawn from the CALIFA survey (S\'anchez, Garc\'ia-Benito
  $\&$ Zibetti 2016), with M$_{\star}$/M$_{\odot}$ = 10$^{9.7}$--10$^{11.4}$
  and z = 0.005--0.03. Unsharp masks of the Sloan Digital Sky Survey
  (SDSS) r'-band mosaics are used to identify the Boxy/Peanut/X
  features. Barlenses are identified in the images using the
  simulation snapshots by Salo $\&$ Laurikainen (2017) as an
  additional guide.  Our decompositions with GALFIT include in
  addition to bulges, disks and bars, also barlenses as a separate
  component.  For 26 of the decomposed galaxies the CALIFA DR2 V500
  grating data-cubes are used to explore the stellar ages and
  metallicities, at the regions of various structure components.}
{We find that $25 \pm 2\%$ of the 1064 galaxies in the whole CALIFA sample 
show either
  X-shape or barlens feature. In the decomposed galaxies with barlenses, on average
  13$\%\pm$2$\%$ of the total galaxy light belongs to this component, leaving
  less than 10$\%$ for possible separate bulge components. Most
  importantly, bars and barlenses are found to have similar cumulative
  stellar age and metallicity distributions.  The metallicities in
  barlenses are on average near solar, but exhibit a large range.  In
  some of the galaxies barlenses and X-shape features appear
  simultaneously, in which case the bar-origin of the barlens is
  unambiguous.  }
{This is the first time that a combined morphological and stellar population analysis
is used to study barlenses. We show that their stars are accumulated 
in a prolonged time period, concurrently with the evolution of the narrow bar.   }


   \maketitle
%

\section{Introduction}
 
Central mass concentrations of galaxies are often thought to have
assembled in early galaxy mergers \citep{toomre1972,negroponte1983},
or to have formed via inward drift of massive clumps at high redshifts \citep{elmegreen2008,bournaud2016}. Such early
formed structures, dynamically supported by velocity dispersion, are called as
classical bulges.  However, it has been pointed out that even
45$\%$ of the bright S0s and spirals have bulges which are actually
vertically thick inner bar components (L\"utticke, Dettmar $\&$ Pohlen 2000;
Laurikainen et al. 2014; Erwin $\&$ Debattista 2017; see also Yoshino $\&$ Yamauchi 2015), 
in a similar manner as the Milky Way bulge
\citep{nataf2010,mac2010,wegg2013,ness2016}. In edge-on view such
inner parts of bars have Boxy, Peanut, or X-shaped morphology
\citep{bureau2006}, and in face-on view a barlens morphology
\citep{lauri2014,atha2015,lauri2017}, i.e. they look like a lens
embedded in a narrow bar \citep{lauri2011}. 
Simulation models of \cite{salo2017} predict that barlens morphology, with the
nearly round appearance, occurs preferably in galaxies with centrally
peaked mass concentrations.  Whether this mass concentration is
triggered by the bar induced inflow of gas and subsequent star
formation, or predates the bar, i.e. is a classical bulge or an inner
thick disk is not yet clear.  Also, although there is strong
observational and theoretical evidence for a bar origin of the
Boxy/Peanuts/barlens bulges in the Milky Way mass galaxies, it is
still an open question how much baryonic mass in the local Universe is
confined into these structures.
 
Historically, bulges were thought to be like
mini-ellipticals. Morphologies and stellar populations of the
bulge-dominated galaxies have indeed supported the idea that their bulges,
largely consisting of old stars, formed early in some rapid event, and
that their disks gradually assembled around them
\citep{kauffmann1993,zoccali2006}. Consistent with this picture is
also that all bulges seem to share the same fundamental plane with the
elliptical galaxies
\citep{bender1992,falco2002,kim2016,costantin2017}.

However, there are many important observations which have challenged
the picture of early bulge formation, related either to a monolithic 
collapse \citep{eggen1962} or galaxy mergers.  At redshifts
z = 1--3 very few galaxies actually have bulges in the same sense as those observed in
the nearby universe. Those galaxies are rather
constellations of massive star forming clumps
\citep{abraham1996,berg1996,cowie1996,elmegreen2005}, which 
have been proposed to gradually coalesce to galactic bulges
\citep{noguchi1999,bournaud2007,elmegreen2008,combes2014,bournaud2016}.
Challenging is also the observation that in the nearby universe
the fraction of galaxies with classical bulges is very low. Most dwarf
sized galaxies and Sc--Scd spirals have rather disk-like pseudo-bulges dominated
by recent star formation \citep{kormendy2010,salo2015}.  Even a large
fraction of the bright Milky Way mass S0s and spirals might lack 
a classical bulge \citep{lauri2010,lauri2014}. 

An important observation was also that 90$\%$ of the stellar mass in
the Milky Way mass galaxies and in galaxies more massive than that has accumulated since z = 2.5, so that bulges actually
formed in lock-step with the disks until z = 1
\citep{dokkum2010,dokkum2013,marc2014}.  Cosmological simulation
models predict that in massive halos the cold and hot gas phases are
de-coupled, so that after in-situ star formation at z $>$ 1.5 the gas
cannot penetrate through the hot halo gas anymore
\citep{naab2007,feldmann2010,johansson2012,qu2017}. Those galaxies
become red and dead at high redshifts, recognized as fairly small
centrally concentrated ``red nuggets''
\citep{daddi2005,trujillo2006,damjanov2011}. Alternatively mass
accretion may continue via accretion of stars produced in satellite
galaxies, leading to massive elliptical galaxies (Oser et al. 2010;
see also Kennicutt $\&$ Evans 2012). 
However, in less massive halos the gas accretion can continue as long
as there is fresh gas in the near galaxy environment.  This prolonged
accumulation of gas into the halos, which gas at the end settles into
the galactic disks, is expected to play an important role in the
evolution of the progenitors of the Milky Way mass galaxies. At the same time as the galactic disks gradually increase in mass also their central mass concentrations increase. This can occur via multiple disk instabilities manifested as vertically thick Boxy/Peanut/barlens structures (Martinez-Valpuesta, Shlosman $\&$ 2006) of bars. Bars are also efficient in triggering gas inflow (Berenzen et al. 1998) thus further accumulating mass in central galaxy regions.
Whether this is the dominant way of making the
  central mass concentrations in the Milky Way mass galaxies is an
  interesting question, which needs to be systematically studied for
  a representative sample of nearby galaxies.

There are many galaxies in which Boxy/Peanut bulges are convincingly
identified. Kinematic analysis tools
have been developed to recognize them both in edge-on \citep{atha1999}
and face-on views (Debattista et al. 2005; see also reviews by
Athanassoula 2016, and Laurikainen $\&$ Salo 2016), which methods are
successfully applied to some individual galaxies.  Good examples are
NGC 98 \citep{mendez2008} and ten more low inclination galaxies
studied kinematically by \citet{mendez2014}. In our terms NGC 98, with
a clear signature of Boxy/Peanut structure in its line-of-sight
  velocity profile (in H4, the fourth moment in Gauss-Hermite series),
  is also a barlens galaxy by its morphology. In edge-on view the
Boxy/Peanuts are easy to detect \citep{bureau2006}, but then the
challenge is to identify also the bars. A characteristic feature of Boxy/Peanut
bulges is cylindrical rotation, which has been used to identify them
in 12 mid-to-high inclination galaxies by \citet{mola2016} using
Integral-Field Unit (IFU) observations, and by
\citet{williams2011,williams2012} using long-slit spectroscopy.
However, the interpretation of cylindrical rotation largely
depends on galaxy orientation \citep{iannuzzi2015,mola2016}, and is
also time-dependent \citep{saha2018}. So far, the most efficient way
of identifying the Boxy/Peanut structures at intermediate galaxy
inclinations has been to inspect their isophotal shapes, by inspecting
their boxyness/diskiness \citep{erwin2013,herrera2017}, or calculating
the higher Fourier modes \citep{ciambur2016}. Barlenses overlap with
these identifications and have been recognized in large galaxy samples
\citep{lauri2011,lauri2014,li2017}.

In spite of the success in identifying the vertically thick inner bar
components at all galaxy inclinations, very little has been done for
estimating their relative masses or stellar populations.  Some
preliminary estimates in the local universe were made by
\citet{lauri2014} (the relative masses) and by \citet{herrera2017}
(colors). In the current paper these issues are addressed for the
Calar Alto Legacy Integral Field Area (CALIFA) survey.  The vertically
thick inner bar components are first recognized in barred
galaxies, and then detailed multi-component bulge/disk/bar/barlens
(B/D/bar/bl) decompositions are made for a sub-sample of barlens
galaxies, following the method by \citet{lauri2014}. For the same
galaxies B/D/bar decompositions have been previously made by
M\'endez-Abreu et al. (2017, hereafter MA2017). However, we show that
the inclusion of barlens component into the decompositions significantly 
modifies the interpretation of mass of possible classical bulges.

\section{What are barlenses and how do they relate to Boxy/Peanut/X-shape structures}

By barlenses (bl in the following) we mean lens-like structures
embedded in bars, covering $\sim$1/2 of the length of the narrow bar \citep{lauri2011},
manifested as Boxy/Peanut or X-shape features at nearly edge-on galaxies.
They are typically a factor of $\sim$4 larger than nuclear disks, nuclear rings, or
nuclear bars.  When the concept of a barlens was invented, it was already suggested 
to be a vertically thick inner bar component, of which evidence was later
shown by the simulation models of \citet{atha2015} and
\citet{salo2017}. Barlenses and Boxy/Peanut/X-shape features are found
in galaxies with stellar masses of M$_{\star}$/M$_{\odot}$ = 10$^{9.7}$--10$^{11.4}$
\citep{lauri2014, herrera2015, li2017}.

\begin{figure}
\centering
\includegraphics[angle=0,width=9.0cm]{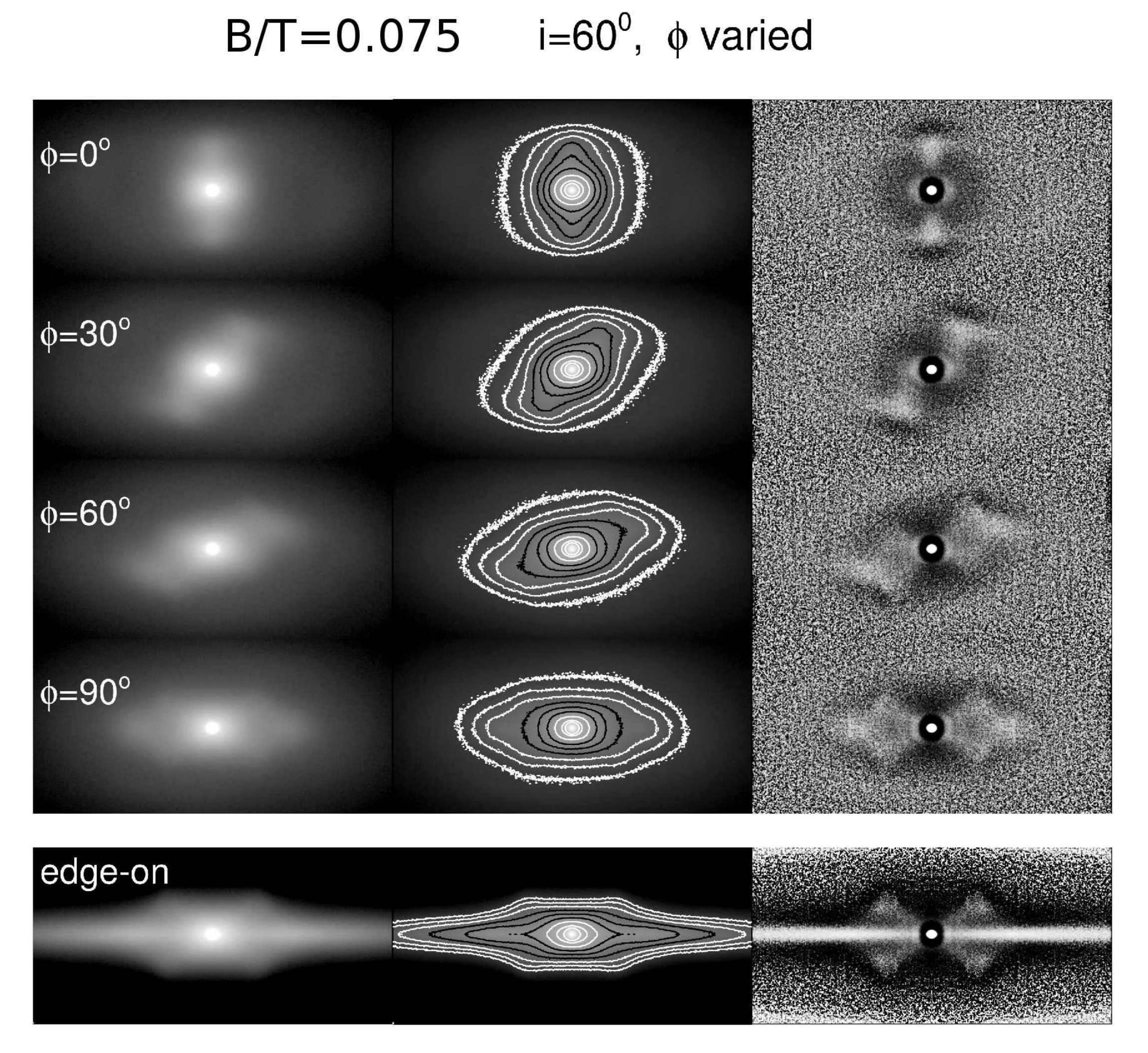} 
\caption{Synthetic simulation images from
  \citet{lauri2017}. In the {\it left panels} the direct image in the magnitude
    scale is shown, the {\it middle panels} show the contours of the same image, and the {\it right panels}
  show the unsharp mask image.  In the four upper lines the galaxy
  inclination is fixed to $i$ = 60$^{\degr}$, whereas the azimuthal
  angle $\phi$ with respect to the bar major axis varies. In the 
    lowest line the same model is seen edge-on.  The simulation
  initial values contain a small classical bulge with bulge-to-total ratio $B/T$ = 0.075. 
  During the simulation a bar with a vertically thick inner barlens component forms.}
\label{synthetic-models}
\end{figure}

Using a sample of 80 barlenses and 89 X-shaped bars in the combined
Spitzer Survey of Stellar Structure in Galaxies (S$^4$G) and Near-IR
S0-Sa galaxy Survey (NIRS0S), \citet{lauri2014} showed that the
distribution of the parent galaxy minor-to-major ($b/a$) axis ratios
of the galaxies with barlenses and X-shape features partially overlap. Together they
form a flat distribution, as expected if X-features and barlenses are
physically the same structures seen at different viewing angles. A
more detailed analysis of the relation between galaxy orientation and
barlens morphology was made by \citet{lauri2017} and
\citet{salo2017}. Synthetic images made from the simulation snapshots
predicted how the barlens morphology gradually changes as a function
of galaxy inclination, in a similar manner as in observations.
  In particular, there is a range of intermediate galaxy inclinations
  where the X-feature and the barlens are visible at the same
time. At intermediate inclinations the barlens morphology 
becomes complex due to a combined effect of galaxy inclination and the
azimuthal viewing angle, which is illustrated in Figure
\ref{synthetic-models}: the model galaxy is shown at the fixed
inclination $i$ = 60$^{\degr}$, seen at four different azimuthal
angles $\phi$ with respect to the bar major axis (four upper panels). In
the lowest panel the same snapshot is shown in edge-on view ($i$ =
90$^{\degr}$ and $\phi$=90$^{\degr}$).  In the original study the
simulation snapshots were viewed from 100 isotropically chosen
directions, varying the inclination and the azimuthal angle. It is
worth noticing that neither the X-features nor the barlenses are
visible when the bar is seen close to end-on ($\phi$=0$^{\degr}$) at
high inclinations. However this situation occurs for $<$ 10$\%$ of all
viewing directions (see Fig. 2 in \citet{lauri2017}). 
We use the simulated barlens morphologies as a guide while identifying
them in the CALIFA sample.

\vskip 0.20cm In this study the following nomenclature is used:
\vskip 0.15cm
\noindent {\it Photometric bulge:} excess central light on top of the disk,
extrapolated to the center.
\vskip 0.1cm

\noindent {\it Bar:} elongated bar component (barlens flux not included).
\vskip 0.1cm

\noindent {\it Barlens (bl):} a lens-like structure covering $\sim$1/2 of barlength.
It is assumed to be a vertically thick inner bar component (same as Boxy/Peanut/X). 
\vskip 0.1cm

\noindent {\it Separate bulge component (or bulge):} some galaxies have a
central peak in the surface brightness profile (embedded in the
barlens), which flux is fitted with a S\'ersic function (see Section 5).
\vskip 0.1cm

\noindent {\it Central region (C):} central galaxy region covering r = 0.3 x
r$_{bl}$ (see Section 7.1). It is measured in a similar manner for all
galaxies.

\section{Sample selection}

As a starting point we use the Calar Alto Legacy Integral Field
  Area (CALIFA) survey of galaxies \citep{sanchez2012}, described in
  detail by \citet{walcher2014}. It consists of galaxies in the mass
  range of M$_{\star}$/M$_{\odot}$ = 10$^{9.7}$--10$^{11.4}$,
  covering the redshift range z = 0.005--0.03.  CALIFA contains a {\it
    mother sample} of 939 galaxies, which are originally selected from Sloan
  Digital Sky Survey data release 7 (SDSS-DR7), and an {\it extension
    sample} of 125 galaxies, which is included in SDSS-DR12
  \citep{alam2016}.  Altogether this makes 1064 galaxies. In the
  public data release DR2 \citep{sanchezGZ2016} 
  IFU data-cubes are given for 667 of these galaxies. From the
sample of 1064 galaxies, and using the SDSS r'-band mosaic images (See
Appendix \ref{appendixA}), we identified 236 galaxies, which have
either a barlens in the original image, or an X-shaped feature in the
unsharp mask image (see Section 4): of these 110 (+9 uncertain) have
barlenses, and 124 (+3 uncertain) have X-shaped bars. In 15 additional
galaxies both features were identified. Altogether this makes 24$\%$
of all CALIFA galaxies (excluding the uncertain cases). Taking into
account also the uncertain cases, and the fact that we are probably
missing some ($\sim 7\%$) of the features due to an unfavorable viewing
angle, we get 26$\%$ as an upper limit. Combining with the statistical
uncertainty due to the sample size indicates about $25\pm 2\%$
frequency of B/P/X/bl features.

The selected barlens and X-shaped galaxy samples are shown in Tables
\ref{appendixB} and \ref{appendixC}, respectively. Shown in the tables
are also the redshifts, absolute r'-band magnitudes, and masses of the
galaxies, as given in the public CALIFA data-release
\citep{sanchezGZ2016}. As our intention in this study is to
  compare the photometric decomposition results with those derived
  using the IFU-observations, not all barlens galaxies were
  decomposed. Instead, for the decomposition analysis a sub-sample of
54 galaxies was selected, including all barlens galaxies which have
V1200 grating CALIFA data-cubes available. Of these galaxies 6 had an uncertain
barlens identification.  Considering only the galaxies with clear barlens
identifications we were able to
do reliable decompositions (with barlens fitted) for 46 of the 48 galaxies,
shown in Table~\ref{table-1}. Shown in the table are also
the Hubble stages from the CALIFA data-release. Notice that in NGC 6004 a barlens was identified, but because of its
low surface brightness it could not be fitted. In NGC 7814 the bulge component has the size 
comparable to the image FWHM, 
for which reason the effective radius is not given.
Of the selected 54 galaxies V500 grating
data-cubes were available for 34 galaxies, of which 8 had uncertain
barlens identifications. Excluding the uncertain cases 26 galaxies
were selected for the stellar populations analysis.  Of these galaxies
8 have also X-features in the unsharp mask images. For all 26 galaxies
reliable barlens decompositions were found.
Our final samples are:

\vskip 0.10cm

\noindent (1) CALIFA sample of (N=1064; unsharp masks)
\vskip 0.10cm

\noindent (2) All barlenses (N=110+9 uncertain)
\vskip 0.10cm

\hskip 0.1cm (2.1) 46 bl-galaxies (new decompositions)
\vskip 0.10cm

\hskip 0.1cm  (2.2) 26 bl-galaxies (analysis of IFU data-cubes).
\vskip 0.10cm

\noindent (3) All X-shaped galaxies (N=124+3 uncertain)

\begin{figure}
\centering
\includegraphics[angle=0,width=8.5cm]{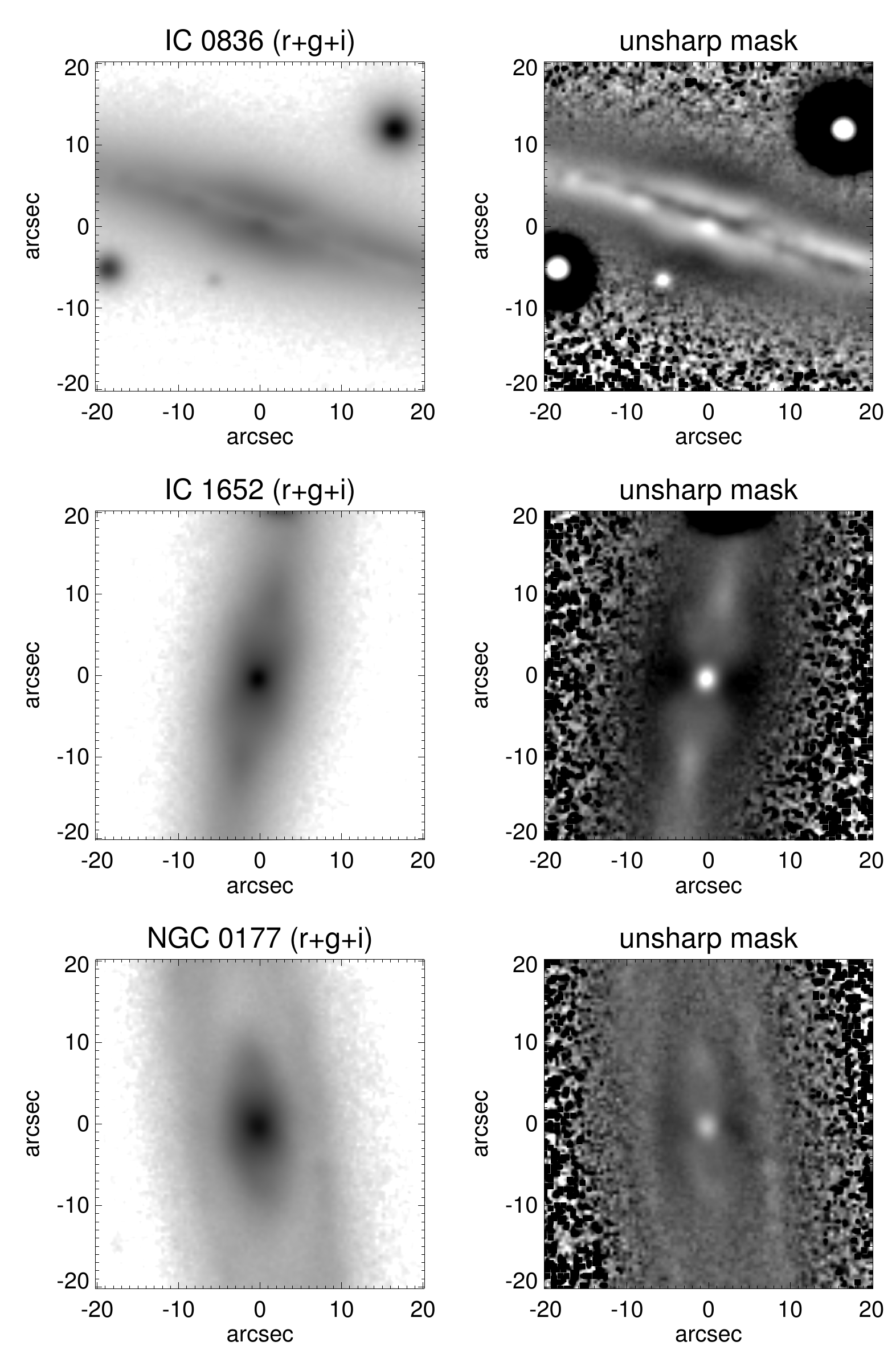}
\caption{Examples of the X-shaped bars. {\it Left panels} show 
the bar regions of the combined r'+g'+i' SDSS mosaic images.
{\it Right panels} show the unsharp mask images made for the r'-band mosaics, using the same image cuts.}
\label{unsharp-X}
\end{figure}

\begin{figure}
\centering
\includegraphics[angle=0,width=9.0cm]{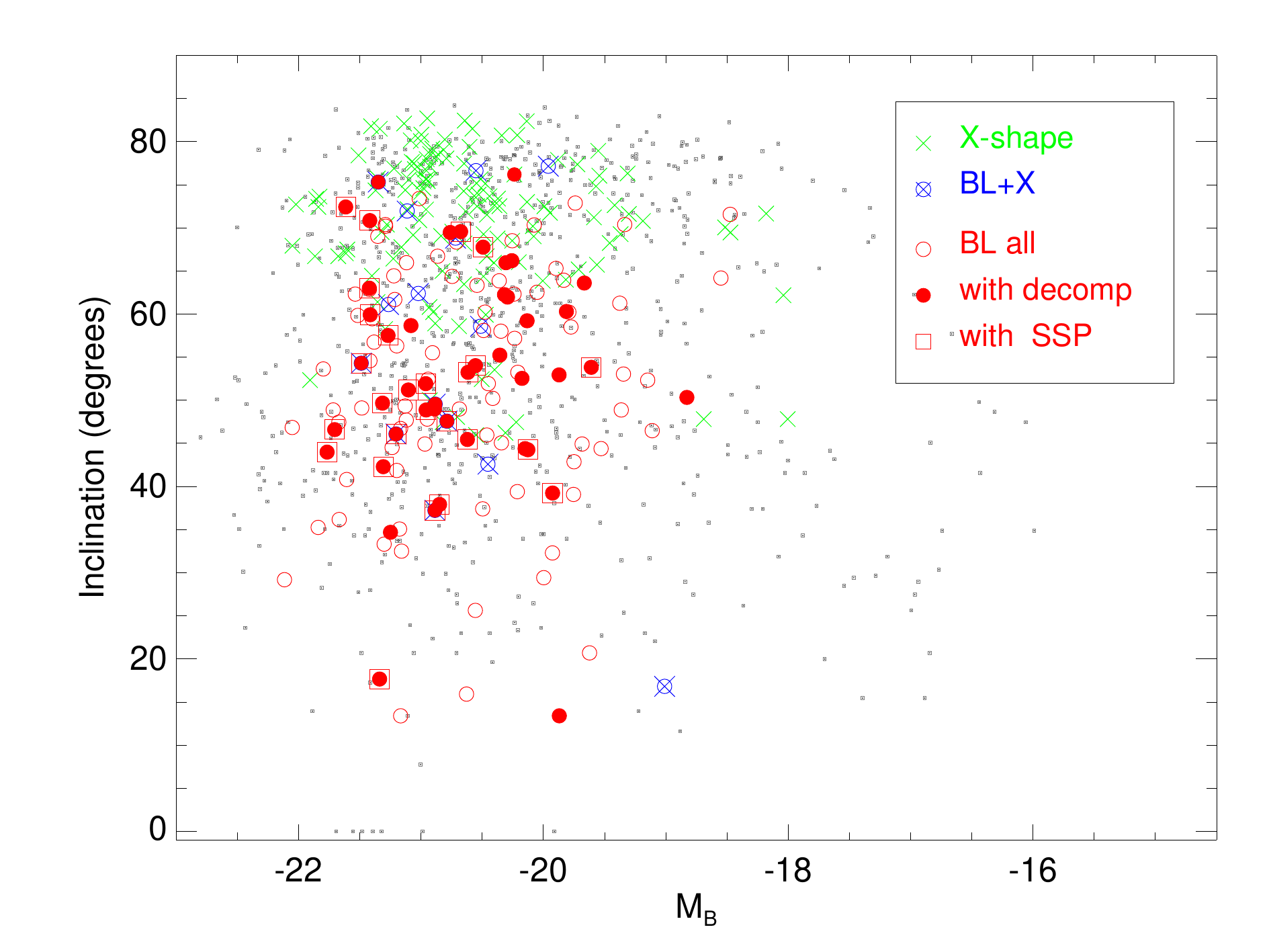}
\caption{The galaxy inclination plotted as a function of the absolute
  B-band galaxy magnitude. The parameter values are from HyperLEDA.
  {\it Grey dots} are for the galaxies in the complete CALIFA sample of 1064
  galaxies, {\it green symbols} show the X-shaped bars, {\it red symbols} the
  barlenses, and {\it blue symbols} the galaxies in which both
  features appeared.
  {\it Filled red circles} denote the barlens galaxies for which decompositions were made
  by us, and {\it open squares} the 26 galaxies, for which also the V500 grating SSP
  data-cubes were available. }
\label{blX-INC_MB}
\end{figure}


\begin{table*}
\caption{The decomposed 46 barlens galaxies. Shown are the main
  parameters of the bulges (columns 4-6) and barlenses (columns 7-9)
  in our decompositions, and those of the bulges by MA2017 (columns
  10-12). The type of model is given in column 3: B = bulge, bl =
  barlens, D = disk, bar = bar, L = outer lens, N = unresolved
  nucleus. The Hubble stage T is taken from CALIFA DR3
  \citep{sanchezGZ2016}.  Marked with (X) in columns (3) are the
  galaxies having also an X-shape feature, and with * in column (1)
  the galaxies for which we analyzed also the stellar populations. The
  effective radius (R$_{\rm e}$) is given in [kpc] using the distance
  from Nasa/IPAC Extragalactic Database. Of the given distances the
  mean values were used (H$_{\circ}$ = 75 km/sec/Mpc). If the distance
  was not given it was calculated from the redshift. Due to the high galaxy inclination
the decomposition for IC 1755 is uncertain.}
\label{table-1}      
\centering          
\begin{tabular}{llllllllllll}
\hline\hline    
 \noalign{\smallskip}
         &    &         & Bulge &      &            & bl      &      &             &  MA2017      &      &            \\ 
Galaxy   & T  & model   & $B/T$& $n$  & $R_e$ [kpc]& $bl/T$ & $n$ & $R_e$ [kpc] &  $B/T$    & $n$  & $R_e$ [kpc]  \\%
 (1)    &(2) &  (3)    & (4)  & (5)  & (6)        & (7)   & (8)  & (9)        & (10)    & (11) & (12)        \\%
\noalign{\smallskip}
\hline
\noalign{\smallskip}
IC 0674*  & 2  &B/D/D/bar/bl &0.08  &1.6   &0.43       &0.14   &0.5   &2.294        &   0.28     &3.6 & 2.321  \\
IC 1199*  & 3  &D/bar/bl   &  -    & -     & -           &0.07   & 2.2  &0.544        &   0.09     &1.9 & 1.100  \\
IC 1755  & 3   &D/bar/bl    &-     &-     &-           &0.11   &1.1   &1.117        &   -        &-   &    -      \\
IC 4566*  & 3  &D/bar/bl   & -    & -    &-           &0.09   & 1.5  &0.652        &  0.09     &1.4 & 0.713  \\
NGC 0036* & 3  &B/D/bar/bl  &0.02  &1.0   &0.11       &0.10   &0.8   &1.618        &  0.24     &4.3 & 3.01 \\
NGC 0171* & 3  &B/D/bar/bl  &0.04  &1.0   &0.34       &0.06   &0.4   &1.183        &  0.08     &1.3 & 0.775 \\
NGC 0180* & 3  &D/bar/bl(X)   &-     &-     &-           &0.06   &1.3   &0.544     &  0.05     &1.1 & 0.544 \\
NGC 0364 &-2   &D/bar/bl   &-     &-     &-           &0.21   &0.9   &0.850        &  0.20     &1.4 & 0.871 \\
NGC 0447 & 1   &B/D/bar/bl  & 0.04 &1.6   &0.22       &0.13   &0.9   &1.183        &  0.19     &2.2 & 1.147 \\
NGC 0776* & 3  &N/D/bar/bl  & -    &-     &-           &0.12   &1.4   &0.563        &  0.11     &1.2 & 0.528 \\
NGC 1093 & 4   &D/bar/bl(X)   &-     &-     &-           &0.04   &1.2   &0.306      &  0.16     &1.9 & 1.360  \\
NGC 1645* & 0  &D/bar/bl   &-     &-     &-           &0.19   &1.0   &0.798        &  0.15     &0.9 & 0.589 \\
NGC 2253* &4   &D/bar/bl   &-     &-     &-           &0.07   &2.1   &0.393        &  0.06     &1.1 & 0.292  \\
NGC 2486 &2    &B/D/bar/bl  &0.09  &0.9   &0.32      &0.11   &0.7   &1.493        &  0.21     &2.2 & 0.814 \\
NGC 2487 &3    &B/D/bar/bl  &0.04  &1.6   &0.14       &0.04   &0.5   &0.726        &  0.09     &3.0 & 0.393  \\
NGC 2540 &4    &D/bar/bl   &-     &-     &-           &0.04   &1.6   &0.127        &   -        &-   & -         \\
NGC 2553 &3    &B/D/bar/bl  &0.09  &0.7   &0.22       &0.21   &0.8   &1.163        &  0.24     &1.7 & 0.629  \\
NGC 3300 &0    &B/D/bar/bl  &0.04  &1.1   &0.18       &0.07   &0.5   &0.636        &  0.09     &1.1 & 0.337  \\
NGC 3687 &3    &D/bar/bl   &-     &-     &-           &0.10   &1.5   &0.316        &  0.07     &1.3 & 0.263 \\
NGC 4003* &0   &D/bar/bl   &-     &-     &-           &0.23   &1.4   &1.241        &  0.23     &1.6 & 1.182  \\
NGC 4210* &3   &D/bar/bl   &-     &-     &-           &0.03   &1.2   &0.281        &  0.03     &0.7 & 0.265  \\
NGC 5000* &4   &B/D/bar/bl(X)  &0.03  &0.7   &0.08       &0.03   &0.6   &0.469     &  0.07     &3.9 & 0.348 \\
NGC 5205* &4   &D/bar/bl   &-     &-     &-           &0.07   &0.8   &0.319        &  0.06     &0.9 & 0.262  \\
NGC 5378* &3   &B/D/bar/bl  &0.04  &1.1   &0.15       &0.14   &1.0   &0.944        &  0.20     &2.4 & 0.623  \\
NGC 5406* &3   &B/D/bar/bl  &0.07  &0.8   &0.14       &0.05   &0.4   &0.500        &  0.12     &1.3 & 0.218 \\
NGC 5657 &4    &D/D/bar/bl  &-     &-     &-           &0.09   &0.5   &0.619        &  0.09     &0.6 & 0.532 \\
NGC 5720* &4   &D/bar/bl   &-     &-     &-           &0.04   &1.2   &0.644        &  0.08     &1.1 & 0.559 \\
NGC 5876 &0    &B/D/bar/bl  &0.06  &0.6   &0.19       &0.26   &1.0   &0.851        &  0.29     &1.5 & 0.595 \\
NGC 5947* &4   &B/D/bar/bl  &0.06  &0.7   &0.19       &0.07   &0.7   &0.693        &  0.13     &2.4 & 0.468 \\
NGC 6004* &4   &B/D/bar    & 0.04 &2.7   &0.18       &-      &-     &-            &  0.03     &1.9 & 0.197  \\
NGC 6186 &3    &B/D/bar/bl/L & 0.16 &0.9   &0.54       &0.06   &0.2   &1.295        &  0.20     &1.0 & 0.661  \\
NGC 6278 &0    &D/bar/bl   &-     &-     &-           &0.29   &1.8   &0.373        &  0.34     &2.4 & 0.492  \\
NGC 6497* &2   &B/D/bar/bl  &0.05  &1.4   &0.18       &0.20   &1.0   &1.478        &  0.26     &3.1 & 1.103 \\
NGC 6941* &3   &D/bar/bl(X)   &-     &-     &-           &0.11   &1.8   &0.941     &  0.09     &1.6 & 0.828 \\
NGC 6945 &-1   &B/D/bar/bl  &0.09  &0.9   &0.20       &0.12   &0.5   &1.018        &   0.25     &3.7 & 0.787   \\
NGC 7321* &4   &D/bar/bl   &-     &-     &-           &0.04   &1.6   &0.384        &  0.05     &1.5 & 0.489 \\
NGC 7563* &1   &B/D/bar/bl  &0.09  &1.5   &0.24       &0.31   &0.9   &1.276        &  0.53     &2.1 & 1.079 \\
NGC 7611 &-3   &D/bar/bl   &-     &-     &-           &0.23   &1.0   &0.283        &  0.23     &1.5 & 0.284 \\
NGC 7623 &-2   &D/bar/bl   &-     &-     &-           &0.31   &2.0   &0.962        &  0.47     &1.9 & 1.262 \\
NGC 7738* &3   &B/D/D/bar/bl &0.05  &0.5   &0.35       &0.19   &0.9   &2.543        &  0.15     &1.5 & 1.158  \\
NGC 7824 &2    &B/D/bar/bl  &0.03  &2.7   &-           &0.24   &0.9   &2.156        &  0.42     &2.3 & 1.794  \\
UGC 01271&0    &B/D/bar/bl  &0.06  &0.5   &0.19       &0.20   &1.0   &1.026        &  0.18     &1.3 & 0.559  \\
UGC 03253*&3   &D/bar/bl   & -     & -     & -           &0.09   &1.6   &0.536        &   0.07     &1.1 & 0.389 \\
UGC 08781*&3   &D/bar/bl(X)   &-     &-     &-           &0.16   &1.8   &0.756      &  0.16     &2.0 & 0.854  \\
UGC 10811*&3   &D/bar/bl   &-     &-     &-           &0.10   &1.5   &0.921        &  0.06     &0.7 & 0.537 \\
UGC 12185& 3   &D/bar/bl(X)   &-     &-     &-           &0.14   &2.5   &0.575        &  0.19     &2.8 & 0.980  \\
\noalign{\smallskip}
\hline
   \end{tabular}
\end{table*}               

 \section{Identification of structures and measuring the sizes of bars and barlenses}

For identification of the X-shape features unsharp mask images of the
r'-band mosaic images were made for the complete CALIFA sample of 1064
galaxies. The way how the mosaics were made is explained in
Appendix A.  For making the unsharp masks the images were
convolved with a Gaussian kernel, and the original images were divided
by the convolved images. A few prototypical X-shape galaxies were
selected, and used to find the optimal parameters for the Gaussian
kernel. The galaxies were individually checked, and if needed, the
convolution process was repeated many times with different kernel
sizes and contrast levels of the images. Having in mind that the
r'-band mosaics are strongly affected by dust obscuration particularly in
the edge-on view, identification of the X-shape feature was accepted
even if it appeared only in one side of the galaxy. All the
  unsharp mask images of the X-shaped bars are shown in the web page {\tt http://www.oulu.fi/astronomy/CALIFA\_BARLENSES}, of which
representative examples are shown in Figure \ref{unsharp-X}.
Although barlenses were primarily identified visually in the original
mosaic images, attention was paid also to the barlens morphologies in
the unsharp mask images.

In this work the sizes and minor-to-major (b/a) axis ratios of barlenses were measured to be used as
  auxiliary data for making reliable structure decompositions.  They
were measured by fitting ellipses to the visually identified
outer edges of barlenses in the original mosaic images, in a similar
interactive manner as in \citet{herrera2017}.  The barlens regions
superimposed with the bar were excluded from the fit.  From the fitted
ellipses the major ($a$) and minor ($b$) axis lengths and position
angles ($PA$) of the barlenses were measured. It was shown by
\citet{herrera2017} that this visual method is as good as if the outer
isophotes of barlenses were followed instead.  The measurements are
shown in Table \ref{appendixD}.  In the table barlengths are also
given, which were visually estimated from the r'-band mosaic images,
by marking the bar ends in the deprojected images. Shown also are the
orientation parameters of the disks, estimated from the outer
isophotes of the galaxies, as in \citet{salo2015}.

The galaxy inclinations of the complete CALIFA sample are shown in
Figure \ref{blX-INC_MB}, plotted as a function of the absolute B-band
galaxy magnitude. On top of those over-plotted are all barlens and
X-shaped galaxies, and the two sub-samples of barlens galaxies. Shown
separately are also the 15 barlens galaxies with X-shape features.  It
appears that the galaxies in our sub-samples are fairly randomly
distributed in magnitude, which means that they are representative
examples of the complete CALIFA sample.  The galaxies with barlenses
and X-shape features have also similar magnitude distributions. 
  Notice that for consistency, in Figure 3  values from HyperLEDA were used for all galaxies, including those for which we made our own measurements.

\section{Multi-component decompositions}

\begin{figure*}
\centering
\includegraphics[angle=0,width=18.0cm]{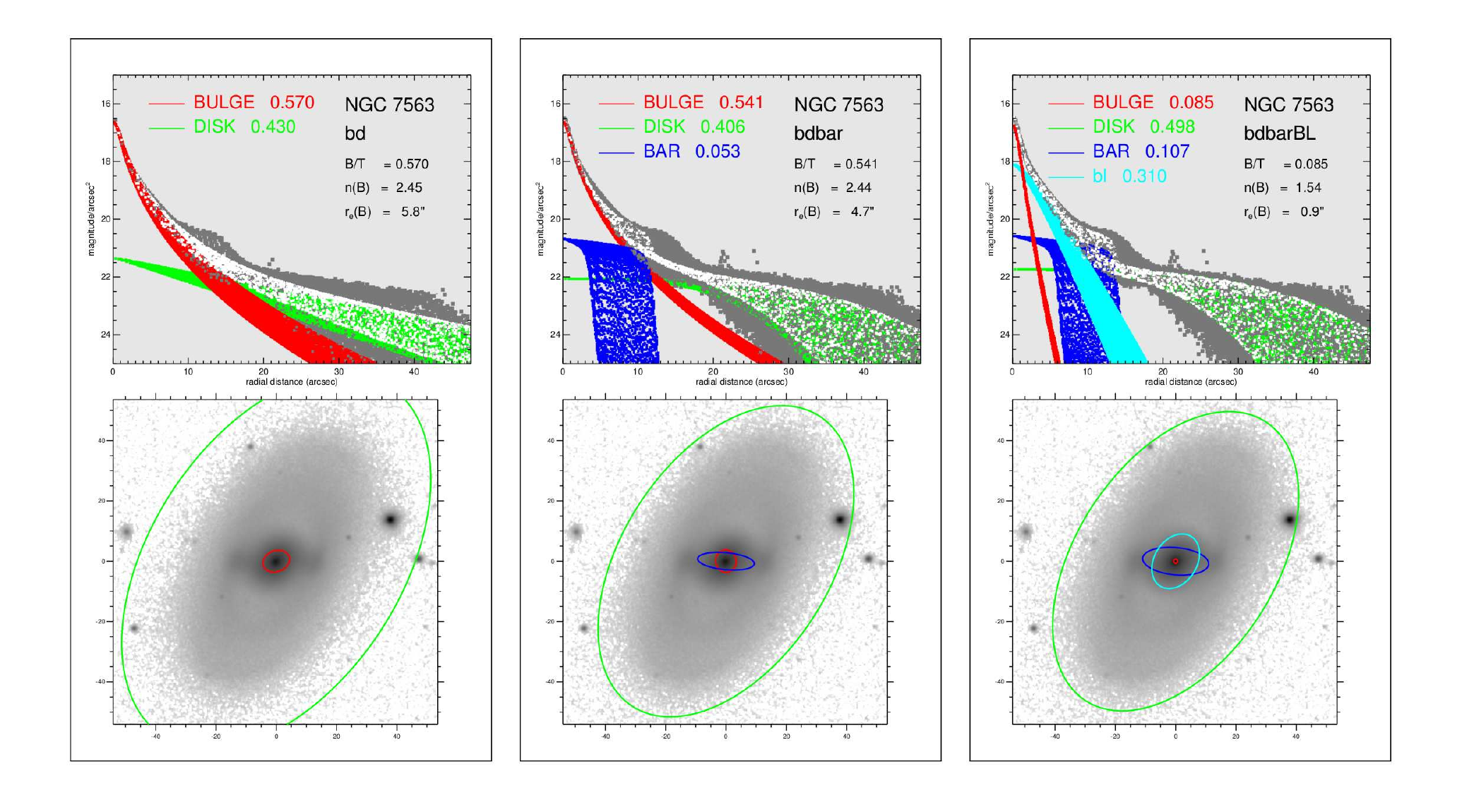}
 \vskip -0.5cm  
\includegraphics[angle=0,width=18.0cm]{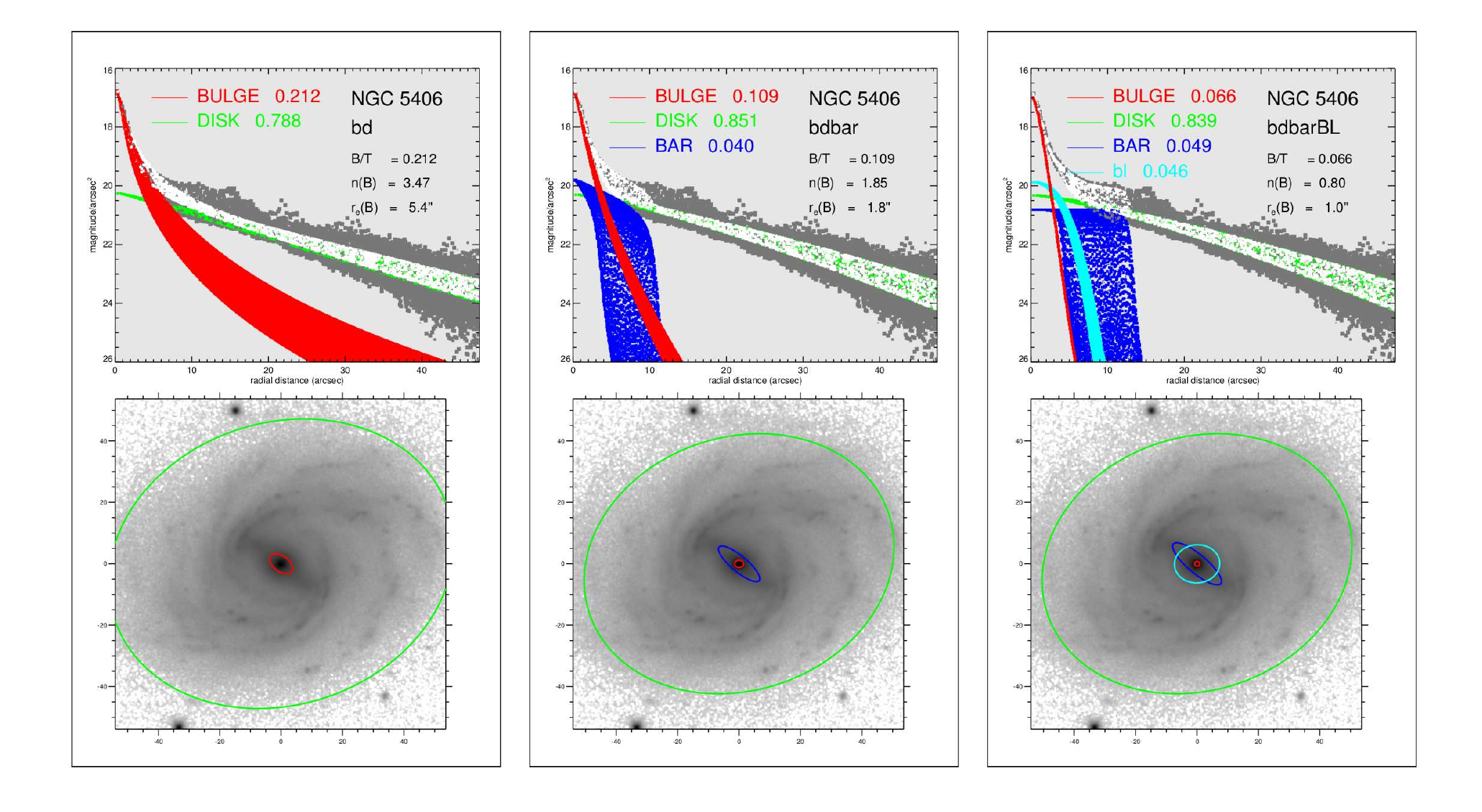}
 \vskip -0.5cm 
\caption{Three decomposition models (B/D, B/D/bar, and B/D/bar/bl) are
  shown for the galaxies NGC 7563 and NGC 5406.  In the {\it upper
    panels} the 2D representations of the surface brightness profiles
  are shown: black dots show the fluxes of the image pixels as a function of distance in the sky plane, and the
 white dots those of the total model images. The colors illustrate the
  fitted models of the different structure components. The {\it lower
    panels} show the r'-band mosaic images: on top of them plotted are
  the effective radii of the fitted models.}
\label{N7563-N5406}
\end{figure*}

\subsection{The method and model functions}

We use the GALFIT code \citep{peng2010} and the GALFIDL software
\citep{salo2015} to decompose the 2D light distributions of the
galaxies to different structure components. Our decomposition strategy
is described in detail by \citet{salo2015}. The Levenberg-Marquardt
algorithm is used to minimize the weighted residual ${\chi^2}_{\nu}$
between the observed and model images. The full model image consists
of the models of the different structure components, each convolved
with the image Point Spread Function (PSF). The SDSS r'-band mosaic
images, with the resolution of 0.396 arcsec/pix, were used.  For each
science frame a mask and a sigma-image mosaic were constructed. The
$\sigma$ image was used to control the weight of pixels in the
decomposition.  The PSF was made in such a manner that the extended
tail beyond the Gaussian core was taken into account.  The PSF FWHM =
0.8 -- 1.4 arcsec, in good agreement with that obtained also by
MA2017. Preparation of the data for the decompositions is explained in
more detail in Appendix \ref{appendixA}.
 
In GALFIT the isophotal shapes of the model components are defined with generalized ellipses
(Athanassoula et al. 1990):
\begin{equation}
r(x',y') = \left(|{x'-x_0}|^{C+2} + \left|\frac{y'-y_0}{q}\right|^{C+2}\right)^{\frac{1}{C+2\phantom{^2}}} .
\end{equation}
\noindent Here $x_0,y_0$ defines the center of the isophote, $x',y'$ denote
coordinates in a system aligned with the isophotal major axis
pointing at the position angle $PA$, and $q = b/a$ is the minor-to-major
axis ratio. The shape parameter $C$ = 0 for pure ellipses, for $C > $
the isophote is boxy, and for $C < $ 0 it is disky. Circular
isophotes correspond to $C$ = 0 and $q$ = 1. The x-axis is along the
apparent major axis of the component. The galaxy center is taken to be
the same for all components. For the radial surface brightness distribution
a S\'ersic function was used for the bulges, barlenses, and disks:
\begin{equation}
\Sigma(r)= \Sigma_{\rm e} \exp \left(-\kappa \left[ (r/R_{\rm e})^{1/n} -1\right]\right),
\end{equation}
\noindent where $\Sigma_{\rm e}$ is the surface brightness at the effective
radius $R_{\rm e}$ (isophotal radius encompassing half of the total flux of
the component).  The S\'ersic index $n$ describes the shape of the
radial profile. The factor $\kappa$ is a
normalization constant determined by $n$.  
The value $n$ = 4 corresponds to the R$^{1/4}$ law,
and $n$ = 1 to an exponential function.
For a bar a modified Ferrers function was used:
\begin{equation}
  \Sigma(r)= 
  \left\{\begin{array}{ll} 
  \Sigma_o \left[1- (r/r_{\rm out})^{2-\beta}\right]^\alpha  & r < r_{\rm out}\\
  0  & r \ge r_{\rm out}
  \end{array} \right.
\end{equation}
\noindent The outer edge of the profile is defined by $r_{\rm out}$,
$\alpha$ defines the sharpness of the outer cut, and the parameter
$\beta$ defines the central slope. $\Sigma_{\circ}$ is the central
surface brightness.  When an unresolved component was identified in
the galaxy center, it was fit with a PSF-convolved point source.

%
\begin{figure}
\centering
\includegraphics[angle=0,width=9.0cm,height=10cm]{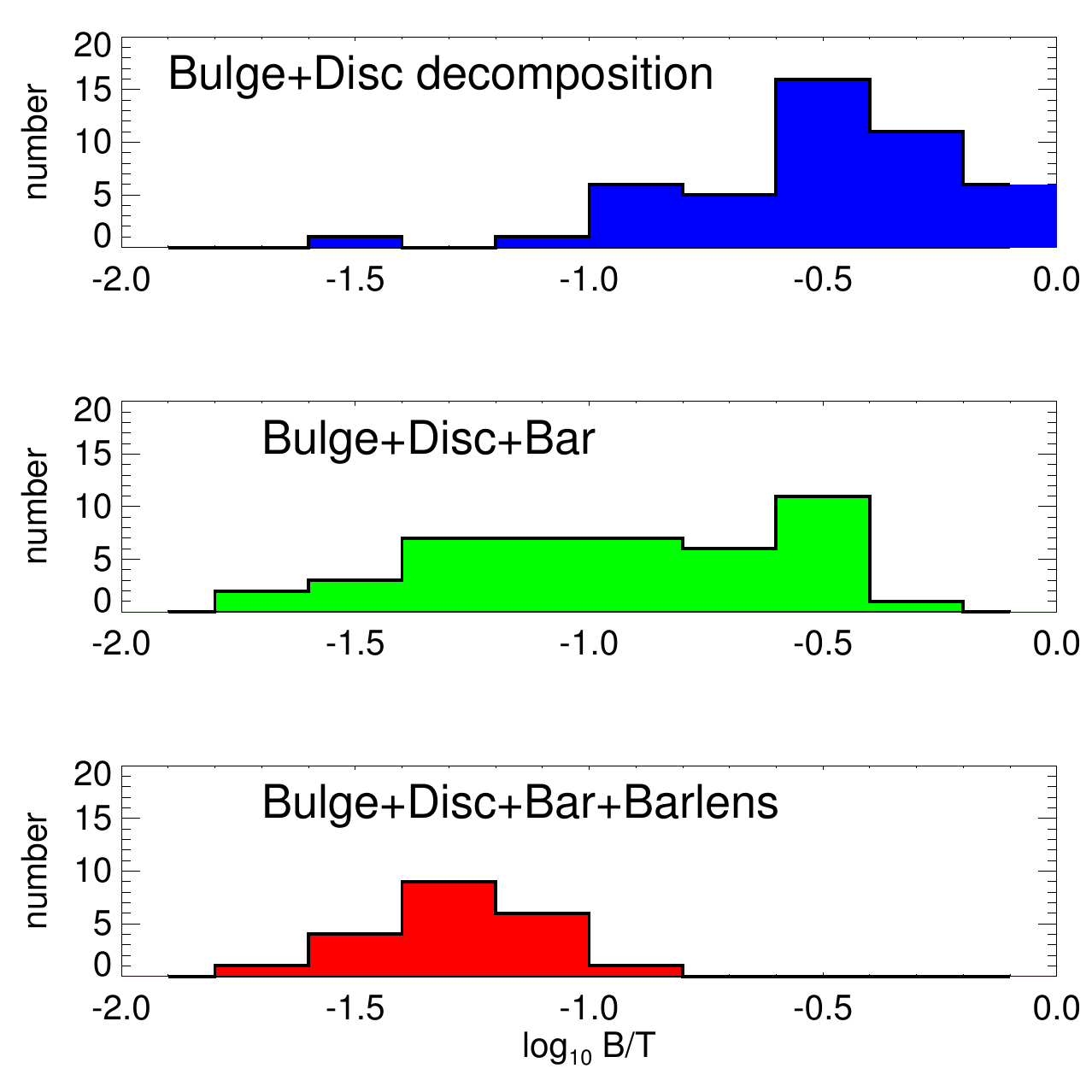}
\caption{The bulge-to-total ($B/T$) flux ratios for the 46 galaxies
  decomposed in this study. The values in three type of models are
  shown: bulge/disk (B/D, {\it upper panel}), bulge/disk/bar (B/D/bar,
  {\it middle panel}) and bulge/disk/bar/bl (B/D/bar/bl, {\it lower
    panel}). The $B/T$-values for the individual galaxies are shown in
  Table 1. Compared to the B/D and B/D/bar models, the number of the B/D/bar/bl models is smaller,
because only half of the decomposed galaxies were fitted with a separate bulge component.
}
\label{BD-BDbar-BDbarBL}
\end{figure}

\subsection{Our fitting procedure}

Our main goal is to extract barlenses from the other structure
components.  We started with single S\'ersic fits in order to
highlight possible low contrast features in the residual images,
obtained by subtracting the model from the original image. For
detecting these features the unsharp mask images were also
useful. Then bulge/disk (B/D) decompositions were made to find the
initial estimate of the parameters of the disk, and also to have an
approximation of the flux on top of the disk. In all of our
decompositions the galaxy centers were fixed. Also the orientation
parameters of the disk were fixed to those corresponding to the outer
galaxy isophotes (see Table C.0). After the initial single S\'ersic 
and bulge-disk decompositions an iterative process was
started, in order to share the light above the disk between bars,
barlenses, and possible separate bulge components (B/D/bar/bl
models). Since the same function was used for the bulges and
barlenses, care is needed to avoid possible degeneracy between the two
components. Therefore, the starting values were selected to be as
close to observation as possible.
The parameters of the disk were kept fixed until good first
approximations for all the other structure components were found. Once
found, releasing the disk parameters in the fitting process 
typically did not change much the parameters of the other components.
Only two galaxies in our sample have nuclear bars or rings visible in the direct images.

\begin{figure}
\centering
\includegraphics[angle=0,width=9.0cm]{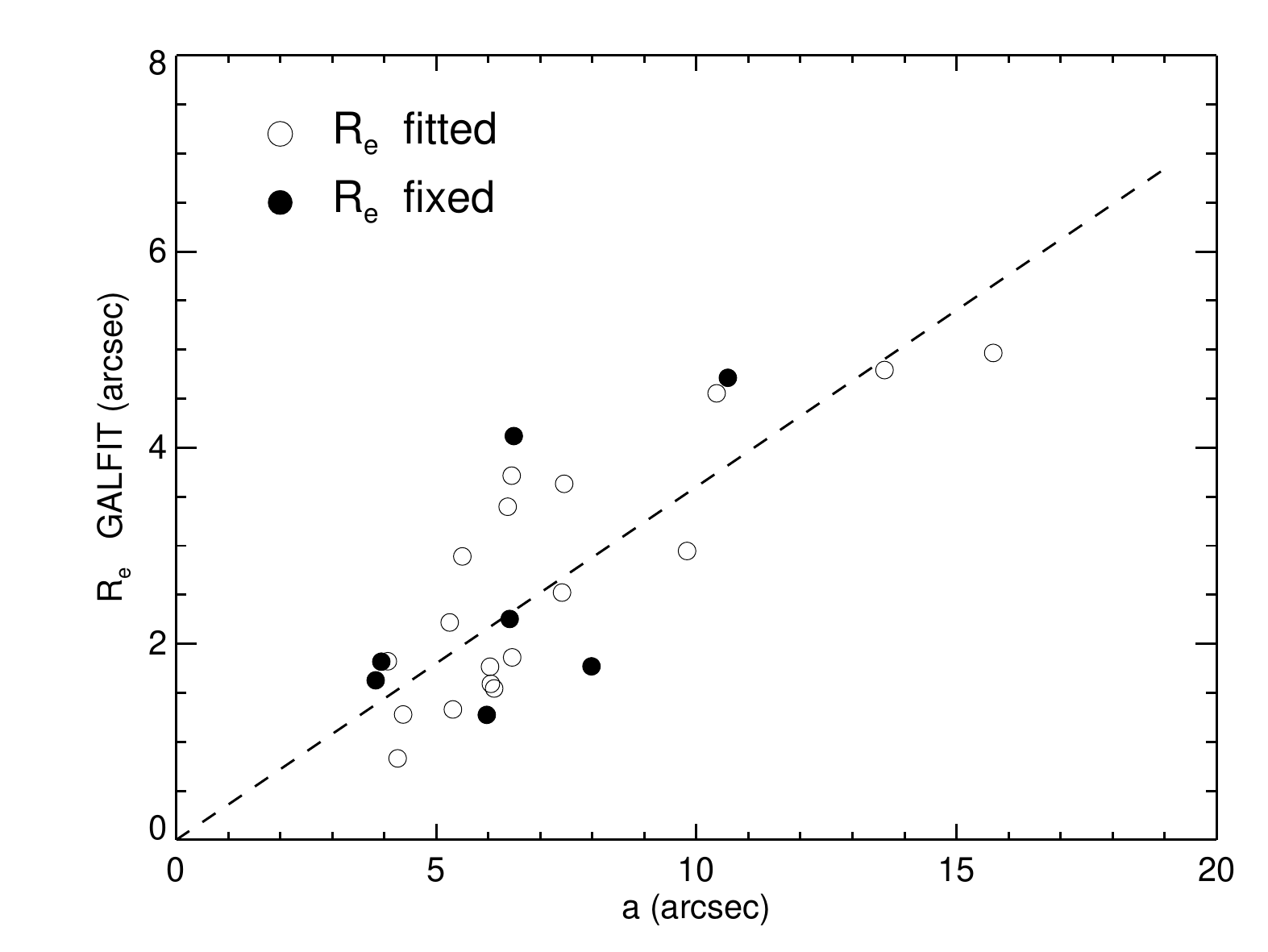}
\caption{The relation between the visually estimated barlens size 
    (semimajor-axis of the fitted ellipse) and the barlens effective
  radius that comes out in our decompositions.  Dashed line shows the
  relation $R_{\rm e}$=0.36a, obtained by a linear fit to the values
  obtained in the decompositions where barlens $R_{\rm e}$ was a free parameter.
  Linear correlation coefficient between $R_{\rm e}$ and $a$ is 0.83.} 
\label{blsizecom}
\end{figure}

\begin{figure}
\centering
\includegraphics[angle=0,width=9.0cm]{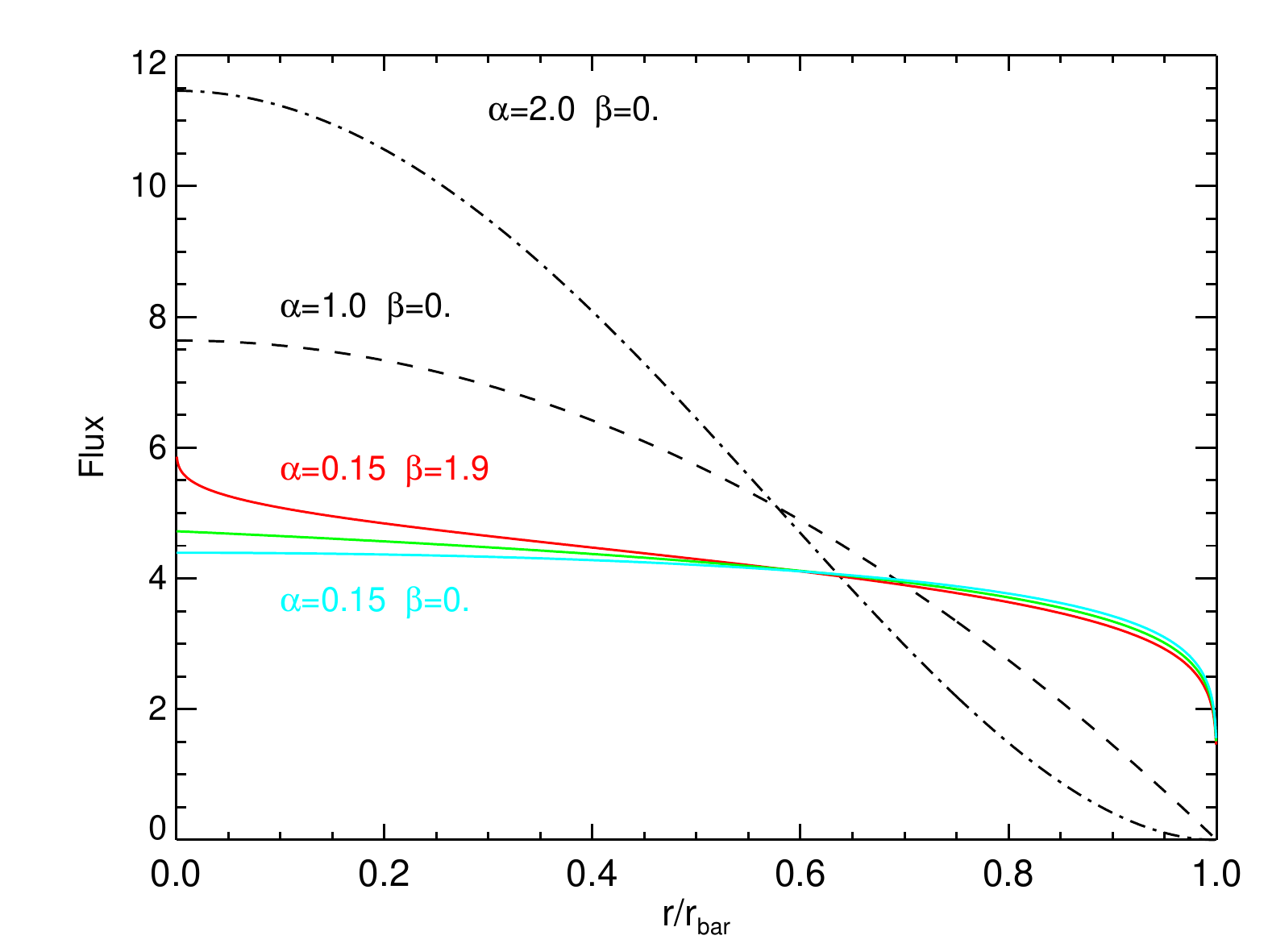}
\caption{The surface brightness profiles for the theoretical Ferrers function 
using different values for the parameters $\alpha$ and $\beta$, related to the sharpness of the outer truncation and the central slope of the profile, respectively. 
Lines indicate different combinations of $\alpha$ and $\beta$:
the red line shows the typical values obtained in the current study. 
The different curves are normalized to correspond to the same total bar flux.}
\label{ferrer}
\end{figure}

In order to avoid degeneracy between the structure components, and
also to reduce the number of free parameters in GALFIT, we utilized
the measured sizes and shapes of bars and barlenses (see Section 4)
when choosing the initial values in the decompositions. In practice,
when using a S\'ersic model for the barlens this means adjusting the
$R_e$ so that the modeled barlens has similar outer isophote size as
the visual estimate ($a$). In some cases, we had to fix the barlens
$R_e$. Moreover, their axial ratio was always fixed to the measured
$b/a$-ratio. Namely, the fact that the barlens flux is superimposed with
that of the bar means that the barlens model, if completely free, can
easily become artificially elongated along the bar major axis.  Figure
\ref{blsizecom} displays the relation between the visually estimated
size of the barlens and the size that comes out from our
decompositions. A linear fit for galaxies where the barlens size was
left as a free parameter indicates $R_e \approx 0.36 a$. No large
deviations from this trend are visible even in the cases where the
barlens size was fixed.

We followed an approach in which any of the parameters of the
structure components can be temporarily fixed, until good starting
parameters were found.  For evaluating our best fitting model a human
supervision was important. In particular, we compared the observed and
model images, the observed and fitted surface brightness profiles (1D
and 2D), as well as the residual images after subtracting the model
from the observed image.

\subsection{How to avoid degeneracy between the fitted components}
   
It is well known that in the structure decompositions the main source
of uncertainty is the choice of the fitted components, and possible
degeneracy of the flux between those components, and not the formal
errors given by the $\chi^2$ minimization.
\footnote{There are also other uncertainties in the decomposition, related to 
sky subtraction, and to $\sigma$ and PSF-images \citep{salo2015}. 
Such errors for the SDSS r'-band mosaic images in the MANGA (Mapping
Nearby Galaxies at Apache Point Observatory) galaxy sample have been estimated
 by Laine et al. (private communication). 
 Due to the PSF and $\sigma_{\rm sky}$ they are $\sim$5$\%$ on $B/T$ and
  $R_{\rm e}$(bulge), and due to the PSF $\sim$15$\%$ on S\'ersic $n$.}.  

The most important factor is how many physically meaningful components are fitted, which is
illustrated in Figure \ref{N7563-N5406}. Shown are the B/D, B/D/bar, and
B/D/bar/bl models for the galaxies NGC 7563 and NGC 5406. The B/D
models for both galaxies fail to re-cover anything which could be
called as a real bulge. Including the bar improves the fit considerably, in
agreement with many previous studies
\citep{lauri2005,lauri2010,gadotti2008,salo2015}. How much the bar
improves the fit depends on the surface brightness profile: for
galaxies with small photometric bulges (NGC 5406) the B/D/bar model
works quite well, but if the photometric bulge is prominent and the
profile is centrally peaked (NGC 7563), GALFIT tries to fit a massive
bulge with a high S\'ersic index. However, the B/D/bar/bl model for
NGC 7563 fits the surface brightness profile more accurately. Most
importantly, the model is consistent with what we see in the image:
the galaxy outside the bar is not dominated by a large spheroidal, but
rather by a dispersed ring with a down-bending surface brightness
profile at the outer edge.  The way how the number of the fitted
parameters affects $B/T$ in the sample decomposed in this study is
summarized in Figure \ref{BD-BDbar-BDbarBL}: an average $B/T$-value in
the B/D-models is $\sim$0.30, in the B/D/bar models it is $\sim$0.15,
and in the B/D/bar/bl-models $\sim$0.06.

Our attempt to handle the bulge/barlens degeneracy was that two
S\'ersic functions were used only when two clear sub-sections with
different slopes appear in the central surface brightness profile. The
bar/barlens degeneracy is reduced by using different fitting functions
for them: Ferrers function for the bar, and S\'ersic function for the
barlens.  However, the bar/barlens separation worked well only if we
did not fix the profile shape parameters of the bar ($\alpha$ and
$\beta$) as is usually done.  In the literature most bar
decompositions have been done with fixed $\alpha \ge$ 2 (i.e. see Salo
et al. 2015; MA2017), whereas in the current study the final models
typically adjusted $\alpha$ to values close to zero (on average
$\alpha \sim$ 0.15; for $\alpha$ close to zero the value of $\beta$
has less significance, see Fig.  \ref{ferrer}). In our decompositions
the bar is quite flat and sharply truncated at the outer edge. We
  further tested how much the small central light concentration
  introduced by a positive $\beta$ value in Ferrers function can
  affect $B/T$. Changing $\beta$ from 1.9 to 0 in the decomposition for
  NGC 7563 changes $B/T$ only from 0.085 to 0.081, and in IC 1755 from
  0.104 to 0.107, which means that the $B/T$ using both $\beta$ values are
  practically the same.  Clearly, the large values of $\alpha$ used
in earlier decompositions stem from the omission of the central
barlens component.

\begin{figure*}
\centering
\includegraphics[angle=0,width=18.0cm]{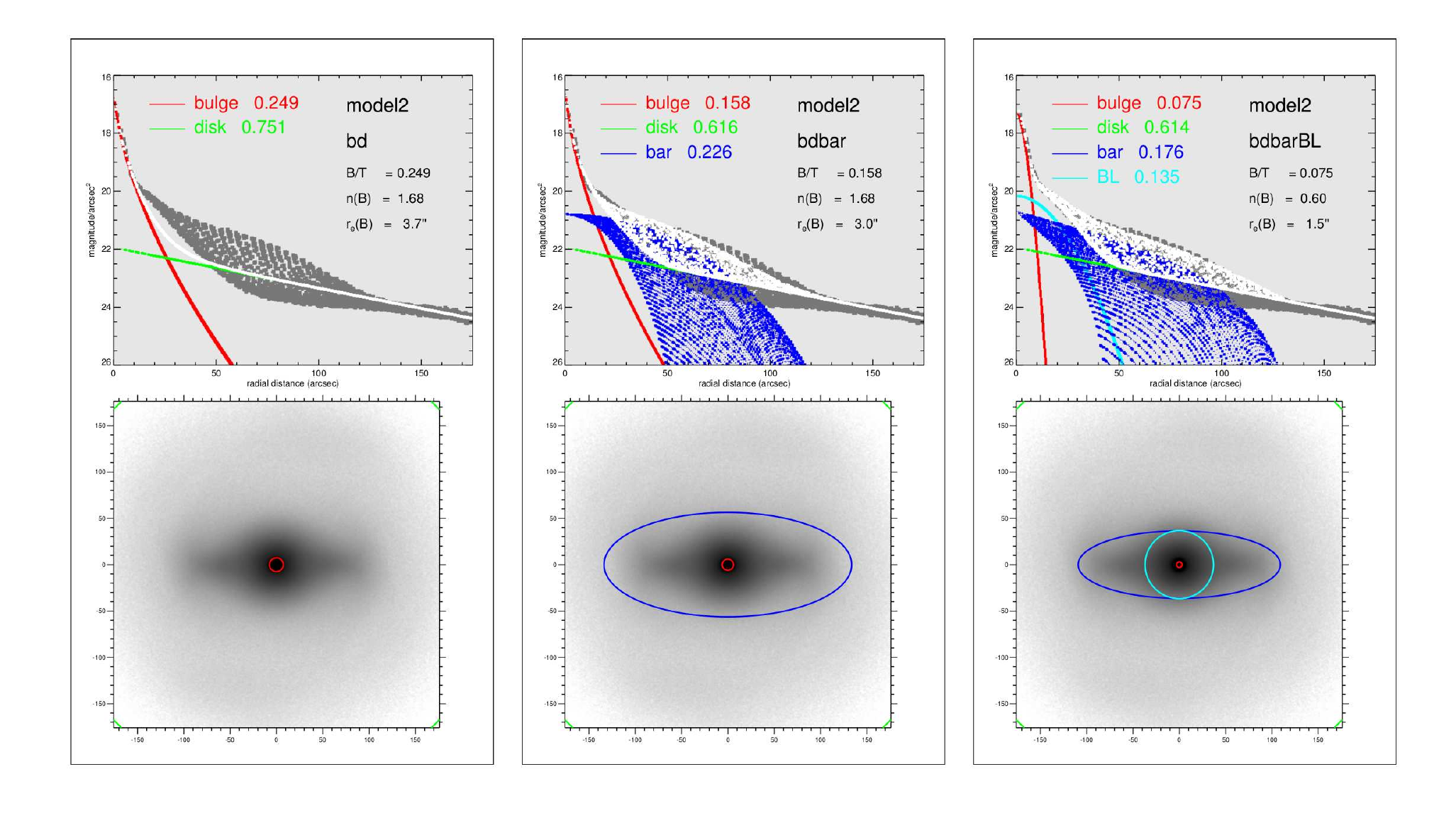}
\caption{Three decompositions (B/D, B/D/bar and B/D/bar/bl) for
  a simulation snapshot taken from  Salo $\&$
  Laurikainen (2017).  The model is explained in the text. The meaning of
  the lines and symbols are the same as in Figure \ref{N7563-N5406}. }
\label{simul}
\end{figure*}

\begin{figure}
\centering
\includegraphics[angle=0,width=9.0cm]{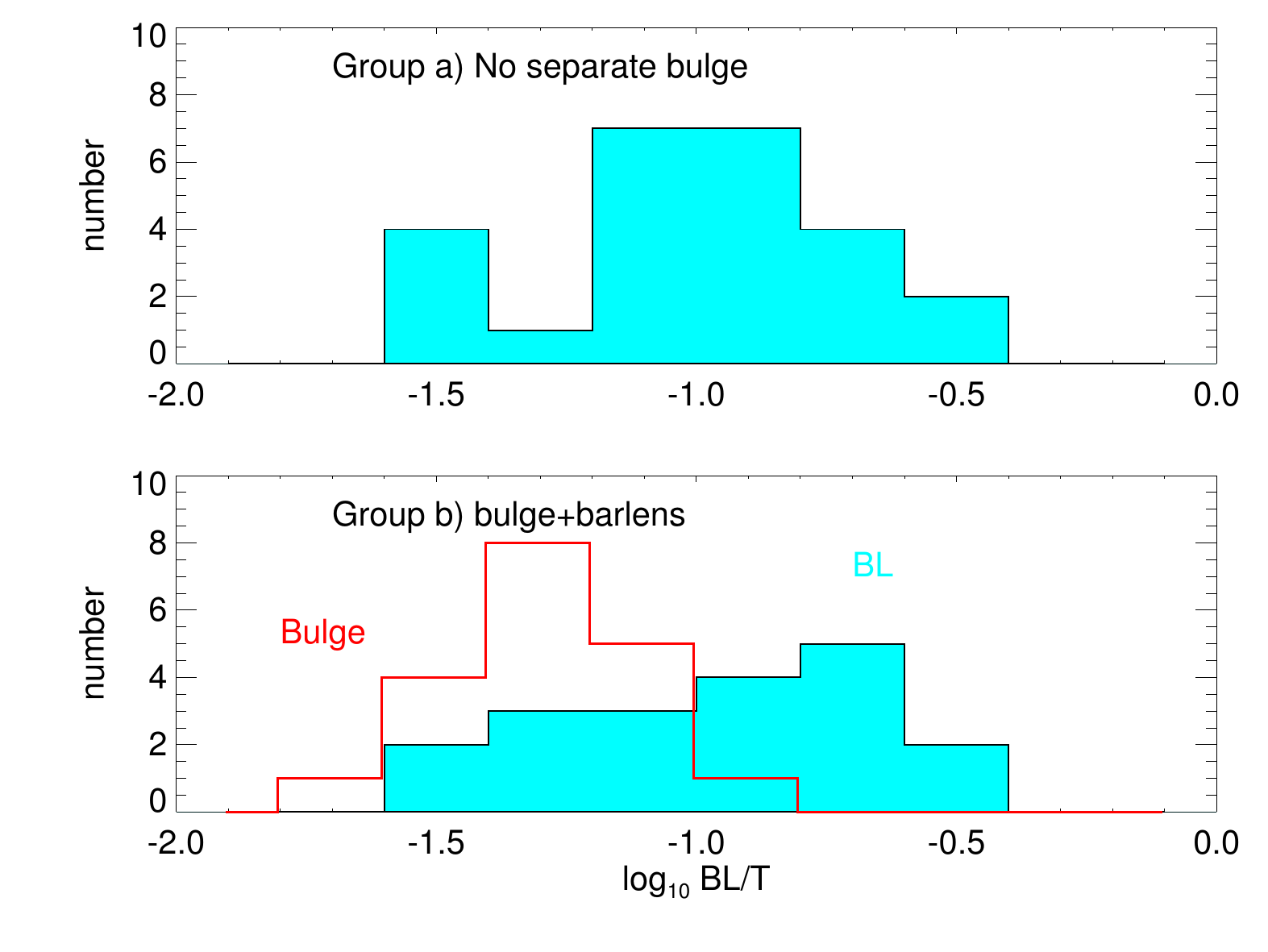}
\caption{The 26 barlens galaxies are divided to groups (a) ({\it upper panel}) and (b) ({\it lower panel}) as explained
in Section 6. The blue histograms indicate the barlens flux fraction $BL/T$.
Additionally, the red histogram (same as in Fig. 5, lowermost frame) in the lower frame shows the relative fluxes of the
separate bulge components in the group (b) galaxies.}
\label{groups_a_b}
\end{figure}


\subsection{Decompositions for a synthetic simulation image}

Similar decompositions as those shown for the observed galaxies in
Figure \ref{N7563-N5406}, were made also for a simulation snapshot,
taken from the N-body simulation model by \citet{salo2017}. These are
stellar dynamical models with no gas or star formation, carried out
with Gadget-2 \citep{springel2005} and using self-consistent initial
galaxy models. For the disc component $5 \cdot 10^6$ particles were
used, and in order to have good enough resolution a gravity softening
$\epsilon$ = 0.01 $h_{\rm R}$ was used, where $h_{\rm R}$ is the scale length of
the disk.  The model mimics a typical Milky Way galaxy, with stellar
mass of M$_{\star}$/M$_{\odot} = 5 \cdot 10^{10}$, a small pre-existing bulge
($B/T$ = 0.075, $R_{\rm e}$ = 0.07 $h_{\rm R}$), an exponential disk,
and a spherical halo. The decomposed snapshot (same as used in Fig. 1) was
taken 3 Gyrs after the bar was formed and then stabilized. The model
developed a vertically thick inner bar component, which is X-shaped in
edge-on view (see the lowest panel in Fig. \ref{synthetic-models}),
and a barlens morphology in more face-on view. During the simulation
the bulge changed very little.

Our decompositions are shown in Figure \ref{simul}. It appears that
the original bulge light fraction $B/T$ is over-estimated both in the
B/D and B/D/bar decompositions (these yield $B/T$=0.25 and $B/T$ =
0.16, respectively), whereas the B/D/bar/bl decomposition recovers
very well the small original bulge ($B/T$ = 0.075) and the particles
representing the barlens. The contribution of the bar is slightly
over-estimated in the B/D/bar model.  The morphology and surface
brightness profile of the simulated barlens galaxy is remarkably
similar to those of the galaxies NGC 7563 and NGC 5406.
 \vskip 0.20cm 

Representative examples of the decompositions for the barlens galaxies
are shown in Figures \ref{decompositions_1}, \ref{decompositions_2},
and \ref{decompositions_3}. The output decomposition files with the parameters
of the different components of all the decomposed galaxies are shown in the
  web page {\tt http://www.oulu.fi/astronomy/CALIFA\_BARLENSES}.

\section{Comparison with MA2017}

\begin{figure*}
\centering
\vskip -5.0cm
\includegraphics[angle=0,width=19.0cm]{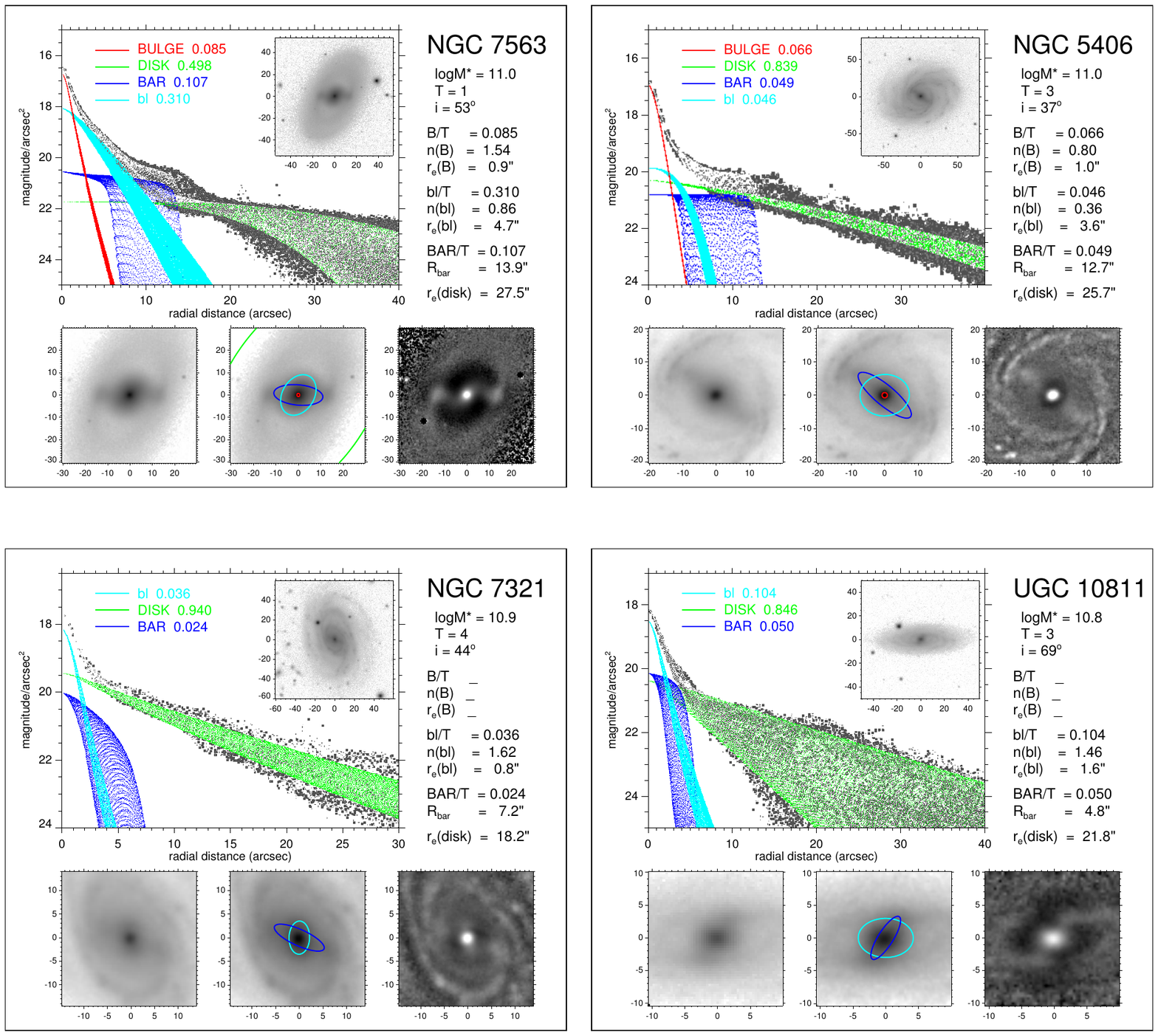}
\vskip -5.0cm
\caption{Examples of our multi-component decompositions. {\it Large
    panel}: black dots show the surface brightnesses (r'
  magnitude/arcsec$^2$) of the pixels in the two-dimensional image,
  white dots the values in the final model image, and the colors the
  values corresponding to the different structure components of the
  model. Bulges, disks and barlenses were fitted with a S\'ersic
  function, and bars with a Ferrers function.  The decomposition
  parameters, galaxy masses (logM$_{\star}$), galaxy inclinations (i)
  and Hubble stages (T), are shown in the top right.  {\it Lower
    panels:} (from left to right) show the bar region of the r'-band
  SDSS mosaic image, the same image with the model components plotted
  on top of that, and the unsharp mask image. The image in the large
  panel shows the full galaxy image.}
\label{decompositions_1}
\end{figure*}

\begin{figure*}
\centering
\vskip -5.0cm
\includegraphics[angle=0,width=19.0cm]{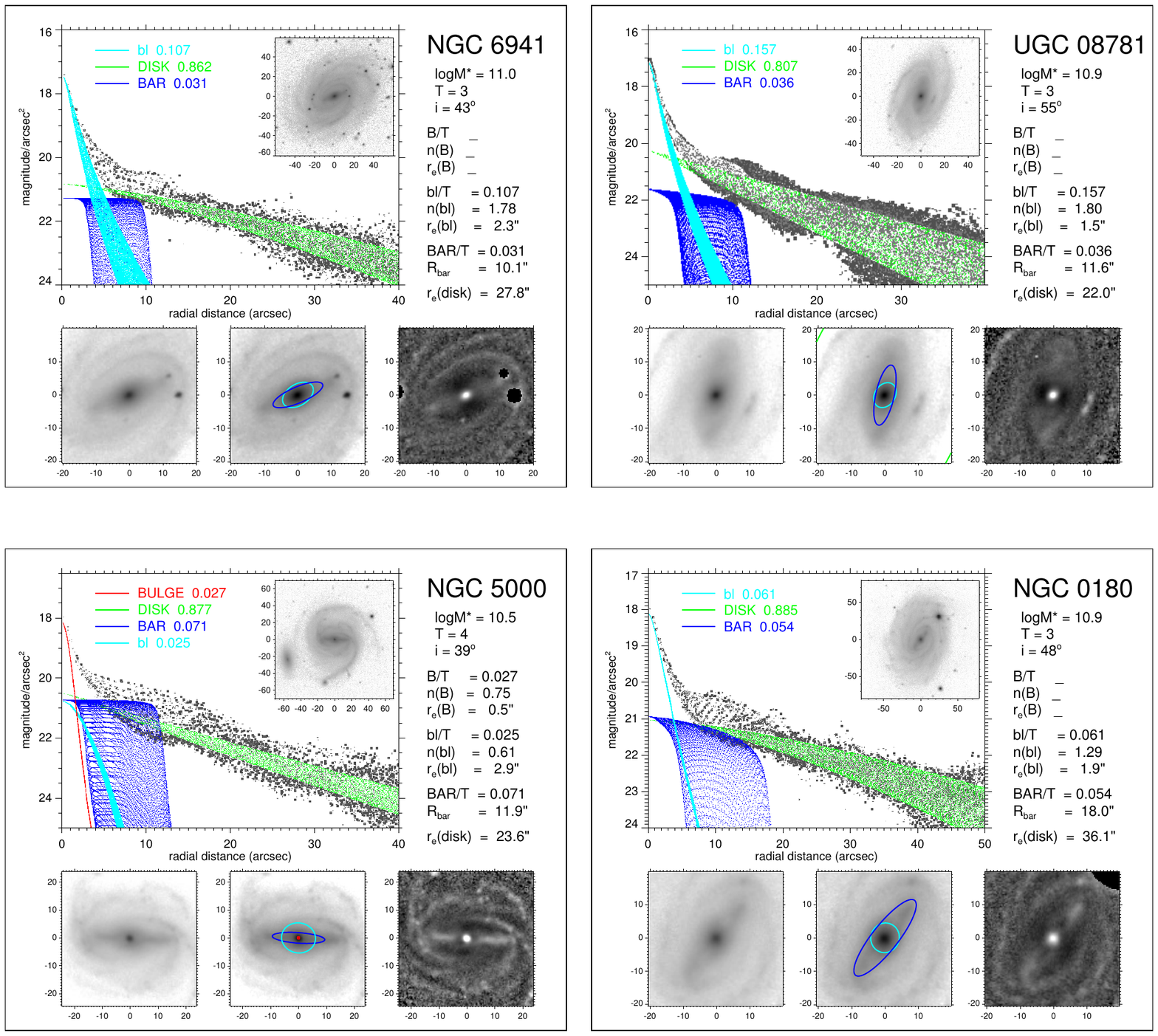}
\vskip -5.0cm
\caption{See Figure \ref{decompositions_1}.}
\label{decompositions_2}
\end{figure*}

\begin{figure*}
\centering
\vskip -5.0cm
\includegraphics[angle=0,width=19.0cm]{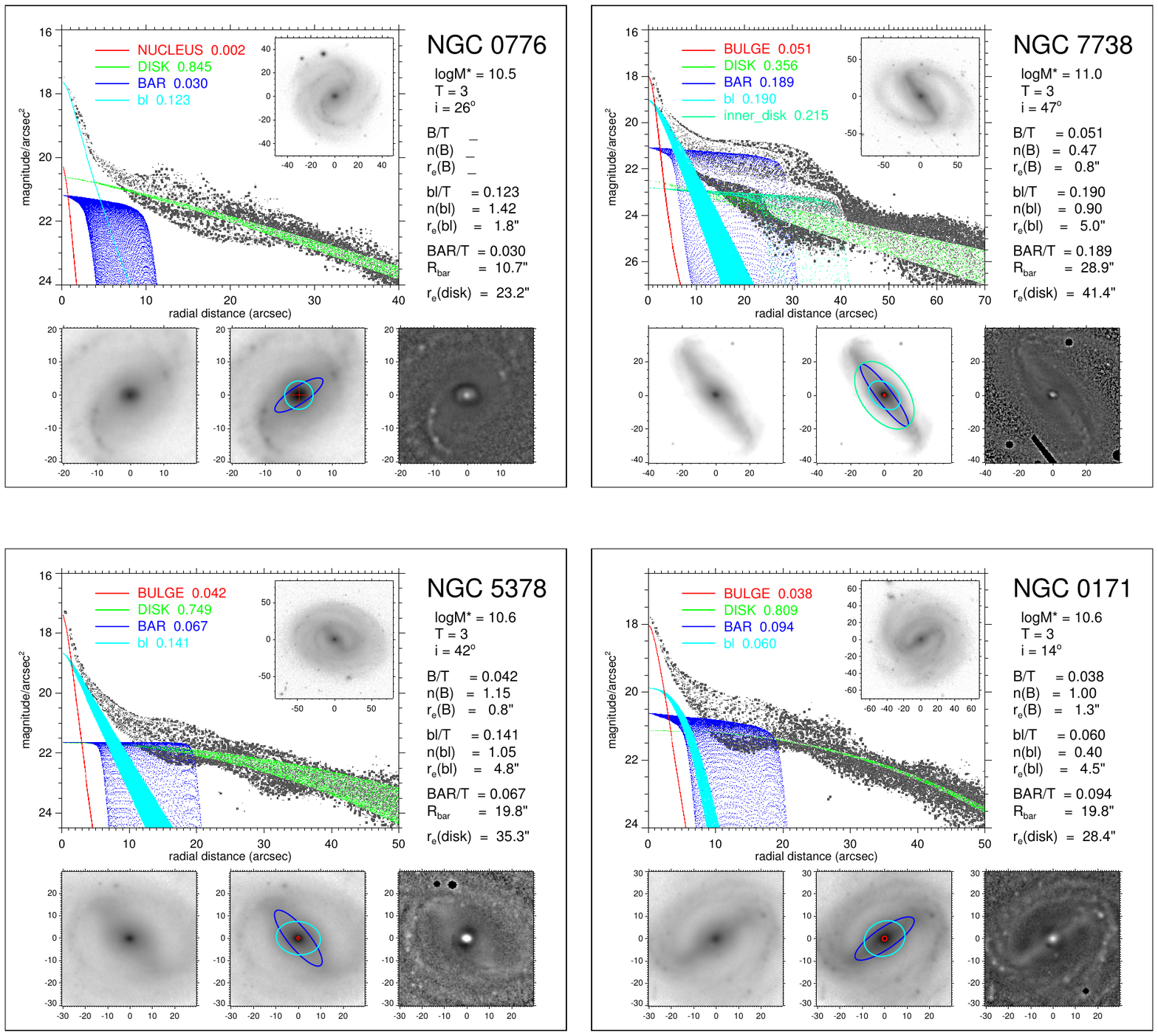}
\vskip -5.0cm
\caption{See Figure \ref{decompositions_1}.}
\label{decompositions_3}
\end{figure*}

The galaxies decomposed by us have been previously decomposed also by
MA2017 using B/D/bar models.  For comparison with their work we
divided our decompositions to two groups: (a) galaxies in which the
barlens had no clear central peak in the surface brightness profile,
and (b) those in which a central peak appeared, and it was fitted with
a separate S\'ersic function. The mean parameter values are given in
Table~\ref{table-2}, where the values given by MA2017 are also
shown. It is worth noticing that this division is to some extent
  artificial, because even those galaxies in which no separate bulge
  component was fitted, might have some low luminosity central
  components, possibly affecting the S\'ersic index.

We used group (a) to test the robustness of our decomposition method
by comparing its results to MA2017.  While we use GALFIT, MA2017 used
GASP2D for their decompositions.  Both codes use Levenberg-Marquardt
algorithm to search for the model parameters.  For non-exponential
disks MA2017 used two truncated disks, while we used a S\'ersic
function with $n<$ 1. In spite of these differences, our comparison
shows that the B/D/bar decompositions in the two studies are in good
agreement: both studies find the mean values of
 $<$bl(bulge)/T$>$ = 0.13, $<n>$ = 1.38, and $<$ $R_{\rm e}>$ = 0.60--0.64. This means that
our method is robust: it is neither user dependent, nor is it
sensitive to the code used, or the way how the underlying
non-exponential disk is fitted.  However, our interpretation of the
flux on top of the disk (after subtracting the bar) is different: we
consider the central S\'ersic component in the decompositions as a barlens,
while MA2017 interpreted it as a separate bulge component.


\begin{table}
\caption{The mean parameter values of barlenses in group (a),
  and of barlenses and separate bulge components in group (b).  For
  comparison the bulge parameters obtained by MA2017 are also shown.
The groups are explained in Section 6. Notice that what is called as a barlens by us
in group (a), is called as a bulge by MA2017.   
}
\label{table-2}      
\centering          
\begin{tabular}{lll}
\hline\hline     
\noalign{\smallskip}
\noalign{\smallskip}
                & This study       & MA2017 \\
\noalign{\smallskip}    
\hline
\noalign{\smallskip}
Group (a):           &               &                \\                
\noalign{\smallskip}
$bl(bulge)/T$  &0.13$\pm$0.02  & 0.13$\pm$0.02  \\ 
S\'ersic $n$      &1.38$\pm$0.10  & 1.38$\pm$0.10 \\
$R_{\rm e}$ [kpc]  &0.60$\pm$0.06  & 0.64$\pm$0.07  \\
                 &               &                  \\
Group (b):           &               &               \\
\noalign{\smallskip}
$B/T$             &0.06$\pm$0.00  & 0.21$\pm$0.02  \\ 
S\'ersic $n$      &1.10$\pm$0.11  & 2.30$\pm$0.22  \\
$R_{\rm e}$ [kpc]  &0.23$\pm$0.03  & 1.23$\pm$0.13  \\
\noalign{\smallskip}
\noalign{\smallskip}    
$bl/T$             &0.13$\pm$0.02  &   \\ 
S\'ersic $n$      &0.71$\pm$0.05  &   \\
$R_{\rm e}$ [kpc]  &0.94$\pm$0.15  &   \\
\noalign{\smallskip}
\hline
   \end{tabular}
\end{table}

For the galaxies in group (b), with both bulge and bl-components in
the decompositions, we found similar barlens parameters ($bl/T$ =
0.13) as we found also for the galaxies in group (a). However, MA2017
find clearly larger values for the bulges, i.e. $<B/T>$ = 0.21 and
$<n>$ = 2.3. In our decompositions less than 10$\%$ of the total
galaxy flux was left for a possible separate bulge component, in
agreement with the previous study by \citet{lauri2014} for barlens
galaxies in the S$^4$G+NIRS0S surveys. In both groups the surface
brightness profiles of barlenses are nearly exponential (the mean
values are $<n>$ = {\bf 1.4} and 0.7 in the groups (a) and (b),
respectively).  The similarity of the barlenses in these two groups
makes sense, because their galaxies have also similar mean Hubble stages
($<T>$ = 2.1$\pm$0.1 and $<T>$ = 2.3$\pm$0.1), and similar mean
  galaxy masses (log M$_{\star}$/M$_{\odot}$ = 10.80$\pm$0.04 and
  10.69$\pm$0.06 in groups (a) and (b), respectively).  The
similarity of the relative barlens fluxes in the two galaxy groups is
illustrated in Figure \ref{groups_a_b}: the $bl/T$-distributions are
very similar once the contribution of the separate bulge component is
taken away.

The reason for the differences in our models and those obtained by MA2017
for the galaxies in group (b) can
be understood by looking at individual galaxies. 
For NGC 7563 three decomposition models were shown in Figure
\ref{N7563-N5406}. It appears that the values $B/T$ = 0.53 and $n$ =
2.1 obtained by MA2017, are equivalent with those of our B/D/bar
model with $B/T$ = 0.54 and $n$ = 2.4.  However, in our final model
$B/T$ = 0.09 and $bl/T$ = 0.31. In this galaxy the unsharp mask image
clearly shows a barlens in favor of our model (see Fig. \ref{decompositions_1}). Also,
the surface brightness profile inside the bar radius is better fitted
in our best model than in the more simple B/D/bar model. Other similar
galaxies in our sample are NGC 5378 and NGC 7738. In NGC 5000 (see
Fig.\ref{decompositions_2}) the central mass concentration is less
prominent, and therefore also the difference in $B/T$ between the two
studies is much smaller ($B/T$ = 0.07 and 0.03 in MA2017 and our study,
respectively).  However, while MA2017 finds S\'ersic $n$ = 3.9 for
this galaxy, we find $n$ = 0.7. It is unlikely that this galaxy has a
de Vaucouleurs' type surface brightness profile, because in the
unsharp mask image X-shape feature appears, which confirms the bar
origin of the bulge.  MA2017 finds fairly large $B/T$ and S\'ersic
$n$-values also for the galaxies NGC 0036 and NGC 1093, which are
doubtful, because these galaxies are late-type spirals with only a small amount
of flux on top of the disk.

\section{CALIFA data-cubes and SDSS colors}

We use the CALIFA data cubes by \citet{sanchezGZ2016} for obtaining
the stellar ages, metallicities, and velocity
dispersions ($\sigma$) for the different structure components.
For the average values of these parameters the SSP.cube.fits cubes were 
used, whereas for calculating the radial profiles of populations of different 
age and metallicity bins we used the SFH.cube.fits.
The Field-of-View (FOV) of the observations is 74"x64",
covering 2--3 $R_{\rm e}$ of the galaxies. The FWHM = 2.5 arcsec
corresponds to 1 kpc at the average distance of the galaxies in the
CALIFA survey. CALIFA has two gratings, V500 and V1200, with the
wavelength ranges of 2745-7500 $\AA$ with $\lambda/\Delta\lambda$ =  
850, and 3400-4750 $\AA$ with $\lambda/\Delta\lambda$ = 1650,        
respectively. We use the V500 grating data-cubes, of which the
pipeline data reductions are fully explained by \citet{sanchez2016b}.
The spectral resolution is 327 km/s. It was shown by Sanchez et al. that
for the $\sigma$-measurements there is one-to-one relation between
the two gratings when $\sigma$ $\ge$ 40 km/sec. For the V1200 grating
\citet{falco2017} estimated 5$\%$ uncertainties for $\sigma>$
150 km/sec, 20$\%$ for $\sigma$ = 40 km/sec, and 50$\%$ for $\sigma$ =
20 km/sec. With the V500 grating this translates to uncertainties of
10$\%$ at $\sigma>$ 150 km/sec, and 40$\%$ at $\sigma$ = 40
km/sec. With the S/N$\sim$50 and having prominent stellar absorption
lines, \citet{falco2017} report that reliable $\sigma$-values can be obtained down to 30 km/sec
within the innermost r$\sim$10'', without binning.

The pipeline (Pipe3D) reductions of the stellar populations and
metallicities are explained by \citet{sanchez2016b}. Their spectral
fitting included the following steps: first a simple Single Stellar
Population (SSP) template is used to fit the stellar continuum, which
is used to calculate the systemic velocity, central $\sigma$, and the
dust attenuation.  After that the emission lines were subtracted from
the original spectrum and more sophisticated SSP-templates were used
for obtaining the stellar populations, metallicities, and star
formation histories. The library covers 39 stellar ages (between 1 Myr
and 13 Gyrs), and 4 metallicities in respect to solar metallicity (log$_{10}$
Z/Z$_\odot$ = -0.7, 0.4, 0.0 and 0.2). The templates used are a
combination of the synthetic stellar spectra from the GRANADA library
\citep{martins2005}, and the libraries provided by the MILES-project
\citep{sanchez2006,falco2011,vazdekis2010}.  The Salpeter
\citep{salpeter1995} initial mass function was used. It has been
estimated by \citet{sanchez2016b} that with S/N $\ge$ 50 the stellar
populations are well recovered within an error of $\sim$0.1dex.

We calculated also the average (g'-r') and (r'-i') colors of the structure components using the SDSS mosaic
images.  As we are interested only in relative values between the
structure components, no extinction corrections were made. The flux
calibration parameters were taken from the image headers.
The colors were calculated from the ratio of total fluxes in different bands, using the measurement regions described in the next subsection.

\subsection{Definitions of the measured regions}

Mean stellar ages and metallicities were calculated for different
structure components of the galaxies, in the regions illustrated in
Figure \ref{regions}, and defined in the following manner:

\begin{figure}
\centering
\includegraphics[angle=0,width=9.0cm]{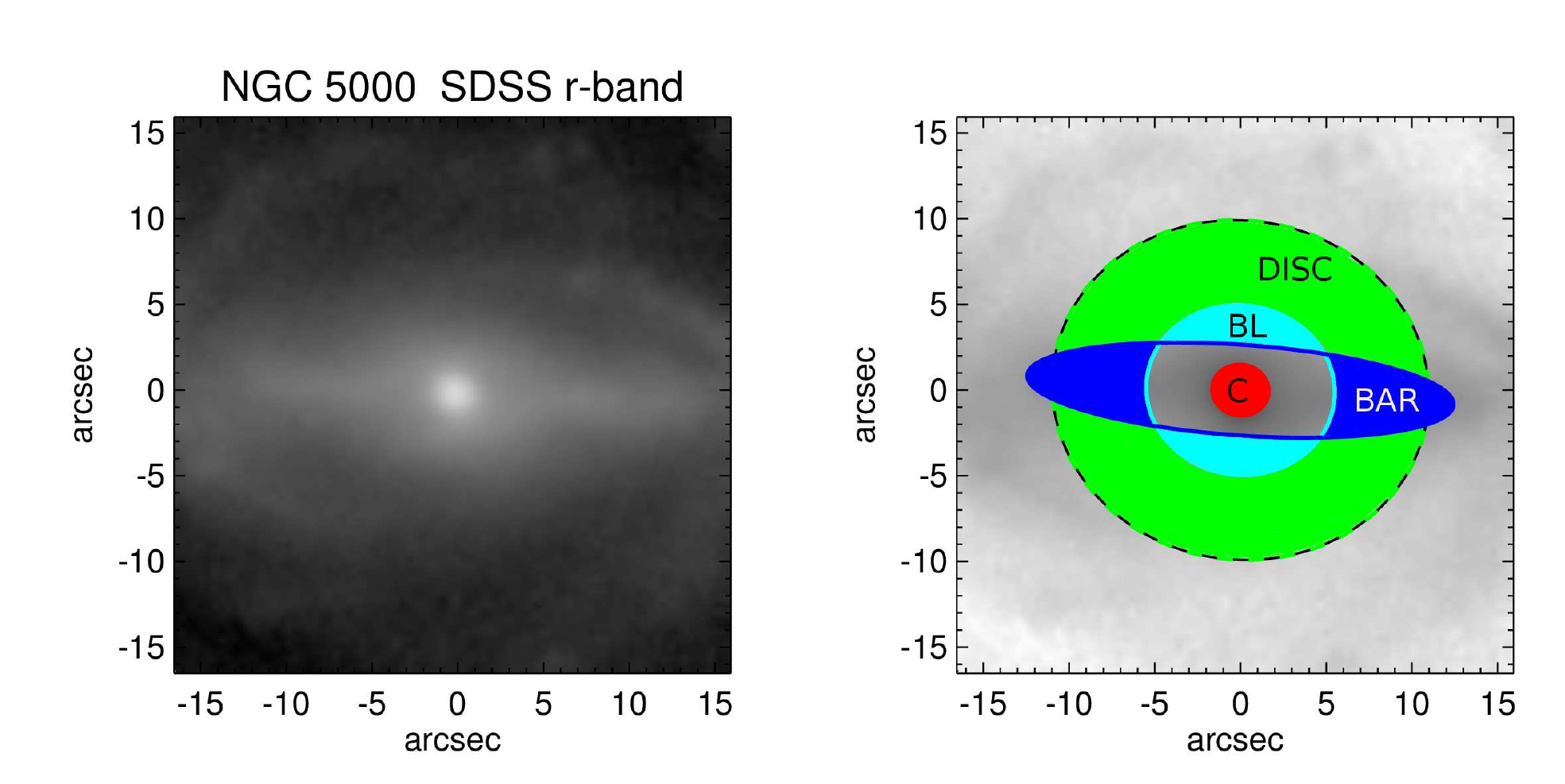} 
\caption{Illustration of the measured galaxy regions, as defined in
  Section 7.1. {\it Left panel} shows the original r'-band mosaic image
  of NGC 5000, to show the bar region, and {\it right panel} shows the
  definitions of the regions. The meaning of the colors are: red =
  galaxy center (C), turquoise = barlens (bl), blue = bar, green =
  disk.}
\label{regions}
\end{figure}

\vskip 0.1cm 
\noindent {\it C } (galaxy center): is defined as an
elliptical region around the galaxy center, having the same position
angle and $b/a$ axis-ratio as the barlens, and an outer radius r = 0.3
$r_{\rm bl}$.  This size is clearly larger than the maximum FWHM =
1.4 arcsec of the SDSS r'-band mosaic images, and larger than the FWHM = 2.5
arcsec of the V500 grating CALIFA data-cubes. The radius was large
enough to cover possible nuclear rings. Note that this parameter is
calculated for all galaxies, independent of whether a separate bulge
component was fitted in the decomposition or not.
\vskip 0.1cm 

\noindent {\it bl} (barlens): an elliptical zone
inside the barlens radius, but excluding the galaxy center C and the
region overlapping with the bar. The $b/a$ axis-ratio and the position angle
were those obtained from our visual tracing of barlenses (see Section 4). 
\vskip 0.1cm 

\noindent {\it bar} (elongated bar): an elliptical region inside r =
$r_{\rm bar}$, excluding the barlens. We used the measured
position angle of the bar, and a fixed axial ratio $b/a$ = 0.25. 

\vskip 0.1cm 
\noindent {\it disk} (disk): we used an elliptical stripe between $r_{\rm bl}$ 
and 2 $r_{\rm bl}$, excluding the zone covered by the
bar.  The ellipticity and position angle were the same as for
the barlens.
\vskip 0.15cm 

Almost similar measurement regions were used in \citet{herrera2017} for
SDSS colors of the S$^4$G-galaxies: in the current study 'C' corresponds to
what was denoted as 'nuc2' in their study, and 'bl' denoted as 'blc' in
Herrera-Endoqui et al..  

\section{Mean stellar populations, metallicities, and velocity dispersions}

\begin{figure}
\centering
\includegraphics[angle=0,width=8.5cm]{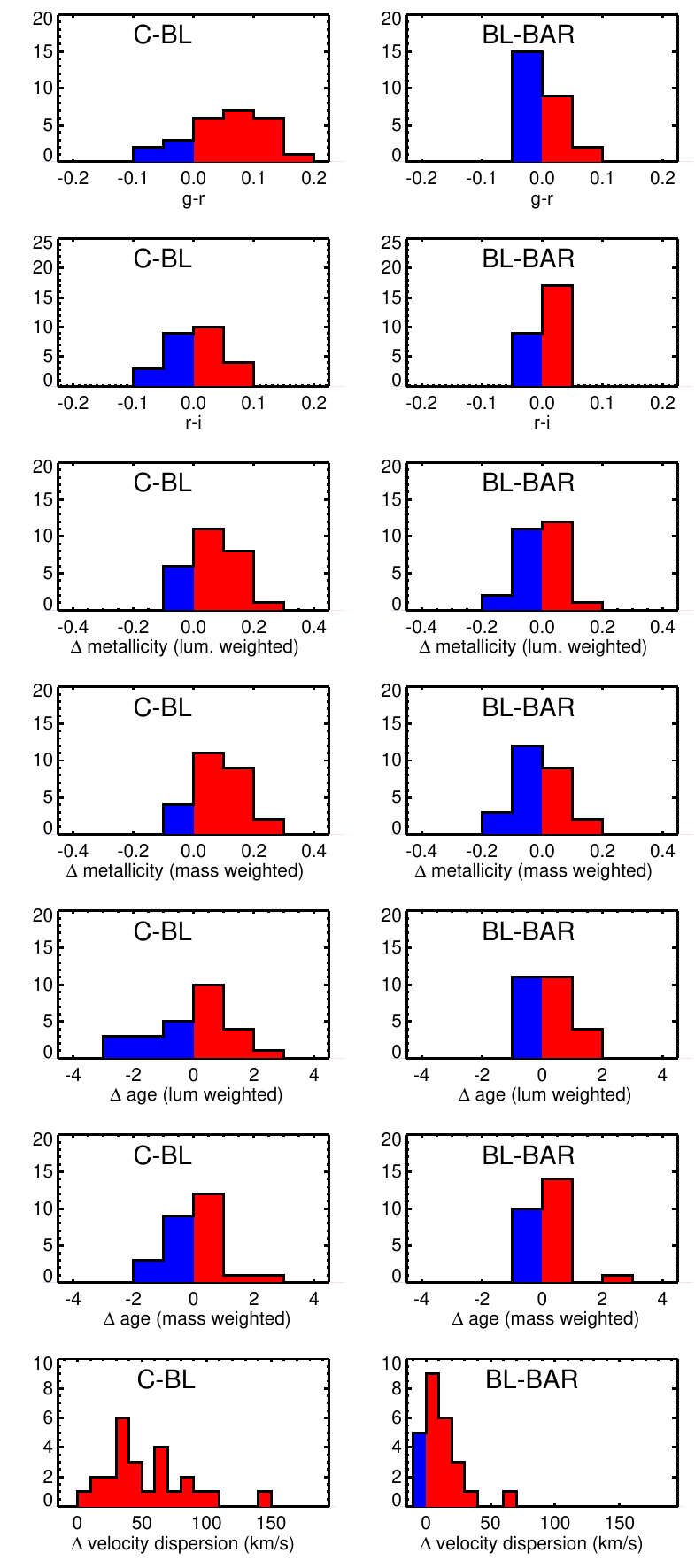}
\caption{The differences of the parameter values between the central
  galaxy regions and barlenses (C-BL), and between barlenses and
  bars (BL-BAR), are shown for the sample of 26 galaxies.  Positive and
  negative deviations are shown with red and blue colors,
  respectively. The parameters are the same as in
  Table 3.}
\label{summary}
\end{figure}

\begin{table*}
\caption{Using the CALIFA V500 data-cubes (e.g. S\'anchez, Garc\'ia-Benito $\&$ Zibetti 2016)
  the mean parameter values are calculated for
  the different structure components. The regions used 
  are shown in Figure \ref{regions}, explained in Section 7.1. Shown separately are also
  the galaxies with (bulge) and without (no bulge) a separately fitted
  bulge component.  The parameters are: stellar velocity
  dispersion ($\sigma$) [km/sec], mass ('m') and light ('l') weighted
  stellar ages [Gyrs], and metallicity in respect to solar
  metallicity (log$_{10}$ Z/Z$_{\odot}$). Shown also are (g'-r') and
  (r'-i') colors obtained from the SDSS mosaic images. 
  The uncertainties are calculated from the sample standard deviation, divided
by square-root of sample size. One of the galaxies did not have an i'-band image.} 
\label{table-5}  
    
\centering          
\begin{tabular}{lccccc}
\hline\hline     
\noalign{\smallskip}
\noalign{\smallskip}
                             &    C        & bl             & bar    &  disk  & N       \\    
\noalign{\smallskip}
\hline
\noalign{\smallskip}
$\sigma$ (all)             & 210$\pm$5  &157$\pm$9   &132$\pm$9 &- & 25 \\ 
$\sigma$ (bulge)           & 212$\pm$7  &147$\pm$15  &127$\pm$12&- & 11 \\ 
$\sigma$ (no bulge)        & 207$\pm$7  &165$\pm$10  &137$\pm$14 &- &  14 \\
 \noalign{\smallskip}

age (m, all)            &8.8$\pm$0.2   &8.7$\pm$0.2  &8.3$\pm$0.4 &8.3$\pm$0.1 & 25 \\
age (m, bulge)         &9.0$\pm$0.4   &8.8$\pm$0.2   &8.7$\pm$0.3 &8.3$\pm$0.2 & 11 \\
age (m, no bulge)      &8.7$\pm$0.2   &8.7$\pm$0.2   &7.9$\pm$0.6 &8.2$\pm$0.2 & 14 \\
 \noalign{\smallskip}

age (l, all)           &5.3$\pm$0.5   &5.4$\pm$0.4   &5.2$\pm$0.4 &4.1$\pm$0.3 & 25 \\
age (l, bulge)         &5.8$\pm$0.7   &5.7$\pm$0.6   &5.6$\pm$0.6 &4.5$\pm$0.5 & 11 \\
age (l, no bulge)      &4.9$\pm$0.7   &5.1$\pm$0.5   &4.8$\pm$0.5 &3.9$\pm$0.3 & 14 \\
 \noalign{\smallskip}

log$_{10}$ $Z/Z_{\odot}$ (m, all)         &-0.06$\pm$0.02  &-0.02$\pm$0.03 &-0.04$\pm$0.03 &-0.12$\pm$0.02 & 25 \\
log$_{10}$ $Z/Z_{\odot}$ (m, bulge)       &-0.04$\pm$0.03  &-0.03$\pm$0.03 & 0.00$\pm$0.03 &-0.09$\pm$0.02 & 11 \\
log$_{10}$ $Z/Z_{\odot}$ (m, no bulge)    &-0.01$\pm$0.04 &-0.03$\pm$0.04 &-0.06$\pm$0.04 &-0.14$\pm$0.04 & 14 \\

 \noalign{\smallskip}
log$_{10}$ $Z/Z_{\odot}$ (l, all)         &-0.10$\pm$0.02 &-0.10$\pm$0.02 &-0.09$\pm$0.02 &-0.19$\pm$0.02 & 25  \\
log$_{10}$ $Z/Z_{\odot}$ (l, bulge)       &-0.08$\pm$0.03 &-0.09$\pm$0.03 &-0.06$\pm$0.03 &-0.17$\pm$0.02 & 11 \\
log$_{10}$ $Z/Z_{\odot}$ (l, no bulge)    &-0.12$\pm$0.03 &-0.10$\pm$0.03 &-0.11$\pm$0.04 &-0.20$\pm$0.03 & 14 \\

 \noalign{\smallskip}
$g'-r'$ (all)            &0.885$\pm$0.013  &0.824$\pm$0.010  &0.825$\pm$0.010 &0.769$\pm$0.011&  25  \\
$g'-r'$ (bulge)          &0.887$\pm$0.022  &0.816$\pm$0.012  &0.815$\pm$0.013 &0.767$\pm$0.016&  11 \\
$g'-r'$ (no bulge)       &0.884$\pm$0.017  &0.830$\pm$0.016  &0.833$\pm$0.014 &0.772$\pm$0.015&  14 \\
 \noalign{\smallskip}

$r'-i'$ (all)            &0.404$\pm$0.017  &0.419$\pm$0.005 &0.400$\pm$0.017 &0.395$\pm$0.017&  24 \\
$r'-i'$ (bulge)          &0.415$\pm$0.006  &0.410$\pm$0.007 &0.410$\pm$0.007 &0.409$\pm$0.009&  10 \\
$r'-i'$ (no bulge)       &0.395$\pm$0.031  &0.426$\pm$0.007 &0.390$\pm$0.031 &0.382$\pm$0.030&  14 \\
 
\hline 
   \end{tabular}
\end{table*}               

The mean stellar ages and metallicities of the structure components
are shown in Table \ref{table-5}. Shown separately are also the
galaxies decomposed with ('bulge') and without ('no bulge') a separate
bulge component (groups (a) and (b) in Section 6, respectively). Both
mass ('m') and light ('l') weighted values are given.  In the same
table the mean stellar velocity dispersions $\sigma$, and the (g'-r')
and (r'-i') colors are also shown. For these parameters the
differences between barlenses and central galaxy regions (C-BL), and
between barlenses and bars (BL-BAR) are illustrated in Figure
\ref{summary}. The average values were obtained by finding all
  spaxels in the region of interest and calculating the density or
  luminosity weighted means of the corresponding ('m' and 'l',
  respectively) data cube values.

\begin{figure}
\centering
\includegraphics[angle=0,width=9.0cm]{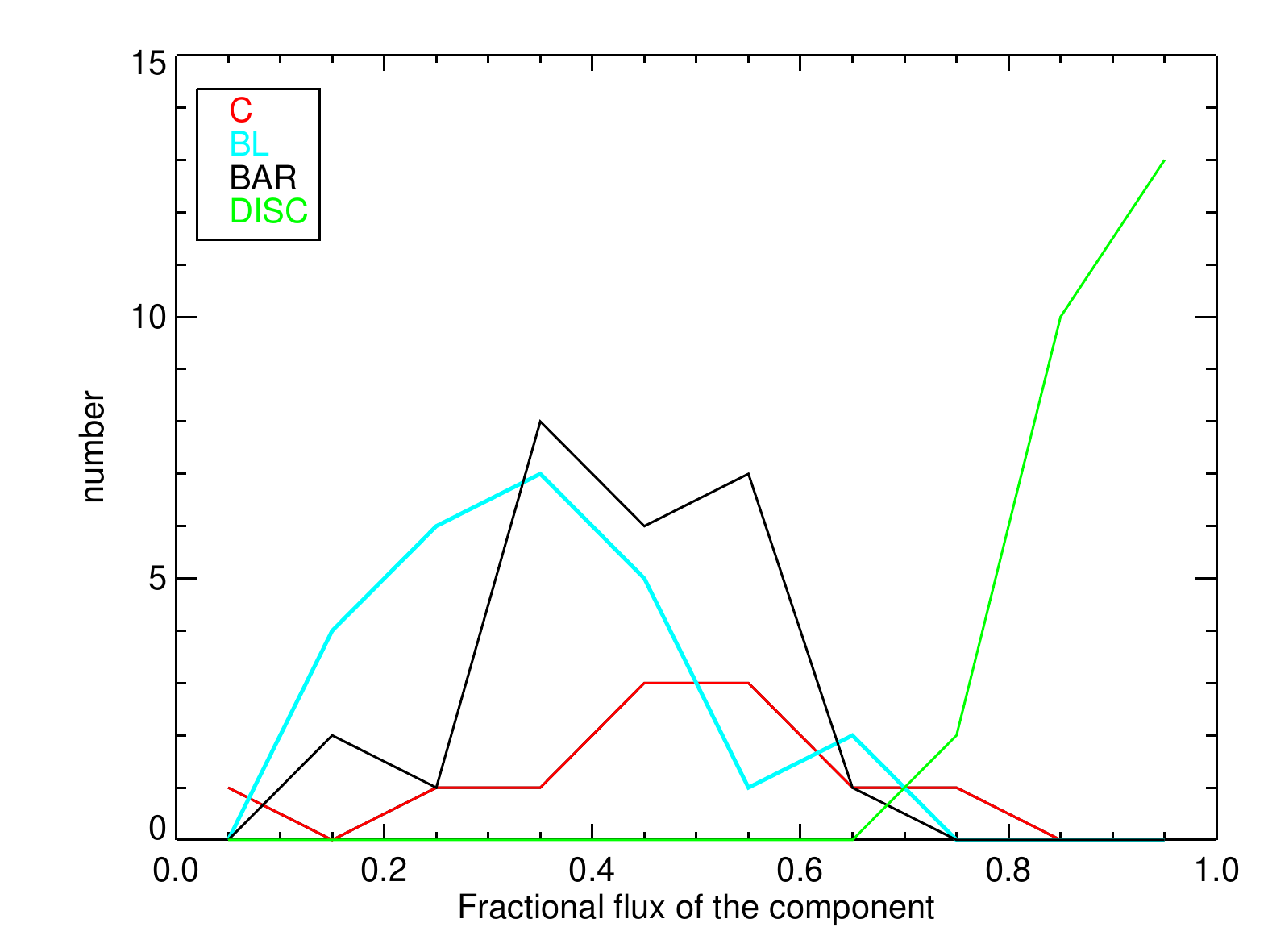}
\caption{We used our decomposition models of Section 5 to estimate the
  amount of contamination in the measurements of average values for
  different structure components. The plot shows the distribution of
  the fractional contribution of the component itself to the total
  flux in the measurement region of that structure component, as defined
  in Section 7.1.
}
\label{fig_contamination}
\end{figure}

The measurement regions inevitably correspond to a superposition of more
than one structure component. In Figure \ref{fig_contamination} we
estimate the amount of contamination, with the help of our decomposition
models: the figure shows for each measurement region how much of the
total flux in the decomposition comes from the structure we intend to
measure.  It appears that for the barlens measurement regions
typically 20$\%$-50$\%$ of the flux is due to the barlens, the rest is
mainly due to the underlying disc. For the bar the disk contamination
is slightly less, the contribution from the bar itself amounting to
30$\%$-60$\%$.

\subsection{Stellar velocity dispersions}

Bars have generally old stellar populations which means that prominent
stellar absorption lines appear in the spectrum. The S/N in the bar
region is also high because many spaxels are averaged. 
We find that bars and barlenses typically have fairly high
velocity dispersions ($\sigma$ = 130--160 km/sec), which values are
practically the same for both components ($\Delta\sigma\sim$ 20
km/sec).  Also, there is practically no difference in $\sigma$ between bars and
barlenses while comparing the galaxies with and without a separate
bulge component. However, $\sigma$ is clearly higher in the galaxy
centers ($\sigma$ = 207$\pm$5 km/s).
Figure \ref{summary} shows that the central galaxy regions have
always higher $\sigma$-values than the surrounding galaxy components.
The radial dependence of $\sigma$ for the CALIFA sample has been shown 
by Falcon-Barroso et al. (2017).  

\subsection{Colours, stellar ages and metallicities}

We find that bars and barlenses have on average similar mean (g'-r')
and (r'-i') colors, confirming the previous result by Herrera-Endoqui
et al. (2017) for the S$^4$G-galaxies. The mean value
(g'-r')$\sim$0.82 is typical for K-giant stars \citep{lenz1998}. The
central galaxy regions have clearly redder (g'-r') colors
($\Delta (g'-r')\sim$0.06), again in agreement with Herrera-Endoqui et
al.. Most probably this is a dust effect, because the stars in the
central galaxy regions are also similar as in barlenses
(light weighted average ages are 8.8$\pm$0.2 and 8.7$\pm$0.2
Gyrs, respectively).

The stellar ages in our analysis show gradients. The mass
    weighted ages of the central regions are $\sim$0.5 Gyr, and the luminosity
    weighted ages $\sim$1.6 Gyr older than for the disks.  These gradients are
    in a qualitative agreement with those obtained for the whole
    CALIFA sample, by \citet{garciabenito2017} for the mass weighted
    ages, and by \citet{gonzalez2014} for the luminosity weighted
    ages. It is remarkable that despite 
    these age gradients, the ages of bars and barlenses, which appear at different
    radial distances in our measurements, are similar.
    Their mass weighted mean ages are $\sim$9 Gyrs,
    and the luminosity weighted mean ages $\sim$5 Gyrs.

We also show a metallicity gradient of log$_{10}$ Z/Z$_{\odot}$
$\sim$0.1 (using both mass and light weighted values) in a sense that
the disks are less metal-rich than the bars and barlenses. The galaxy
centers are more metal-rich than the rest of the galaxy, both using
the mass and luminosity weighted indices. These gradients are again in
a good qualitative agreement with those obtained by Gonzalez-Delgado
et al. (2015) for the whole CALIFA survey. As the ages, also the
metallicities are similar for bars and barlenses, their mean luminosity
weighted values being log$_{10}$ Z/Z$_{\odot}$ = -0.09$\pm$0.02 and
-0.10$\pm$0.02, respectively.

Figure \ref{summary} illustrates the similarity of all the measured
parameter values of bars and barlenses, and also the way how the central 
galaxy regions in many parameters are at least marginally different from bars and barlenses.

\section{Radial profiles of stellar ages and metallicities}

\begin{figure*}
\centering
\includegraphics[angle=0,width=16.0cm]{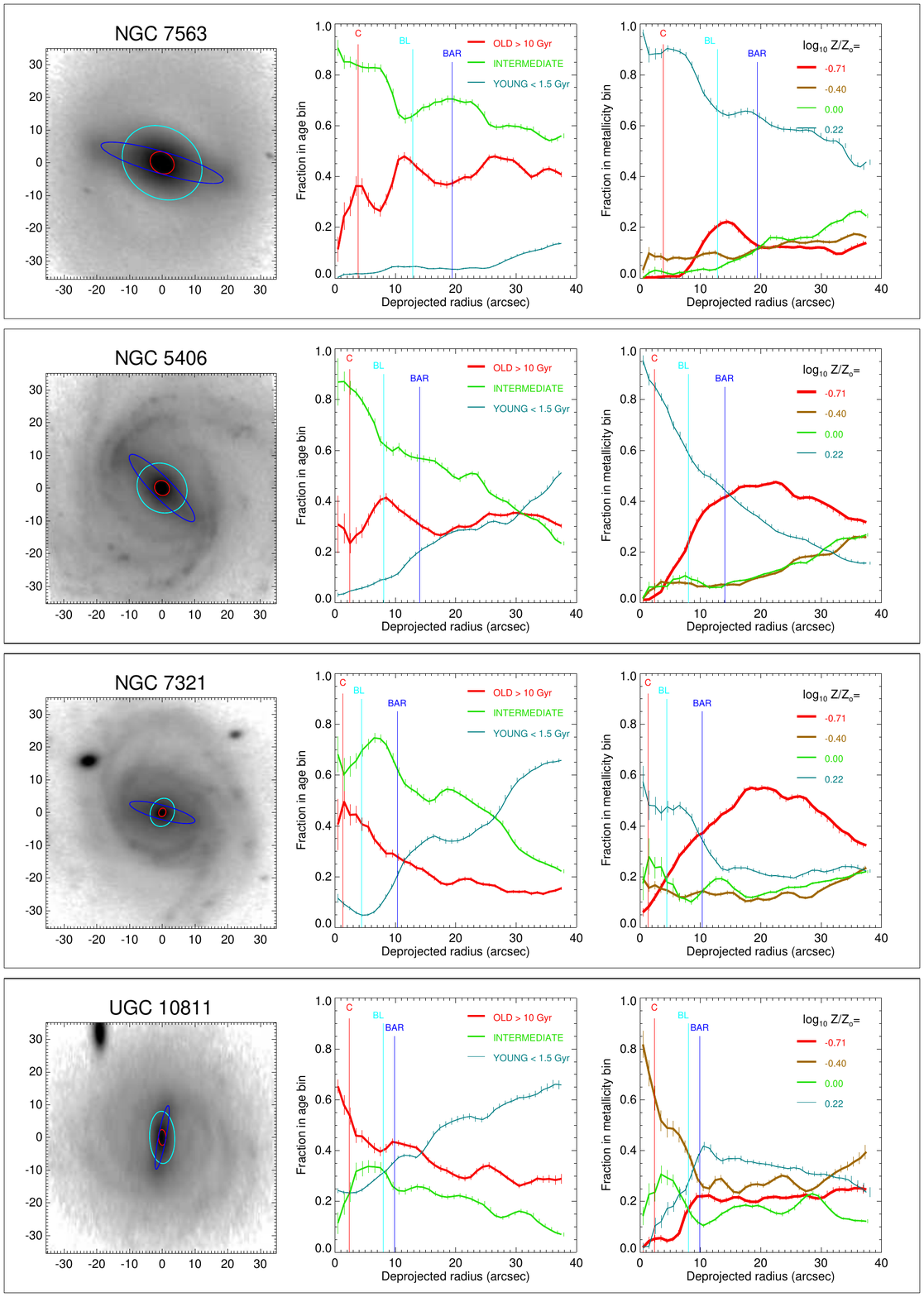}
\caption{{\it Left:} deprojected SDSS r'-band mosaic
  image. Over-plotted are the elliptical zones (see Section 7.1) used in the measurement
  of structure averages, here shown deprojected to disk plane (original measurements were
done in non-deprojected images). {\it Middle:} Fractions of
  stars in the three stellar age bins, as a function of the
  deprojected radial distance. The profiles are constructed using the
  V500 grating data-cubes from Sanchez et al. (2016), loaded from CALIFA database
  (SFH.cube.fits). They correspond to averages over mass and
  luminosity weighted star formation histories.  The vertical lines
  show the deprojected semi-major axis lengths of the bars,
  barlenses, and the central galaxy regions.  {\it Right:} Fractions of stars in four
  different metallicity bins, as given in the CALIFA data-cubes. The
  meaning of the vertical lines are the same as in the middle panel.
}
\label{age_met_profiles_1}
\end{figure*}

\begin{figure*}
\centering
\includegraphics[angle=0,width=16.5cm]{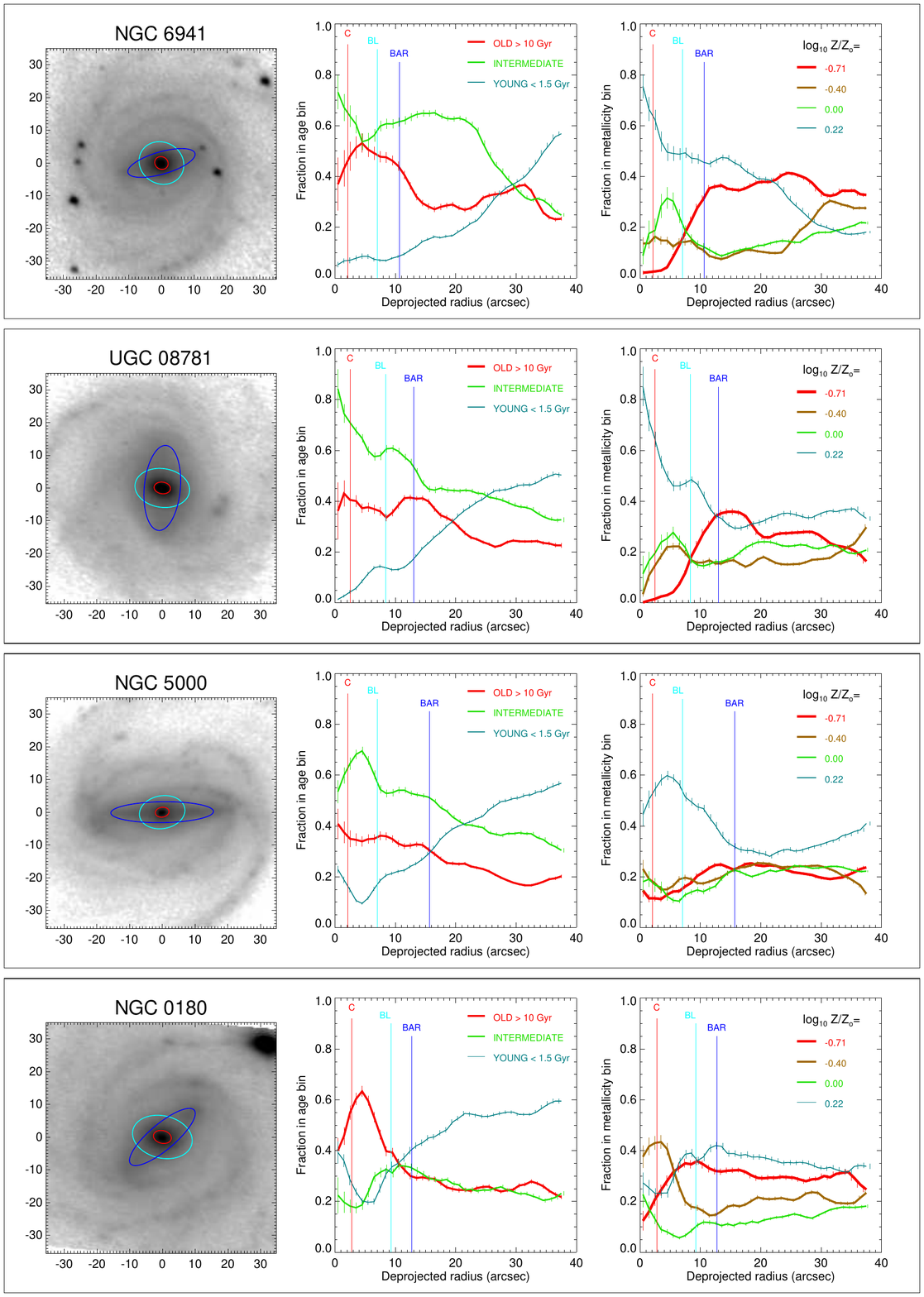}
\caption{See Figure \ref{age_met_profiles_1}.}
\label{age_met_profiles_2}
\end{figure*}

\begin{figure*}
\centering
\includegraphics[angle=0,width=16.5cm]{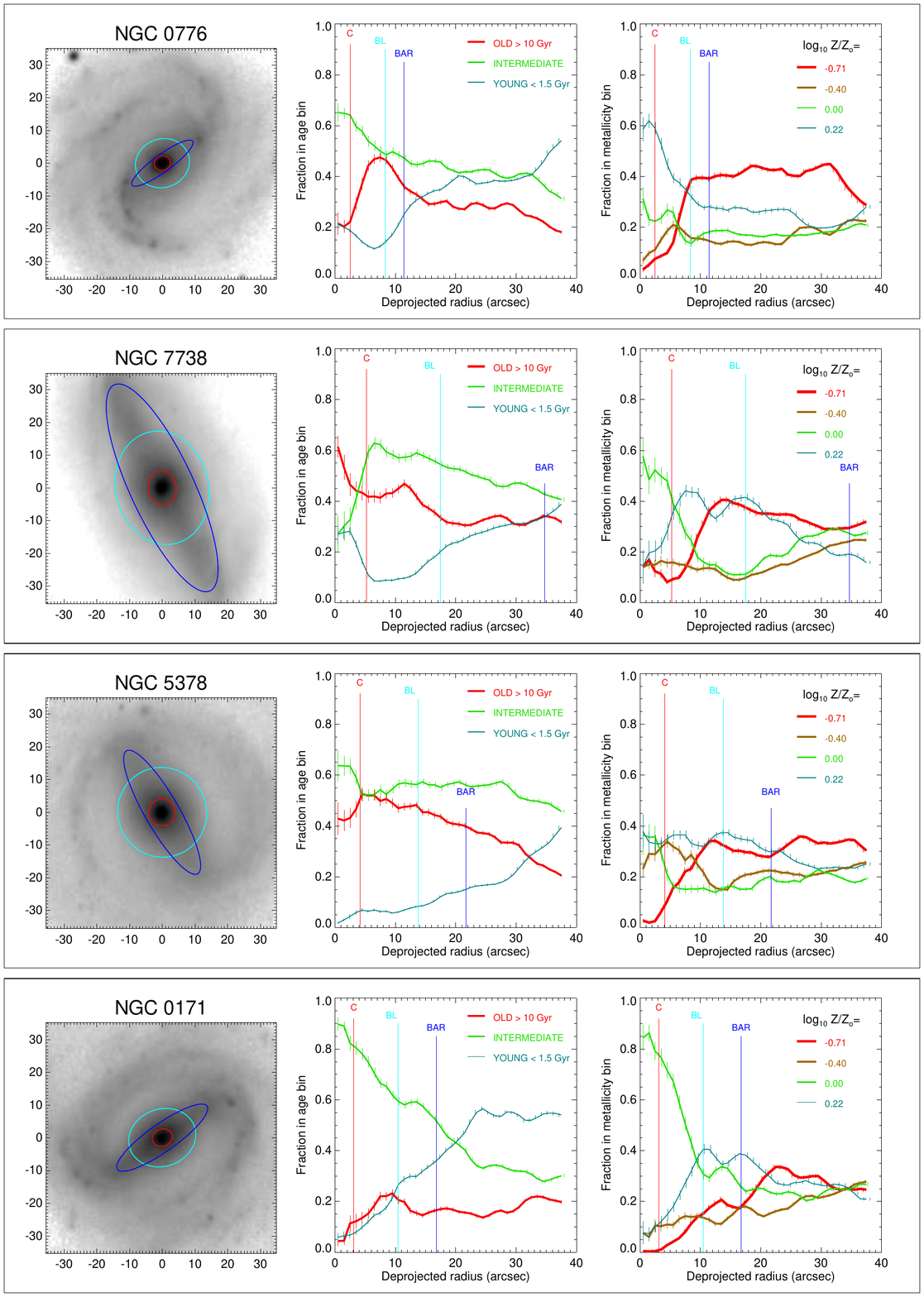}
\caption{See Figure \ref{age_met_profiles_1}.}
\label{age_met_profiles_3}
\end{figure*}

We use the CALIFA SFH cubes to analyze the radial distribution of
  different age and metallicity populations.  For the analysis 
  we have selected typical barlens galaxies, galaxies with
dust-lanes, and barlens galaxies in which X-shape features also
appear. The age and metallicity profiles are shown in Figures
\ref{age_met_profiles_1}, \ref{age_met_profiles_2}, and
\ref{age_met_profiles_3}.  The decompositions for the same galaxies
were shown in Figures \ref{decompositions_1}, \ref{decompositions_2}
and \ref{decompositions_3}.  The profiles are azimuthally averaged in
a few arcsecond bins after deprojecting the galaxies to face-on. This
means that in the barlens regions and in the galaxy centers the
stellar parameters are well captured, but the bar regions might
be slightly contaminated by younger stellar populations of the
disks. The four metallicity bins are as given in the CALIFA data-cubes
(SFH.cube.fits). The original stellar age bins we have collected
to three bins corresponding to young (age $<$ 1.5 Gyr), intermediate
age (1.5 $<$ age $<$ 10 Gyrs), and old (age $>$ 10 Gyrs) stars. 
According to the headers of SFH.cube.fits files, the spaxel values of the SFH data cubes correspond to
luminosity fractions of different age and metallicity bins. Our comparisons
indicated that the mean ages and metallicities
calculated from the SFH cube distributions are close to the mean of metallicity and luminosity averaged mean ages and metallicities, as given in the SSP cubes.

\vskip 0.20cm


\subsection{Typical barlens galaxies}

\noindent {\it NGC 7563 (T = 1)}: This is the type of galaxies in
which barlenses were originally recognized \citep{lauri2011}. The
surface brightness profile (Fig. \ref{decompositions_1}) shows a possible separate bulge component,
and a nearly exponential barlens, which dominates the photometric
bulge. The bar and the barlens are dominated by metal-rich
(log$_{10}$ Z/Z$_{\odot}$ = 0.20) intermediate age stars. There is also a
contribution of very old stars ($>$ 10 Gyrs), but their relative
fraction decreases within the barlens, and drops in the galaxy
center. The bar is surrounded by a dispersed ringlens, which is also
dominated by metal-rich stars, but containing also an increasing
fraction of less metal-rich stars.  It appears that the density peak
in the surface brightness profile is not made of old metal-poor stars
early in the history of this galaxy. 

\vskip 0.15cm 
\noindent {\it NGC 5406 (T = 3.5)}: The distributions of the oldest
and intermediate age stars in the bar/barlens regions are as in NGC
7563, i.e. the fraction of the oldest stars drops at the galaxy center
compared to the barlens region. Outside the barlens the fraction of
younger stars increases due to the prominent spiral arms. However, the
disk is dominated by very metal-poor stars (log$_{10}$ Z/Z$_{\odot}$ = -0.7),
whose fraction starts to drop in the bar region so that in the galaxy
center those stars have disappeared. So, also in this galaxy the bar
and the barlens have had repeated episodes of stars formation which
have enriched the gas in metals. In the disk outside the bar that has
happened in a less extent than in NGC 7563.  The galaxy center has a
higher velocity dispersion ($\sigma$ = 214 km/sec) than the bar or the
barlens ($\sigma$ = 128 and 142 km/sec, respectively), most probably
related to higher stellar density.

\vskip 0.15cm
\noindent {\it NGC 7321 (T = 3)}: This galaxy has qualitatively
similar stellar age and metallicity distributions as NGC 5406, but 
the photometric bulge is less prominent. The fraction
of the most metal-poor stars (log$_{10}$ Z/Z$_{\odot}$ = -0.7) starts to drop
already at r$\sim$18'', corresponding to the high surface brightness
region extending well outside the bar. There is clearly migration of stars
inside the high surface brightness disk. 

\vskip 0.15cm
\noindent {\it UGC 10811 (T = 2)}: This galaxy (and NGC 0180) is
exceptional in our sample, in a sense that the barlens is dominated by
the oldest ($>$ 10 Gyrs) fairly metal-poor (log$_{10}$ Z/Z$_{\odot}$ = -0.40)
stars, which fraction increases toward the galaxy center. The photometric bulge is dominated by the
barlens, which has a nearly exponential surface brightness
profile ($n$ = 1.5). The disk outside the bar is dominated by young metal-rich
stars (log$_{10}$ Z/Z$_{\odot}$ = 0.20), but has also many other metallicities. 
Most probably the bar was formed early, but galaxy modeling is needed 
to interpret how the mass was accumulated to the barlens.

\vskip 0.15cm
\noindent {\it NGC 0776 (T = 3)}: The barlens dominates the bar in
such a level that the morphology approaches a non-barred galaxy
(i.e. has a barlens classification ``f'' by
Laurikainen $\&$ Salo 2017). However, in spite of that the stellar and
metallicity properties are very similar as in such prototypical
barlens galaxies as NGC 5406. The unsharp mask image shows a nuclear
ring, showing also a significant contribution of the young stellar
population. 
 
\subsection{Galaxies with X-shapes}

\noindent {\it NGC 6941 (T = 3), UGC 8781 (T = 3)}, and NGC 5000 (T =
4): In these galaxies the barlenses show also X-shape features in the
unsharp-mask images, which confirms that they are vertically thick
inner bar components. Therefore, it is interesting that also in these
galaxies the bar/barlens regions have similar age and metallicity
distributions as the prototypical barlens galaxies discussed above,
i.e. they are dominated by metal-rich (log$_{10}$ Z/Z$_{\odot}$ = 0.20)
intermediate age stars, with a significant contribution of the oldest
stars (age $>$ 10 Gyrs).  In UGC 08781 and NGC 5000 the fraction of
the oldest stars is similar in the galaxy center and in the
bar/barlens region, whereas in NGC 6941 their fraction drops in the
galaxy center. In these galaxies the metallicity starts to drop
toward the galaxy center already at the edge of the bar.
\vskip 0.15cm

\noindent {\it NGC 0180 (T = 3)}: The bl/X is dominated by
old (age $>$ 10 Gyrs) metal-poor (log$_{10}$ Z/Z$_{\odot}$ = -0.40) stars, in a
similar manner as the barlens in UGC 10811. In the galaxy center the
fraction of the oldest stars drops. The mean velocity dispersions of
the bar, bl/X, and the galaxy center are $\sigma$ = 118, 137, and 181
km/sec, respectively. 
Very old metal-poor stars with high random motions are generally
interpreted as manifestations of merger built classical bulges. 
However, in NGC 0180 the barlens is dominated by an X-shape feature,
which challenges that interpretation. 

\subsection{Barlenses with dust lanes} 

\noindent {\it NGC 7738 (T = 3)}: This is a prototypical barlens
galaxy, similar to NGC 4314 \citep{lauri2014}. The arc-like dust
features in the unsharp mask image are illustrative, because they hint
to the fact that the whole high surface brightness disk surrounding
the bar probably form part of the bar structure.
Two dust-lanes penetrate through the barlens ending up to the
galaxy center, where young stars (age $<$ 1.5 Gyr) appear at r$\sim$ 5
arcsec. This galaxy is enriched in metallicity particularly in the
barlens region. Clearly, fresh
gas has penetrated through the barlens fairly recently triggering
central star formation. Most probably star formation has occurred also
in the barlens, leaving behind a metal-rich stellar population, but
that star formation has been ceased already a long time ago (lack of young stars 
in the barlens).

\vskip 0.15cm

\noindent {\it NGC 5378 (T = 3)}: Two dust-lanes appear,
one penetrating through the barlens ending up to the galaxy center,
and another weaker one following the outer edge of the barlens.
As in NGC 7738, also in this galaxy particularly the barlens and the bar are places
of metallicity enrichment.    
\vskip 0.15cm

\noindent {\it NGC 0171 (T = 3)}: This is a barlens galaxy
seen nearly face-on. The unsharp mask image shows an elongated feature along
the bar major axis at low surface brightness levels.  A nuclear ring
is manifested as an obscuration by dust. The metal enrichment 
has occurred particularly at the edges of the bar and the barlens.
However, contrary to the other barlens galaxies the fraction of the metal-poor 
stars (log$_{10}$ Z/Z$_{\odot}$ = -0.70) starts to drop already at r$\sim$24 arcsec,
where the two-armed prominent spiral arms start. 
It seems that mixing of different stellar ages and metallicities
appear in a large galaxy region, starting well outside the bar via the
prominent spiral arms.

\section{Cumulative age and metallicity distributions}

\begin{figure*}
\centering
\includegraphics[angle=0,width=17.5cm]{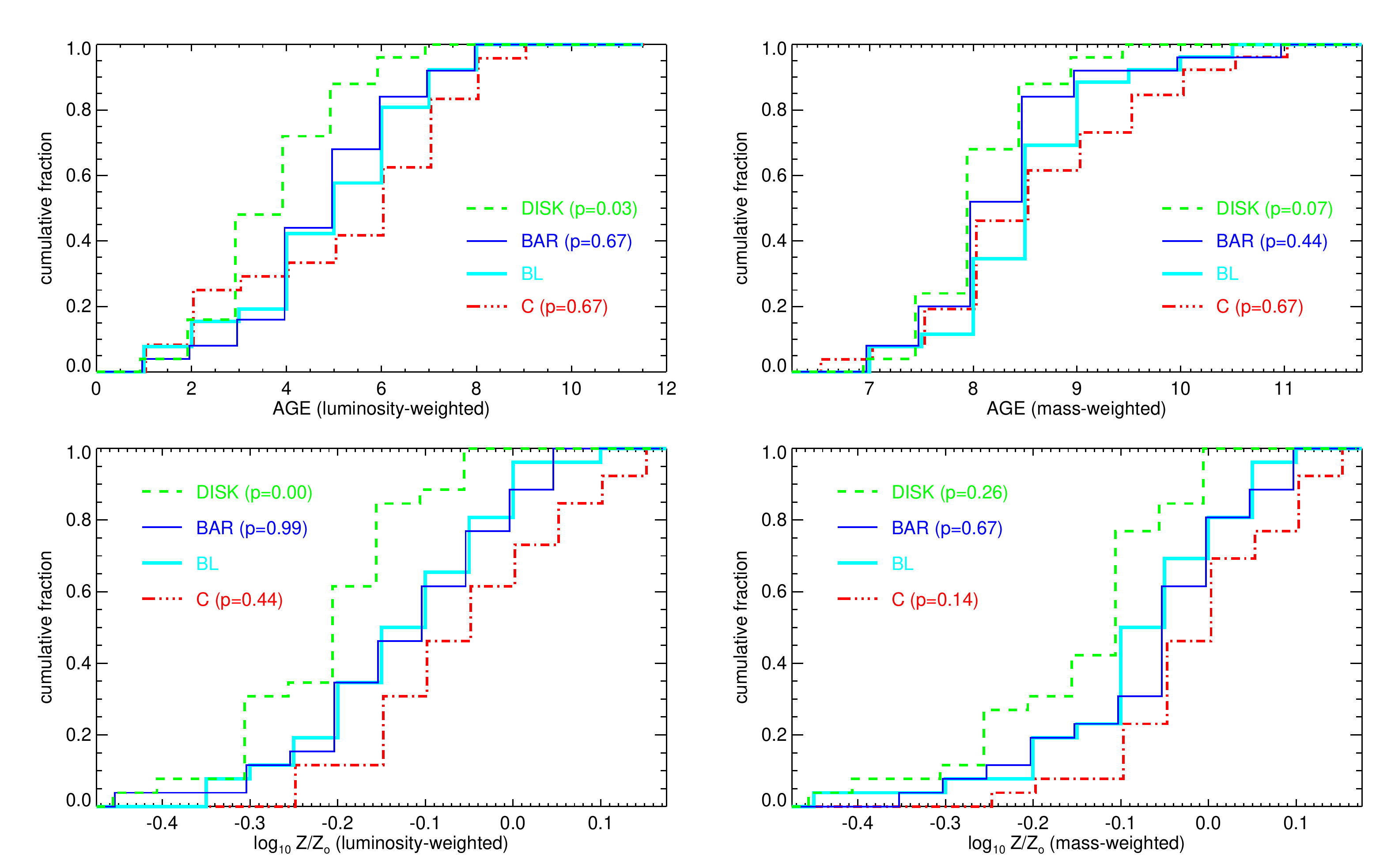}
\caption{Cumulative distributions of the average ages and
  metallicities in the different structure components, measured in
  regions as defined in Figure \ref{regions}. Included are the 26
  barlens galaxies in our sample. The V500 data-cubes by
  \citet{sanchezGZ2016} (SSP.cube.fits) were used, loaded from the
  CALIFA database. Distributions are based on median values of the
  pixels covered by the structure components; practically identical
  distributions are obtained when using flux-weighted means. Labels
  indicate the p-values in KS-tests comparing the distribution with
  that of barlenses ($p<0.05$ indicates a statistically significant
  difference).}
\label{stellar-pop_lum}
\end{figure*}

Above we have discussed the mean stellar ages and metallicities, and
looked at their radial distributions in individual
galaxies. In order to make a more clear comparison between the
structure components, cumulative distributions of the mean stellar
ages and metallicities were derived,
shown in Figure \ref{stellar-pop_lum}. Both luminosity and mass weighted
distributions are shown. While constructing these distributions the
measurements in the zones as defined in Figure
\ref{regions} were used.

It appears that bars and barlenses have remarkably similar age and
metallicity distribution. The luminosity weighted mean stellar ages
are typically 4--8 Gyrs, and the mass weighted indices show stars
older than 8 Gyrs (with a few exceptions). Using the mass weighted
indices, the oldest stars in barlenses are as old as the oldest stars
in the galaxy centers ($\sim$11 Gyrs), which are not much older than
those in bars (i.e. $\sim$10 Gyrs). The metallicities are near solar,
but vary from slightly sub-solar (log$_{10}$ Z/Z$_{\odot}$ = -0.3) to
slightly over-solar metallicities (log$_{10}$ Z/Z$_{\odot}$ = 0.1). In
some galaxies the central regions are dominated by younger stars of
3-6 Gyrs, which can be explained by more recent star formation in
possible nuclear rings, which are not well resolved in the used
data-cubes. The disks within the bar radius have typical luminosity
weighted stellar ages of 3--6 Gyrs, and the oldest stars are $\sim$9
Gyrs old. The disks are on average more metal-poor than the bars and
barlenses, whereas the central galaxy regions are more metal-rich. 

The KS-tests find no significant differences between the age
and metallicity distributions of bars and barlenses. Judging by eye the center regions
and barlenses seem to deviate more: however according to the KS-test 
their differences are not statistically
significant. On other hand the disks have clearly different
luminosity-weighted stellar age and metallicity distributions than
barlenses: p=0.031 and 0.002, respectively, though the difference is no
more statistically significant when the distributions from mass-weighted
indices are compared.  It is worth to keep in mind that our number of
galaxies is fairly small for this kind of statistical tests; also the
overlap of flux between different components tends to dilute possible
underlying differences.

In summary, based on our analysis the stellar age and
metallicity distributions of bars and barlenses are very similar,
and therefore barlenses must have been formed in tandem with the bars.
Barlenses have a range of stellar ages and metallicities which means that
their masses must have been accumulated in several
episodes of star formation. An important fraction of those stars were
formed early in the history of the galaxy.

\section{Discussion}

In the previous sections we have analyzed the photometric bulges,
i.e. the excess mass/flux on top of extrapolated disk profile.  In
which way this mass has accumulated in galaxies, is an important
question in cosmological models of galaxy formation and evolution. The
photometric bulge can consist of a {\it classical bulge}, which is a
spheroidal formed in non-dissipative processes (but see also
Falcon-Barroso et al. 2018, Hopkins et al. 2009), a {\it disky
  pseudo-bulge} (or simply a pseudo-bulge) which is a small
dissipatively formed inner disk, or it can be a {\it Boxy/Peanut/bl}
structure, related to the inner orbital structure of bars.  If the
classical bulge is small all type of structures can form part of the
same photometric bulge.  Not only the Boxy/Peanut/bl, but also the
elongated {\it bar} can comprise an important part of the photometric
bulge. Only the classical bulges are real separate bulge components,
not related to the evolution of the disk.

However, distinguishing the origin of the photometric bulges has
turned out to be complicated. Depending on which emphasis is given to
each analysis method different answers are obtained, and in particular
not much has been done to investigate how should the Boxy/Peanut/bl
bulges appear in the different analysis methods.  As examples of such
controversial results three recent papers are discussed below.  All
papers use mostly CALIFA IFU-observations for kinematics, and in the
structure decompositions the bar flux, if present, is taken away from
the photometric bulge.

\subsection{Interpretation of photometric bulges in three recent studies}

$\bullet$ {\it \citet{neumann2017}} studied 45 non-barred galaxies with a large
range of Hubble types. Their conclusion was that using the Kormendy
relation (log$_{10}$ $R_{\rm e}$ vs. $\mu_e$) and the concentration index
($C_{20,50}$), pseudo-bulges can be distinguished from classical
bulges with 95$\%$ confidence level. In the Kormendy relation they
appeared as outliers toward lower surface brightnesses.  Other
parameters like $B/T$, S\'ersic $n$, and the central $\sigma$-profile
appeared as expected for the two type of bulges.  They found that even 60$\%$ of the
bulges in their sample were ``classical bulges''.  

$\bullet$ {\it \citet{costantin2017}} studied 9 low mass late-type spirals. In
spite of the low galaxy masses, the bulges of these galaxies followed
the same fundamental plane and the Faber-Jackson relation as the
bulges of bright galaxies. In the Kormendy relation they appeared as
low surface brightness outliers as expected for their low galaxy
masses. For these similarities in the photometric scaling relations,
Costantin et al. concluded that there is only a single population of
bulges, which cannot be disk-like systems, i.e. all bulges are
``classical''. 

$\bullet$ {\it \citet{mendez2018}} studied 28 massive S0s, but did not
find any correlation between the photometric ($B/T$, S\'ersic $n$) and
kinematic (the angular momentum $\lambda$ parameter, and
V$_{rot}$/$\sigma$) parameters of the bulges. They ended up to the
conclusion that perhaps all bulges were formed dissipatively. For
massive S0s this happened already at high redshift, for example after
major mergers. The authors reach the opinion that identification 
of bulges photometrically is not meaningful at all.

How to understand these controversial results? It appears that in the
sample by \citet{neumann2017}
a large majority of the galaxies they classified as having
pseudo-bulges have Sc--Scd Hubble types (i.e. have low galaxy masses),
and a similar fraction of their classical bulges have S0--Sb types
(i.e. have high galaxy masses). Having in mind that the Kormendy
relation strongly depends on galaxy mass
\citep{ravikumar2006,costantin2017}, the low surface brightness
outliers in the Kormendy relation, interpreted as pseudo-bulges by
Neuman, Wisotzki and Choudhury, have a natural explanation, which is
reflected also to the concentration parameter $C_{20,50}$: i.e. bulges in
the low mass galaxies have often disk-like properties (e.g. Fisher
$\&$ Drory 2008, 2016).  It is worth noticing that the bulges of the
low mass galaxies in their study have also other indices of
pseudo-bulges, i.e. recent star formation or spiral arms penetrating
into the central galaxy regions.
In the study by Costantin et al.  only the scaling relations
were used to distinguish the type of bulges.  However, the bulges of
the low mass galaxies in their study, interpreted as classical bulges, have recent star formation
or spiral arms penetrating into the central galaxy regions, which are
actually characteristics of disk-like pseudo-bulges.

\citet{mendez2018} made Schwarzschild models, which allow to build up
galaxies by weighting the stellar orbits using the observed
gravitational potential derived from observation.  This makes possible to calculate the
kinematic parameters as in real galaxies, to look at the galaxies at
different viewing angles, and to approximate possible disk
contamination on the bulge parameters. The fact that using these
corrected parameters they did not find any correlation between the
photometric and kinematic parameters of bulges is interesting, because
bright S0s are known to have the most massive bulges in the nearby
universe. We will come back to this question in the next section. 

In conclusion: it seems that the bulges identified as disky
pseudo-bulges in the Kormendy relation, are generally low mass
galaxies which can be recognized as such also via specific
morphological, photometric, or star formation properties. However,
lacking these indicators does not necessarily mean that the bulges
are classical, not even in case when they follow the same scaling
relations (Kormendy, Faber-Jackson, fundamental plane) with the bright
ellipticals.  The scaling relations were introduced to describe
virialized systems such as the elliptical galaxies. Fast rotating
systems can also have fairly large velocity dispersions, and it is not
well studied what kind of deviations from the scaling relations of
ellipticals are expected for such structures as the Boxy/Peanut/bl
components.

\subsection{Nature of barlenses in the CALIFA survey}

\begin{figure}
\centering
\includegraphics[angle=0,width=9.0cm]{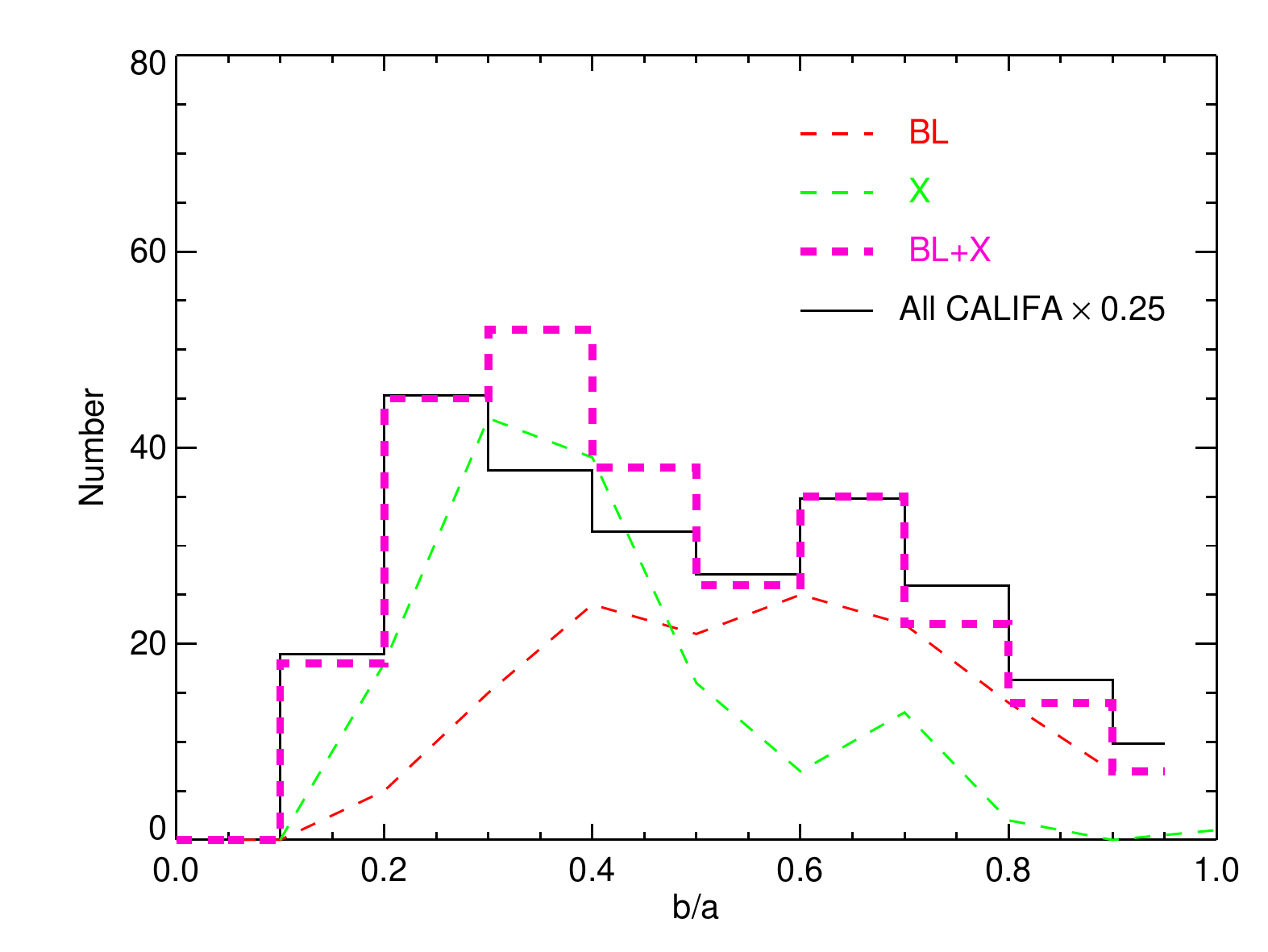}
\caption{The distributions of the minor-to-major ($b/a$) axis
  ratios of the galaxies hosting barlenses (bl; our sample 2) and
  X-shape features (X; our sample 3) in the CALIFA sample.  The
  combined bl+X distribution is compared with that of the
  complete sample of CALIFA galaxies (our sample 1), scaled by a factor of 0.25.
The histograms for both bl and X
include the 15 galaxies were both features appear. In the combined 
bl+X histogram these galaxies are included only once.
The b/a values are from our measurements when available, otherwise from HyperLEDA. However, use of HyperLEDA inclinations would yield very similar distributions.}
\label{fractions}
\end{figure}

\begin{figure}
\centering
\includegraphics[angle=0,width=9.0cm]{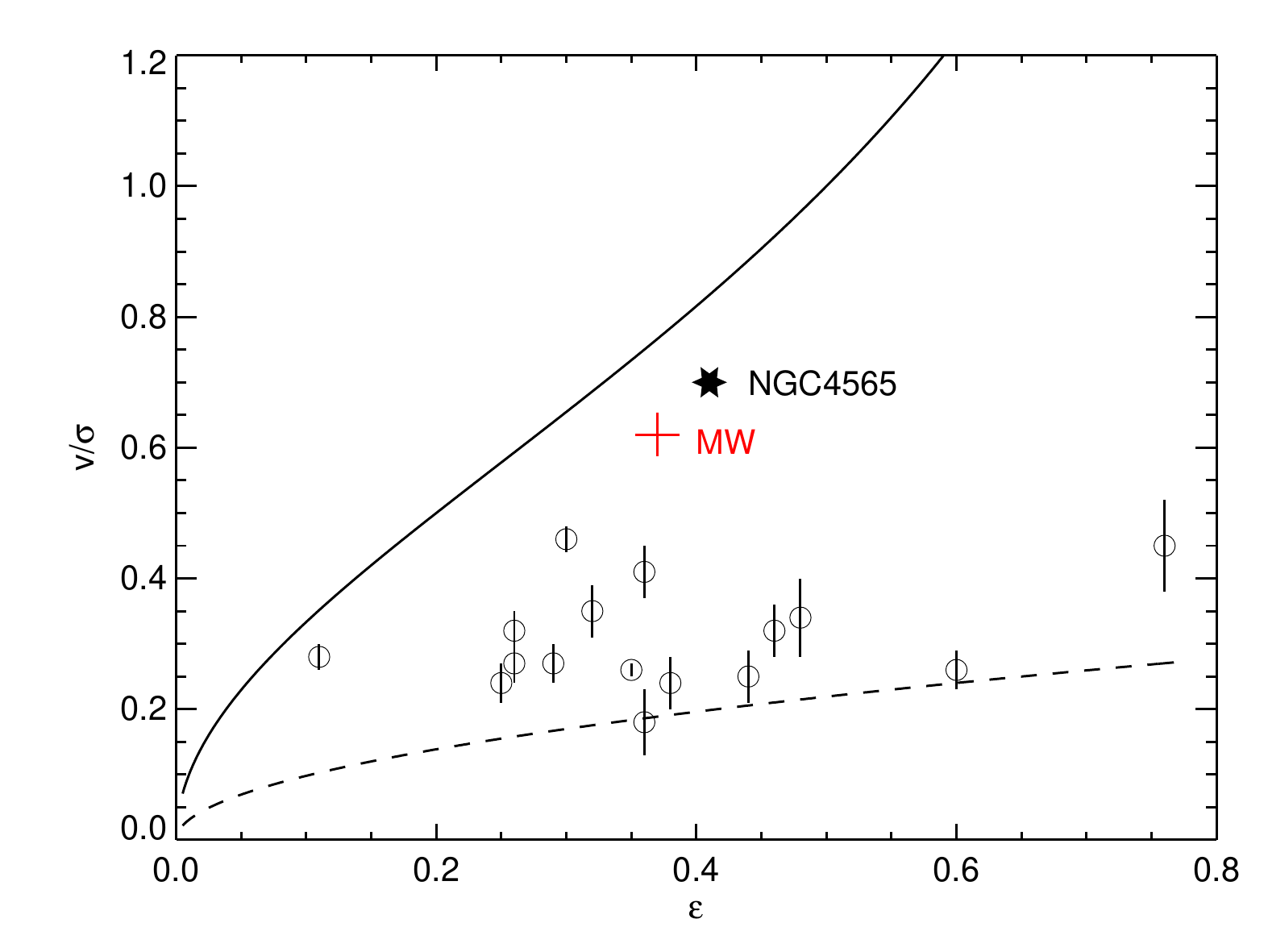}
\caption{The 16 barlens galaxies common with Mendez-Abreu et
  al. (2018) and this work are shown in the V/$\sigma$ -- $\epsilon$
  plane. The parameters refer to values measured within one $R_{\rm
    e}$ of the bulge (see Table 2 in Mendez-Abreu et al. 2018). They
  are corrected for pixelation and for resolution effect (as explained
  in the original paper). The solid line shows the expected relation
  for rotationally flattened oblate spheroids seen edge-on, following the
  approximation V/$\sigma =\sqrt{\epsilon/(1-\epsilon)}$ given in
  Kormendy (1982). The dashed line (V/$\sigma$ =0.31$\sqrt{\epsilon}$) shows the region where the slowly
  rotating bright ellipticals fall in the study by Emsellem et
  al. (2011; see their Fig. 6a). Marked in the figure are also the MW bulge (red cross)
  and NGC 4565 (star) using the values from Rich et al. (2008).}
\label{vmaxsigma}
\end{figure}

In the current study a different approach was taken. We first
identified all the vertically thick inner bar components
(Boxy/Peanut/X/bl) in the CALIFA survey, then made detailed B/D/bar/bl
decompositions for 48 barlens galaxies (out of which 46
were considered reliable), and in a half of them
studied also the stellar populations and metallicities of the
different structure components.

A number histogram of the minor-to-major ($b/a$) axis ratios of the
CALIFA galaxies is shown in Figure \ref{fractions}, and on top of that
histograms of the galaxies with barlenses and X-shape features are
also shown.  As expected, the X-shaped galaxies reside preferably in
highly inclined galaxies and the barlenses in more face-on systems,
with a significant overlap between the two. An important fact is that
the combined distribution of barlenses and X-shapes forms a very
similar histogram as obtained for the complete CALIFA sample. As the
fraction of the vertically thick inner bar components should not
depend on galaxy orientation, the distributions are consistent with
the interpretation that barlenses and X-shape features are
manifestations of the same structure seen at different viewing angles.
The same conclusion for the S$^4$G+NIRS0S galaxies (z $<$ 0.01) was
previously made by \citet{lauri2014}. Taking into account that in
5-10$\%$ of the cases the geometry is not favorable for distinguishing
such components, and assuming that half of the galaxies are barred,
$\sim$50$\%$ of the barred galaxies in CALIFA are estimated to have
vertically thick inner bar components. This is practically the same
percentage as the 46$\%$ found by \citet{lauri2014} for the
S$^4$G+NIRS0S galaxies. The number is consistent also with that
  obtained by Li et al. (2018), who found that 38$\%$ of barred
  galaxies with inclinations $i$ $<$ 70$^\circ$ in the Carnegie-Irvine
  Galaxy Survey, have either a barlens or a Boxy/Peanut. The galaxies
  they studied are at similar distances and have similar host galaxy
  masses as those in S$^4$G+NIRS0S.

Nearly 60$\%$ (16/28) of the kinematic sample of M\'endez-Abreu et al. (2018)
form part of our sample of barlens galaxies, of
which 10 were decomposed by us. In their study the B/D/bar
decompositions from MA2017 were used. 
All the bulges in \citet{mendez2018} follow the same Kormendy relation
as the bright elliptical galaxies (see their Fig. 8), including the 16
barlens galaxies. However, if we use the (V/$\sigma$)
  -- $\epsilon$ plane diagnostics \footnote{(V/$\sigma$) is the
    luminosity weighted rotation velocity within the bulge $R_{\rm e}$, and
    $\epsilon$ = (1 - axial ratio) within the same radius.}, using the
  kinematic parameters of bulges within 1 $R_{\rm e}$, given by
  \citet{mendez2018}, the 16 barlens galaxies fall above the bright
  elliptical galaxies. These galaxies are typically slowly rotating systems (see Fig. \ref{vmaxsigma})
which, being heated systems have lost their opportunity to become fast rotating systems
anymore (see \citet{emsellem2011}).  Barlenses appear in the same
region with the fast-rotating ellipticals, which adds one more
complexity for the interpretation of this diagram, i.e. the vertically
thick inner bar components (barlenses) can have similar kinematic
properties as those classical bulges formed by wet major or minor
mergers (see Naab et al. 2014).  This diagram also shows that
barlenses are not
oblate systems as expected in case of disky pseudo-bulges
\citep{pfenniger1990,friedli1995}.  

\begin{figure}
\centering
\includegraphics[angle=0,width=9.0cm]{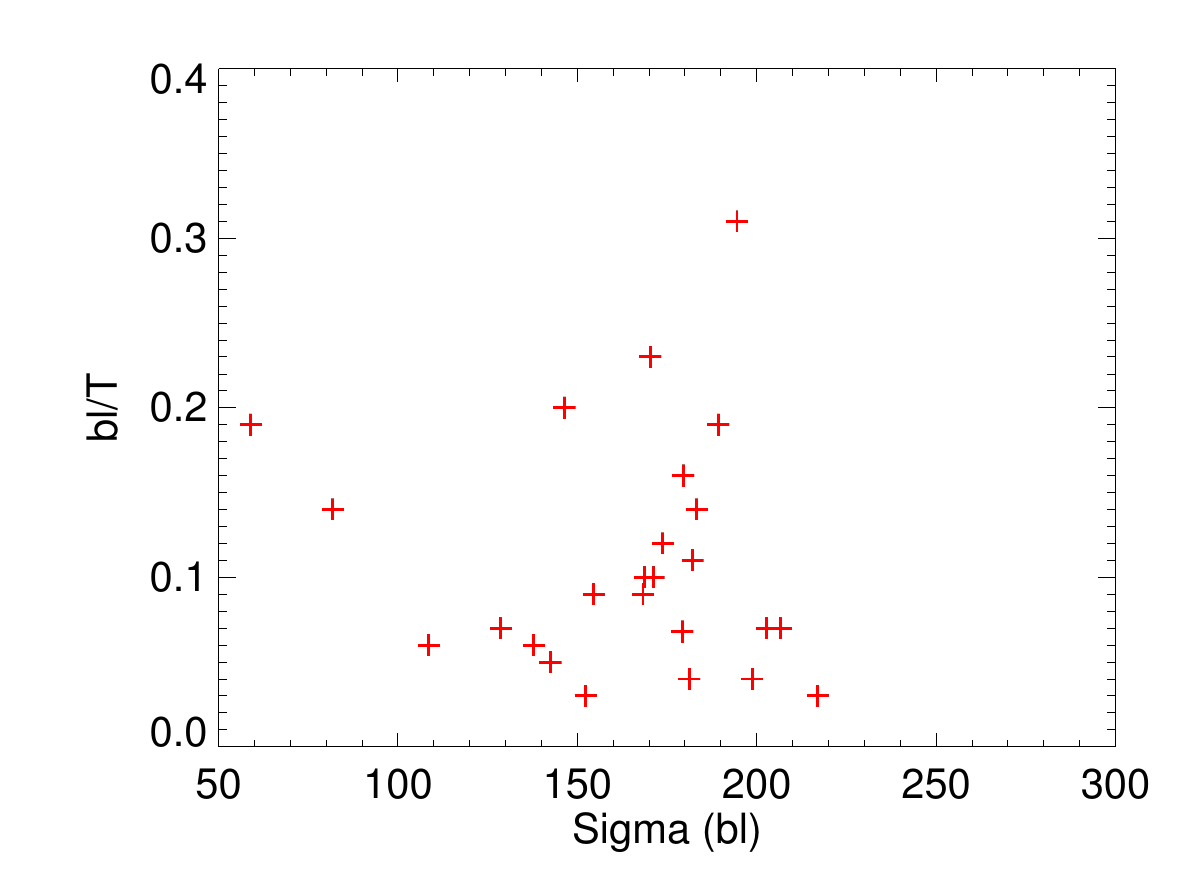}
\caption{The relative flux of barlens ($bl/T$) 
as a function of velocity dispersion $\sigma$ of the same component.
Plotted are the barlenses in our sub-sample of 26 galaxies.}
\label{blT_sigma}
\end{figure}

A novelty of our study was to make a hypothesis that the photometric
bulges in the barred CALIFA galaxies are largely dominated by
barlenses, of which clear examples with identified barlenses were
studied. Although our starting point was morphological, also the
studied physical parameters were found to be consistent with this
picture. Taking this view, the inconsistencies in the literature, using the different
analysis methods applied to barlens galaxies, become more
understandable.  We found that not only the (g'-r') and (r'-i')
colors, but particularly the cumulative distributions of stellar ages
and metallicities are very similar in bars and barlenses. The stars in
barlenses were accumulated in a large time period in several episodes
of star formation, manifested in a range of stellar ages of (4--11
Gyrs) and metallicities. It seems that barlenses were gradually
increased in mass, in tandem with the rest of the bar. The bar origin
of the barlens was further confirmed for a few galaxies showing
X-shape features in the unsharp mask images. The mass weighted stellar
ages of barlenses are also similar as those obtained by
\citet{perez2017} for the X-shaped bar in NGC 6032 (mean
age $\ge$ 6 Gyrs).  Prominent dust lanes in some of the barlens
galaxies in our sample show how gas can penetrate through the barlens
possibly triggering star formation in the galaxy center, and at some
level also in the barlens itself. The relative fluxes of barlenses
($bl/T$) in our decompositions do not correlate with the stellar
velocity dispersions measured in the same galaxy regions (see Fig. \ref{blT_sigma}).

Our finding that the old and intermediate age stars dominate bars and
barlenses is consistent with the kinematic analysis of the CALIFA
survey by \citet{zhu2018}. The kinematic decompositions by Zhu et
al. uses the parameter $\lambda_{\rm z}$ (orbit angular momentum relative to
circular orbit with the same energy) as a dividing line: the orbits
are defined cold when $\lambda_{\rm z}>$ 0.8, hot when $\lambda_{\rm z} <$ 0.1, and
warm in between these two $\lambda_{\rm z}$-values.  In their study the
photometrically identified bulges (flux on top of the disk) are
dominated both by hot and warm orbits, corresponding to our old and
intermediate age stars. Only in the most massive galaxies with M$_{\star}$
$>$10$^{11}$ (not included in our sample) the bulges are dominated by
hot orbits. 

If the photometric bulges in the Milky Way mass disk galaxies were
dominated by bars and barlenses there is no reason why they should
appear in the same location with the elliptical galaxies in the
V/$\sigma$ - $\epsilon$ plane. Neither should they behave in a similar
manner as the star forming pseudo-bulges in the disk
plane. Barlenses can be dynamically hot, have fairly high effective
surface brightnesses ($\mu_{\rm e}$) for given effective radii ($R_{\rm
  e}$), and also to have fairly old stellar populations, which in some
analysis methods can mislead the interpretation of their origin.

\subsection{Comparison with the Milky Way bulge}

The Milky Way (MW) bulge is known to have strong evidence for being of
bar origin. The bulge is X-shaped and cylindrically rotating, as
detected in the distribution of the red clump giant stars
\citep{wegg2013,ness2016}. From our point of view it is important that
the MW bulge can be considered also as a barlens: in face-on view the
bulge has been suggested to have a similar morphology as the barlens in NGC 4314, a galaxy
seen nearly face-on (see the review by Bland-Hawthorn $\&$ Gerhard
2016). In CALIFA, for example NGC 7563 has a similar galaxy/barlens morphology.

The MW bulge is dominated by old red clump giant stars
\citep{mcwilliam1994,zoccali2008}, but has also stars with
intermediate ages \citep{clarkson2011,bensby2011}, which is
qualitatively in agreement with what we see also in barlenses. The
main body of the stars have slightly sub-solar metallicities and ages between
9--13 Gyrs, and a small contribution of stars younger than 5 Gyrs. The
X-feature in the MW bulge has been detected in the distribution of the
red clump stars \citep{mac2010,nataf2010,wegg2013,ness2016}, but not
in that of the very old RR Lyrae stars (e.g. Wegg $\&$ Gerhard 2013),
which has provoked a discussion of a possible additional classical
bulge in the MW. The MW bulge has also a nuclear disk at r $<$ 200 pc,
showing recent star formation \citep{laundhart2002}. Nuclear disks
with stellar ages of $<$ 4 Gyrs appear also in some of the barlens
galaxies in our sample. In fact, many barlenses in the local universe
have embedded star forming nuclear rings or disks
\citep{lauri2011}. However, in CALIFA sample the resolution is not
ideal for detecting those.

In the MW bulge the high-metallicity (HM) stars are concentrated to
the X-feature and close to the Galactic plane
\citep{zoccali2008,gonzalez2013,johnson2011}, whereas the low
metallicity (LM) RR Lyrae stars form a more round component
\citep{decany2013}, which is also centrally peaked
\citep{pietrukowicz2015}. Together the LM and HM stars make the
observed metallicity gradient, so that the metallicity decreases
toward higher galactic latitudes. For barlenses we do not have
information about the vertical metallicity distributions, but as we
are looking at the galaxies in fairly face-on view, most probably we
see a superposition of both components, manifested as a large range of
stellar metallicities. Both in the MW bulge and in the barlenses the
oldest metal-poor stars form a minority in their stellar populations.
As shown in Figure \ref{vmaxsigma}, 
in the V/$\sigma$ - $\epsilon$ plane the MW bulge appears
slightly below the oblate line, and is clearly above the bright
elliptical galaxies \citep{rich2008,minniti2008}. In fact, this is
what we see also for barlenses in our sample. It is interesting
that in that diagram the MW bulge, seen almost end-on ($\phi$ =
27$^\circ$; Wegg $\&$ Gerhard 2013), appears almost in the same location with
NGC 4565, which is considered as a twin of the MW: NGC 4565 is seen
nearly end-on and has an X-shaped bulge \citep{kormendy2010,lauri2014}.

\subsection{Explaining the Boxy/Peanut/X/bl structures} 

The backbone of the vertically thick inner bar components consist of
both stable periodic orbits, and unstable periodic orbits linked to
chaos (see the review by Athanassoula 2016). It has been suggested
\citep{abbott,wegg2013,portail2015} that the Boxy/X-shaped bulge of
the MW consists of a superposition of various types of bar orbits, a
majority of them being non-resonant box orbits (constituting 60$\%$ of
bar orbits). Only the banana type X1 and the resonant boxlet
(i.e. fish/brezel) orbits make the X-shape feature.  Most probably
barlenses are a superposition of similar orbital families. In the
simulation models the X-features are like horns which are extended
both in xy and xz-directions \citep{atha2015,salo2017}. When the
galaxy inclination decreases the banana type orbits, visible in edge-on view, gradually become
over-shadowed by the more circular or chaotic orbits possibly making
the barlens appearance. In the overall morphology this is shown with
the simulation models by \citet{salo2017}, and is manifested also in
the CALIFA sample as a gradual changing of the X-shape features into
barlenses toward lower galaxy inclinations (Fig. \ref{fractions}). If
the fraction of the central concentration is not high enough or even if the 
resolution in the simulation models is insufficient, e.g. due to large gravity softening, the
vertically thick inner bar component is centrally pinched at all
galaxy inclinations \citep{salo2017}. Such pinched structures in
face-on view are shown also in the simulation models by
\citet{saha2018}. However, in observations such pinched structures 
in nearly face-on view are rare.

In the interpretation of the MW bulge a critical point has been to
explain why the fairly round component with old metal-poor RR Lyrae
stars exists, and what causes the vertical metallicity
gradient. Depending on the model, either a small or no classical bulge
has been suggested, superimposed with the Boxy bulge. In the
simulation models of \citet{shen2010} the relative flux of the
classical bulge is 8--10$\%$ at most, which is consistent with that we
obtained in the barlens galaxies: i.e $\le$ 10$\%$ of the total galaxy
flux appears in a component, which could be interpreted as a possible
classical bulge.

Detailed models for the MW bulge have shown that it is possible to explain,
not only the X-feature, but also the more round LM component without
invoking the concept of a classical bulge. 
{\citet{perezvillegas2017}} explained the old metal-poor population of the MW
  bulge as an inward extension of the slowly rotating
  metal-poor stellar halo. The mix of stars in their N-body models is
  due to the gravitational interaction between the bar and the
  Boxy/Peanut when they form and evolve. 
In the chemo-dynamical models by \citet{atha2017} the MW bulge
  morphology is coupled with the kinematics, metallicity, and stellar
  ages. The different metallicity bins have specific velocity
  dispersions and different locations in the bulge. 
The LM population makes the more
  round component including the thick disk, stellar halo, and a
  possible small classical bulge, 
whereas the HM population contributes only to the X-feature. The
coupling of stellar ages and metallicity, with the kinematics and
morphology in the MW bulge, has been discussed in detail by
\citet{portail2015}. 

It seems that although detailed information of the stellar populations
of the MW bulge is available, it has not been possible to
unambiguously show that there exist also a small classical bulge
embedded in the Boxy bulge. At some level the same concerns our analysis of
barlenses in this study. In half of the studied galaxies separate
bulge components were fitted, which components are potential small
classical bulges. However, instead of being dominated by the oldest
stars ($>$ 10 Gyrs) as expected for merger built structures, in many
barlens galaxies the fraction of the oldest stars even drops in the galaxy
center. The most prominent barlenses in the simulation models form in
centrally peaked galaxies \citep{salo2017}, but a question remains
which forms first, the central mass concentration or the
barlens. Also, in half of the barred MW mass galaxies in the CALIFA
sample no vertically thick inner bar components were identified. It
needs to be further investigated in which physical conditions they
form in galaxies. An interesting observation pointed out in this study was a
drop of the most metal-poor stars, both in the galaxy center and 
in the barlens region, and in some galaxies also in the elongated bar.  In
the simulation models by \citet{perez2017} some migration of stars
occurred within the Boxy/Peanut, but the stars formed in the galaxy
center and in the outer part of the bar stayed in their original
locations.

\vskip 0.20cm
\subsection{Bars in the context of galaxy formation and evolution} 
\vskip 0.10cm

It has been shown by \citet{scannapieco2012} that realistic bars can form
in hydrodynamical cosmological simulations. The disk and even the dark
matter halo keep growing while the bar forms.  They re-run one of
their simulations with higher resolution using the Tree-PM SPH code,
including star formation and chemical enrichment. The model ended up
to form bars which have morphologies, sizes, and surface brightness
profiles similar to those observed at z = 0. \citet{atha2017} studied
the bar/bulge regions of galaxies starting from proto-galaxies
composed only of dark matter and gas. In their models the stars which
ended up mainly to a classical bulge and to a stellar halo, were
formed in a short time period (1.4 Gyr) before the galaxy merger. The
stars born during the merger at high redshift contributed to a thick
disk and a spheroid, whereas the gas accreted from the halo formed a
thin disk, including the bar and the Boxy/Peanut bulge. So, based on these
models the stellar populations of the classical bulges are expected
to have clearly older stellar populations than the Boxy/Peanut/bl bulges.

The galactic halos can interact with the environment, and
therefore continuously evolve. Although a small fraction of
stars in the halos might be accreted as small satellites, most of the
halo stars were formed in starbursts, which stars are $\sim$1 Gyr
younger, have higher metallicities, and are less $\alpha$-enhanced
than the accreted stars \citep{tissera2018}. According to the
simulation models
\citep{hirschmann2013,hopkins2014,stinson2013,woods2014} gas accretion
to the disk is de-coupled from the halo assembly, which allows
continuous gas accretion keeping the star formation rate in the
disk constant \citep{christensen2016,oppenheimer2010,ubler2014}. As a
consequence, the structures formed of the disk stars are expected to
have a large range of stellar ages and metallicities, which is also what we see
in bars and barlenses in this study.

The Boxy/Peanut/bl structures are suggested to be triggered by the
vertical buckling instability
\citep{raha1991,dubinski2003,merritt1994}, or via trapping of disk
stars at vertical resonances
\citep{combes1981,combes1990,quillen2002}. If the bar buckling is the
dominant mechanisms, most probably several buckling events have
occurred during the lives of the galaxies \citep{valpuesta2006} mixing
the stellar populations, which again is consistent with
our observation that barlenses are dominated by stars in a large range
of stellar ages and metallicities. In the simulation models by
\citet{perez2017} there is a time delay of a few Gyrs between the bar
formation and the first buckling event.  There are many morphological
features and parameters which link barlenses to bars. For example, in
the models by \citet{atha2015} and \citet{salo2017}, barlenses
formed in the simulation models have very similar relative sizes as
the barlenses in observations, which is the case also with the galaxies in the CALIFA sample
(r(bl)/r(bar) $\sim$ 0.5).  In the simulation models by \citet{collier2017} the
evolved strong bars have also ansae, i.e. flux concentrations at the
two ends of the bar, which features are observed in nearly half of the
barlens galaxies \citep{lauri2013}.

In which way the bars evolve depends also on the properties of their
halos. In axisymmetric halos bars loose angular momentum to the halos,
and as a consequence they grow in size, temporarily weaken, and after
growing again after the second buckling, bars will stabilize
\citep{valpuesta2006,perez2017}. However, if the halos are spinning
fast ($\lambda >$ 0.06), the bars are predicted to become hot
\citep{collier2017}. Being dominated by chaotic orbits the bar cannot
reform anymore, leaving behind only an oval-shaped slow rotating
structure.  In the models by Collier, Shlosman $\&$ Heller the bars
formed in slowly spinning halos ($\lambda \le$ 0.03) are long and fast
rotating, and they have typically offset dust-lanes, which are
manifestations of standing shocks in the gas flow \citep{atha1992}. We
find such dust-lanes in some of the barlenses in the CALIFA sample
(see Fig. \ref{decompositions_3}). On the other hand, barlenses like
the one observed in NGC 0776 (see Fig. \ref{age_met_profiles_3}) might
be a consequence of the evolution in a fast spinning halo. This
example (and other similar cases shown by Laurikainen $\&$ Salo 2017) opens a
possibility that the photometric bulges even in many galaxies
classified as non-barred, might have been formed in a similar manner
as in barred galaxies. Possible formation of nearly axisymmetric boxy bulges was
first suggested by \citet{rowley1988} and later by \citet{patsis2006}.

\section{Summary and conclusions}

The Calar Alto Legacy Integral Field Area Survey (CALIFA) of 1064
galaxies was used to identify the vertically thick inner bar
components, which are X-shaped in edge-on view, and have a barlens
morphology in less inclined galaxies. A sub-sample of 46 barlens
galaxies was successfully decomposed to different structure components using the
r'-band mosaic images of SDSS-DR12. A sub-sample of 26 galaxies
was further analyzed using the publicly available CALIFA IFU data-cubes
with V500 grating.  CALIFA consists of galaxies with the mass range of
M$_{\star}$/M$_{\odot}$ = 10$^{9.7}$--10$^{11.4}$, at redshifts z =
0.005--0.03. While identifying and fitting the Boxy/Peanut/X/bl
structures the simulation models by \citet{salo2017} were used as a
guide.

This is the first time that barlens galaxies are studied combining the
photometric multi-component decompositions with the IFU stellar
population analysis. Using GALFIT, we made new multi-component
bulge/disk/bar/barlens (B/D/bar/bl) decompositions for obtaining the
relative fluxes of the different structure components. In comparison
to the previous decompositions made by MA2017, a novelty of our study
was that also barlenses were fitted with a separate function. In order
to reduce the number of free parameters in the decompositions, the
sizes and ellipticities of bars and barlenses were measured from direct images. The
CALIFA data-cubes were properly binned to cover the
different structure components, for which bins mass and luminosity
weighted stellar ages and metallicities and stellar velocity
dispersions ($\sigma$) were calculated. The properties of the
structure components were studied, based on the mean values, obtaining
cumulative fractions of the stellar ages and metallicities, and
showing the radial profiles of the parameters for individual
galaxies.
\vskip 0.25cm 
\noindent The main results are summarized below:
\vskip 0.2cm

\begin{itemize}
\item We found that the distribution of the minor-to-major ($b/a$) axis ratios 
of the host galaxies is similar
  for the combined bl+X sample, and for the complete CALIFA
  sample (see Fig. \ref{fractions}).  This supported the idea that
  barlenses are the face-on counterparts of the X-shaped bars.
  Assuming that half of the galaxies are barred, $\sim$50$\%$ of the bars in
  the complete CALIFA sample are estimated to have vertically thick inner bar components.
\vskip 0.20cm

{\it Multi-component decompositions:}
\vskip 0.1cm
\item In our decompositions bulges, barlenses, and disks were fitted with a S\'ersic
  function, and bars with a Ferrers function. A comparison to MA2017
  showed that our B/D/bar decomposition method is robust: it is not
  sensitive to the algorithm used, or how the the underlying disk was
  fitted (with S\'ersic $n <$ 1 or with two truncated disks). 
\vskip 0.15cm

\item In the final models the bar radial profile shape parameters were
  not fixed, leading to values of $\alpha\sim$0.15 in the Ferrers
  function, which was important in our B/D/bar/bl models. This
  corresponds to relatively flat bars with sharp outer truncations,
  the central bright component of the bar being accounted by the
  bl-component.
\vskip 0.15cm

\item We showed that fitting the central density peak and the barlens flux with
  separate S\'ersic functions is of critical importance. Doing so we
  found that on average 13$\%$ of the total galaxy light is associated to barlenses,
  and $\le$10$\%$ for possible separate bulge components. Both
  components were found to be nearly exponential or to have S\'ersic
  $n<$ 1. We compared B/D, B/D/bar and B/D/bar/bl decompositions: the
typical $B/T$-value decreases from $B/T\sim$0.3, to $\sim$0.15, and to $\sim$0.06
in the three models, respectively.    
\vskip 0.15cm

\item A simulation snapshot was decomposed in a similar manner as real
  galaxies. The snapshot was taken from the N-body simulations of \citet{salo2017},
  in which a barlens formed during the galaxy evolution, and a small
  pre-existing bulge was present at the beginning of the simulation. Our
  decomposition retrieved well both the flux of the pre-existing
  bulge, and the distribution of the particles forming the barlens structure.
\vskip 0.2cm

{\it Using the CALIFA V500 data-cubes:}
\vskip 0.1cm

\item Barlenses were found to have similar cumulative fractions of
  stellar ages and metallicities as bars, both using the mass and
  light weighted indices (see Fig. \ref{stellar-pop_lum}). It means
  that barlenses were accumulated in tandem with the bars, in a large time
  period. Also their mean stellar velocity dispersions are very similar ($\Delta \sigma \sim$ 20 km/sec).
\vskip 0.15cm

\item The mass and light weighted stellar ages of bars and barlenses
  are on average $\sim$9 and $\sim$5 Gyrs, respectively. The range
  of light weighted ages is 4--8 Gyrs. The oldest stars are $\sim$ 11
  Gyrs, which are as old as the stars in the galaxy centers. In the
  centrally peaked barlenses the stars are on average $\sim$1 Gyr
  older than in the non-centrally peaked barlenses. 
  \vskip 0.15cm

\item The mean metallicities of bars and barlenses are near solar, but
  there is a range of metallicities between log$_{10}$ Z/Z$_{\odot}$ = -0.3 --
  +0.1. The galaxy centers are more metal-rich than barlenses, whereas
  the disks are less metal-rich. In barlenses the fraction of the most
  metal-poor stars (log$_{10}$ Z/Z$_{\odot}$ = -0.7) rapidly drops, which
  in some galaxies starts already in the bar region.
\end{itemize}

\vskip 0.25cm We have shown that the photometric bulges (i.e. flux on
top of the disk) in barred CALIFA galaxies are dominated by 
vertically thick inner bar components (i.e by barlenses), and not by any
separate bulge components. The obtained stellar ages and metallicities
of barlenses are in a qualitative agreement with those seen in the MW
bulge. We also discussed that in the traditional methods for
distinguishing the different type of galactic bulges, the vertically
thick inner bar components are often ignored, which can lead to
contradictory interpretations of their origin.

\begin{acknowledgements}
We acknowledge the anonymous referee of valuable comments. 
This study uses data provided by the Calar Alto
Legacy Integral Field Area (CALIFA) survey (http://califa.caha.es/)."
"Based on observations collected at the Centro Astron\'omico Hispano
Aleman (CAHA) at Calar Alto, operated jointly by the
Max-Planck-Institut f\"ur Astronomie and the Instituto de Astrof\'isica de
Andaluc\'ia (CSIC). Laurikainen and Salo acknowledge financial support from 
the Academy of Finland (grant n:o 297738), and the European Union’s Horizon
2020 research and innovation programme under the Marie Skłodowska-Curie
grant agreement No 721463 to the SUNDIAL ITN network.

\end{acknowledgements}


\begin{thebibliography}{}

\bibitem[\protect\citeauthoryear{Abbott et al.}{2017}]{abbott} Abbott, C.G., Valluri, M., Shen, J., Debattista, V. P. 2017, MNRAS, 470, 1526
\bibitem[\protect\citeauthoryear{Abraham et al.}{1996}]{abraham1996} Abraham, R.G., van den Bergh, S., Glazebrook, K., Ellis, R.S., Santiago, B. X., Surma, P., Griffiths, R.E. 1996, ApJS, 107, 1
\bibitem[\protect\citeauthoryear{Alam et al.}{2016}]{alam2016} Alam, S., Albareti, F. D., Allende, P.C., Anders, F., Anderson, S. F., Anderton, T., Andrews, B. H., Armengaud, E., Aubourg, E., Bailey, S. et al. 2016, ApJS, 219, 12
\bibitem[\protect\citeauthoryear{Athanassoula}{1992}]{atha1992} Athanassoula, E. 1992, MNRAS, 259, 358
\bibitem[\protect\citeauthoryear{Athanassoula $\&$ Bureau}{1999}]{atha1999} Athanassoula, E., Bureau, M. 1999, ApJ, 522, 699
\bibitem[\protect\citeauthoryear{Athanassoula et al.}{2015}]{atha2015} Athanassoula, E., Laurikainen, E., Salo, H., $\&$ Bosma, A. 2015, MNRAS, 454, 3843 (A+2015)
\bibitem[\protect\citeauthoryear{Athanassoula}{2016}]{atha2016} Athanassoula E. 2016 in 'Galactic Bulges', Astrophysics and Space Science Library, Volume 418. ISBN 978-3-319-19377-9. Springer International Publishing Switzerland, 2016, p. 391, eds. E. Laurikainen, R.F. Peletier, D.A. Gadotti
\bibitem[\protect\citeauthoryear{Athanassoula, Rodionov $\&$ Pranzos}{2017}]{atha2017} Athanassoula, E., Rodionov, S.A., Pranzos, N. 2017, MNRAS, 467, L46   
\bibitem[\protect\citeauthoryear{Bender, Burstein $\&$ Faber}{1992}]{bender1992} Bender, R., Burstein, D., Faber, S. M. 1992, ApJ, 399, 462
\bibitem[\protect\citeauthoryear{Bensby et al.}{2011}] {bensby2011} Bensby, T., Ad\'en, D., Mel\'endez, J., et al.  2011, AA, 533, 134
\bibitem[\protect\citeauthoryear{Berenzen et al.}{1998}] {berenzen1998} Berenzen, I., Heller, C.H., Shlosman, I., Fricke, K.J. 1998, MNRAS, 300, 49
\bibitem[\protect\citeauthoryear{Bertin $\&$ Arnouts}{1996}]{bertin1996} Bertin, E., Arnouts, S. 1996, APS, 117, 393
\bibitem[\protect\citeauthoryear{Bizyaev et al.}{2014}]{bizyaev2014} Bizyaev, D. V., Kautsch, S. J., Mosenkov, A. V., Reshetnikov, V. P., Sotnikova, N. Ya., Yablokova, N. V., Hillyer, R. W. 2014, ApJ, 787, 24
\bibitem[\protect\citeauthoryear{Bland-Hawthorn $\&$ Gerhard}{2016}]{bland2016} Bland-Hawthorn, J., Gerhard, O. 2016, ARA$\&$A, 54, 529
\bibitem[\protect\citeauthoryear{Bournaud, Elmegreen $\&$ Elmegreen}{2007}]{bournaud2007} Bournaud, F., Elmegreen, B. G., Elmegreen, D. M. 2007, ApJ, 670, 237
\bibitem[\protect\citeauthoryear{Bournaud et al.}{2014}]{bournaud2014} Bournaud, F., Perret, V., Renaud, F., et al.  2014, ApJ, 780, 57  
\bibitem[\protect\citeauthoryear{Bournaud}{2016}]{bournaud2016} Bournaud, F. 2016, in 'Galactic Bulges', Astrophysics and Space Science Library, Volume 418. ISBN 978-3-319-19377-9. Springer International Publishing Switzerland, 2016, p. 355, eds. E. Laurikainen, R.F. Peletier, D.A. Gadotti
\bibitem[\protect\citeauthoryear{Bureau et al.}{2006}]{bureau2006} Bureau, M., Aronica, G., Athanassoula, E., Dettmar, R.-J., Bosma, A., Freeman, K.C. 2006, MNRAS, 370, 753
\bibitem[\protect\citeauthoryear{Buta et al.}{2015}]{buta2015} Buta, R., Sheth, K., Athanassoula, E. et al. 2015, ApJS, 217, 32
\bibitem[\protect\citeauthoryear{Christensen et al.}{2016}]{christensen2016} Christensen, C.R., Dav\'e, R., Governato, F., Pontzen, A., Brooks, A., Munshi, F., Quinn, T., Wadsley, J. 2016, ApJ, 824, 57
\bibitem[\protect\citeauthoryear{Ciambur $\&$ Graham}{2016}]{ciambur2016} Ciambur, C., Graham, A. 2016, MNRAS, 459, 1276
   \bibitem[\protect\citeauthoryear{Clarkson et al.}{2011}]{clarkson2011}  Clarkson, W.I., Sahu, K.C., Anderson, J. et al. 2011, ApJ, 735, 37
\bibitem[\protect\citeauthoryear{Collier, Shlosman $\&$ Heller}{2017}]{collier2017} Collier, A., Shlosman, I., Heller, C. 2017, astro-ph: 1712.02802
\bibitem[\protect\citeauthoryear{Costantin et al.}{2017}]{costantin2017} Costantin, L., M\'endez-Abreu, J., Corsini, E. M., Morelli, L., Aguerri, J. A. L., Dalla Bonta, E., Pizzella, A. 2017, AA, 601, 84
\bibitem[\protect\citeauthoryear{Combes}{1981}]{combes1981} Combes F., Sanders R. H., 1981, A$\&$A, 96, 164
\bibitem[\protect\citeauthoryear{Combes et al.}{1990}]{combes1990} Combes F., Debbasch F., Friedli D., Pfenniger D., 1990, AA, 233, 82
\bibitem[\protect\citeauthoryear{Combes}{2014}]{combes2014} Combes, F. 2014, AA, 571, 82
\bibitem[\protect\citeauthoryear{Cowie et al.}{1996}]{cowie1996} Cowie, L. L., Songaila, A., Hu, E. M., Cohen, J. G. 1996, AJ, 112, 839 
\bibitem[\protect\citeauthoryear{Daddi et al.}{2005}]{daddi2005} Daddi, E., Renzini, A., Pirzkal, et al.  2005, ApJ, 626, 680
\bibitem[\protect\citeauthoryear{Damjanov, Abraham $\&$ Glazebrook }{2011}]{damjanov2011} Damjanov, I., Abraham, R. G., Glazebrook, K.  2011, ApJ, 739, 44
\bibitem[\protect\citeauthoryear{Debattista et al.}{2005}]{debattista2005} Debattista, V. P., Carollo, C. M., Mayer, L., Moore, B. 2005, ApJ, 628, 678
\bibitem[\protect\citeauthoryear{D\'ek\'any et al.}{2013}]{decany2013} D\'ek\'any, I., Minniti, D., Catelan, M., Zoccali, M., Saito, R. K., Hempel, M., Gonzalez, O. A. 2013, ApJ, 776, 19
\bibitem[\protect\citeauthoryear{D\'iaz-Garc\'ia et al.}{2016}]{simon2016} D\'iaz-Garc\'ia, S., Salo, H., Laurikainen, E., Herrera-Endoqui, M. 2016, AA, 587, 160
\bibitem[\protect\citeauthoryear{Eggen, Lynden-Bell $\&$ Sandage}{1962}]{eggen1962} Eggen, O.J. Lyndel-Bell, D., Sandage, A.R. 1962, ApJ, 136, 748
\bibitem[\protect\citeauthoryear{Elmegreen, Elmegreen $\&$ Ferguson}{2005}]{elmegreen2005} Elmegreen, D.M., Elmegreen, B. G., Ferguson, T. E. 2005, ApJ, 623, 71
\bibitem[\protect\citeauthoryear{Elmegreen, Bournaud $\&$ Elmegreen}{2008}]{elmegreen2008} Elmegreen, B. G., Bournaud, F., Elmegreen, D. M. 2008, ApJ, 688, 67
\bibitem[\protect\citeauthoryear{Emsellem et al.}{2011}]{emsellem2011} Emsellem, E., Cappellari, M., Krajnovi\'c, D., Alatalo, K., Blitz, L., Bois, M. et al. 2011, MNRAS, 414, 888
\bibitem[\protect\citeauthoryear{Erwin $\&$ Debattista}{2013}]{erwin2013} Erwin, P., Debattista, V.P. 2013, MNRAS, 431, 3060
\bibitem[\protect\citeauthoryear{Erwin $\&$ Debattista}{2017}]{erwin2017} Erwin, P., Debattista, V.P. 2017, MNRAS, 468, 2058
\bibitem[\protect\citeauthoryear{Falc\'on-Barroso, Peletier $\&$ Balcells}{2002}]{falco2002} Falc\'on-Barroso, J., Peletier, R. F., Balcells, M. 2002, MNRAS, 335, 741
\bibitem[\protect\citeauthoryear{Falc\'on-Barroso et al.}{2011}]{falco2011} Falc\'on-Barroso, J., van de Ven, G., Peletier, R. F. 2011, MNRAS, 417, 1787
\bibitem[\protect\citeauthoryear{Falc\'on-Barroso et al.}{2017}]{falco2017} Falc\'on-Barroso, J., Lyubenova, M., van de Ven, G. et al. 2017, AA, 597, 48
\bibitem[\protect\citeauthoryear{Feldmann et al.}{2010}]{feldmann2010} Feldmann, R., Carollo, C. M., Mayer, L., Renzini, A., Lake, G., Quinn, T., Stinson, G. S., Yepes, G. 2010, ApJ, 709, 218
\bibitem[\protect\citeauthoryear{Ferr\'e-Mateu et al.}{2017}]{ferremateu2017} Ferr\'e-Mateu, A. Trujillo, I., Mart\'in-Navarro, I. Vazdekis, A. Mezcua, M., Balcells, M. Dom\'inguez, L. 2017, MNRAS, 467, 1929
\bibitem[\protect\citeauthoryear{Fisher $\&$ Drory}{2008}]{fisher2008} Fisher, D.B., Drory, N. 2008, AJ, 136, 773
\bibitem[\protect\citeauthoryear{Fisher $\&$ Drory}{2016}]{fisher2016}  Fisher, D.B., Drory, N. 2016, in 'Galactic Bulges', Astrophysics and Space Science Library, Volume 418. ISBN 978-3-319-19377-9. Springer International Publishing Switzerland, 2016, p. 41, eds. E. Laurikainen, R.F. Peletier, D.A. Gadotti
\bibitem[\protect\citeauthoryear{Friedli $\&$ Benz}{1995}]{friedli1995} Friedli, D., Benz, W. 1995, AA, 301, 649
\bibitem[\protect\citeauthoryear{Gadotti}{2008}]{gadotti2008} Gadotti, D.A. 2008, MNRAS, 384, 420
\bibitem[\protect\citeauthoryear{Garc\'ia-Benito et al.}{2017}]{garciabenito2017} Garc\'ia-Benito, R., Gonzalez Delgado, R.M., P\'erez, E., Cid Fernandez, R., Cortijo-Ferrero, C., et al. 2017, AA, 608, 27  
\bibitem[\protect\citeauthoryear{Gonz\'alez et al.}{2013}]{gonzalez2013} Gonz\'alez, O. A., Rejkuba, M., Zoccali, M., Valent, E., Minniti, D., Tobar, R. 2013, AA, 552, 110
\bibitem[\protect\citeauthoryear{Gonz\'alez Delgado et al.}{2014}]{gonzalez2014} Gonz\'alez Delgado, R. M., P\'erez, E., Cid Fernandes, R., Garc\'ia-Benito, R. et al. 2014, AA, 562, 47
\bibitem[\protect\citeauthoryear{Gonz\'alez Delgado et al.}{2015}]{gonzalez2015} Gonz\'alez Delgado, R. M., Garc\'ia-Benito, R., P\'erez, E., Cid Fernandes, R., de Amorim, A. L. et al. 2015, AA, 581, 103
\bibitem[\protect\citeauthoryear{Herrera-Endoqui et al.}{2015}]{herrera2015} Herrera-Endoqui, M., D\'iaz-Garc\'ia, S., Laurikainen, E., Salo, H. 2015, AA, 582, 86
\bibitem[\protect\citeauthoryear{Herrera-Endoqui et al.}{2017}]{herrera2017} Herrera-Endoqui, M., Salo, H., Laurikainen, E., $\&$ Knapen, J. 2017, AA, 599, 43
\bibitem[\protect\citeauthoryear{Hirschmann et al.}{2013}]{hirschmann2013} Hirschmann, M., Naab, T., Dav\'e, R. et al. 2013, MNRAS, 436, 2929
\bibitem[\protect\citeauthoryear{Hopkins et al.}{2009}]{hopkins2009} Hopkins, P.F., Cox, T.J., Dutta, S.N., Hernquist, L., Kormendy, J., Lauer, T.R., 2009 ApJ, 694, 842
\bibitem[\protect\citeauthoryear{Hopkins et al.}{2014}]{hopkins2014} Hopkins, P.F., Kere\'s, D., Onorbe, J., Faucher-Gigu\'ere, C-A, Quataert, E., Murray, N., Bullock, J.S. 2014, MNRAS, 445, 581
\bibitem[\protect\citeauthoryear{Iannuzzi $\&$ Athanassoula}{2015}]{iannuzzi2015} Iannuzzi, F., Athanassoula, E. 2015, MNRAS, 450, 2514
\bibitem[\protect\citeauthoryear{Jacob et al.}{2010}]{jacob2010} Jacob, J.C., Katz, D.S., Berriman, G.B., Good, J., Laity, A.C., Deelman, E., Kesselman, C., Singh, G., Su, M.H., Prince, T.A., Williams, R. 2010, Astrophysics Source Code Library
\bibitem[\protect\citeauthoryear{Johansson, Naab $\&$ Ostriker}{2012}]{johansson2012} Johansson, P. H., Naab, T., Ostriker, J. P. 2012, ApJ, 754, 115
\bibitem[\protect\citeauthoryear{Johnson et al.}{2011}]{johnson2011} Johnson, C. I., Rich, R. M., Fulbright, J. P., Valenti, E., McWilliam, A. 2011 ApJ, 732, 108
\bibitem[\protect\citeauthoryear{Kauffmann, White $\&$ Guiderdoni}{1993}]{kauffmann1993} Kauffmann, G., White, S.D.M., Guiderdoni, B. 1993, MNRAS, 264, 201
\bibitem[\protect\citeauthoryear{Kennicutt $\&$ Evans}{2012}]{kennicutt2012} Kennicutt, R.C., Evans, N.J. 2012, ARA$\&$A, 50, 531
\bibitem[\protect\citeauthoryear{Kim et al.}{2016}]{kim2016} Kim, K., Oh, S., Jeong, H., Arag\'on-Salamanca, A., Smith, R., Yi, S.K. 2016, ApJS, 225, 6
\bibitem[\protect\citeauthoryear{Kormendy}{1982}]{1982SAAS...12..115K} Kormendy, J.\ 1982, Morphology and dynamics of galaxies; Proceedings of the Twelfth Advanced Course, Saas-Fee, Switzerland, March 29-April 3, 1982 (A84-15502 04-90).~Sauverny, Switzerland, Observatoire de Geneve, 1983, p.~113-288., 12, 113 
\bibitem[\protect\citeauthoryear{Kormendy $\&$ Fisher}{2008}]{kormendy2008} Kormendy, J., Fisher, D. 2008, ASPC, 396, 297
\bibitem[\protect\citeauthoryear{Kormendy et al.}{2010}]{kormendy2010} Kormendy, J., Drory, N., Bender, R., Cornell, M. E. 2010, ApJ, 723, 54
\bibitem[\protect\citeauthoryear{Kunder et al.}{2016}]{kunder2016} Kunder, A., Rich, R.M., Koch, A. et al. 2016, ApJ, 821, 25 
\bibitem[\protect\citeauthoryear{Launhardt, Zylka $\&$ Mezger}{2002}]{laundhart2002} Launhardt, R., Zylka, R., Mezger, P. G. 2002, AA, 384, 112
\bibitem[\protect\citeauthoryear{Laurikainen et al.}{2005}]{lauri2005} Laurikainen, E., Salo, H., Buta, R., Knapen, J. 2005, MNRAS, 362, 1319
\bibitem[\protect\citeauthoryear{Laurikainen et al.}{2010}]{lauri2010} Laurikainen, E., Salo, H., Buta, R., Knapen, J. H., Comer\'on, S. 2010, MNRAS, 405, 1089
\bibitem[\protect\citeauthoryear{Laurikainen et al.}{2011}]{lauri2011} Laurikainen, E., Salo, H., Buta, R., Knapen, J. H. 2011, MNRAS, 418, 1452
\bibitem[\protect\citeauthoryear{Laurikainen et al.}{2013}]{lauri2013} Laurikainen, E., Salo, H., Athanassoula, E., Bosma, A., Buta, R., Janz, J.  2013, MNRAS, 430, 3489
\bibitem[\protect\citeauthoryear{Laurikainen et al.}{2014}]{lauri2014} Laurikainen, E., Salo, H., Athanassoula, E., Bosma, A. 2014, MNRAS, 444, 80 
\bibitem[\protect\citeauthoryear{Laurikainen $\&$ Salo}{2016}]{lauri2016} Laurikainen, E., Salo, H. 2016, in 'Galactic Bulges', Astrophysics and Space Science Library, Volume 418. ISBN 978-3-319-19377-9. Springer International Publishing Switzerland, 2016, p. 77, eds. E. Laurikainen, R.F. Peletier, D.A. Gadotti
\bibitem[\protect\citeauthoryear{Laurikainen $\&$ Salo}{2017}]{lauri2017} Laurikainen E., Salo, H. 2017, AA, 598, 10
\bibitem[\protect\citeauthoryear{Lenz et al.}{1998}]{lenz1998} Lenz, D.D., Newberg, J., Rosner, R., Richards, G.T., Stoughton, C. 1998, ApJS, 119, 121
\bibitem[\protect\citeauthoryear{Li, Ho $\&$ Barth}{2017}]{li2017} Li, Z-Yu, Ho, L. C., Barth, A. J. 2017, ApJ, 845, 87 
\bibitem[\protect\citeauthoryear{L\"utticke, Dettmar $\&$ Pohlen}{2000}]{Lutticke2000} L\"utticke, R., Dettmar, R.-J., Pohlen, M. 2000, AAS, 145, 405
\bibitem[\protect\citeauthoryear{MacWilliam $\&$ Zoccali}{2010}]{mac2010} MacWilliam, A., Zoccali, M. 2010, ApJ, 724, 1491
\bibitem[\protect\citeauthoryear{Marchesini et al.}{2014}]{marc2014} Marchesini, D., Muzzin, A., Stefanon, M., Franx, M.  2014, ApJ, 794, 65
\bibitem[\protect\citeauthoryear{Martinez-Valpuesta, Shlosman $\&$ Heller}{2006}]{valpuesta2006} Martinez-Valpuesta, I., Shlosman, I., Heller, C. 2006, ApJ, 637, 214
\bibitem[\protect\citeauthoryear{Martins et al.}{2005}]{martins2005} Martins, L. P., Gonz\'alez D., Rosa M., Leitherer, C., Cervi\"no, M., Hauschildt, P. 2005, MNRAS, 358, 49
\bibitem[\protect\citeauthoryear{McWilliam $\&$ Rich}{1994}]{mcwilliam1994} McWilliam, A., Rich, R. M. 1994, ApJS, 91, 749
\bibitem[\protect\citeauthoryear{M\'endez-Abreu et al.}{2008}]{mendez2008} M\'endez-Abreu, J., Corsini, E. M., Debattista, V. P., De Rijcke, S., Aguerri, J. A. L., Pizzella, A. 2008, MNRAS, 389, 341
\bibitem[\protect\citeauthoryear{M\'endez-Abreu et al.}{2014}]{mendez2014} M\'endez-Abreu, J., Debattista, V. P., Corsini, E. M., Aguerri, J. A. L. 2014, AA, 572, 25
\bibitem[\protect\citeauthoryear{M\'endez-Abreu et al.}{2017}]{mendez2017} M\'endez-Abreu, J., Ruiz-Lara, T., S\'anchez-Menguiano, L. et al.  2017, AA, 598, 32
\bibitem[\protect\citeauthoryear{M\'endez-Abreu et al.}{2018}]{mendez2018} M\'endez-Abreu, J., Aguerri, J. A. L., Falc\'on-Barroso, J. et al. 2018, MNRAS, 474, 1307
\bibitem[\protect\citeauthoryear{Merritt $\&$ Sellwood}{1994}]{merritt1994} Merritt D., Sellwood J. A., 1994, Astrophys. J., 425, 551
\bibitem[\protect\citeauthoryear{Minniti $\&$ Zoccali}{2008}]{minniti2008} Minniti D., Zoccali M. 2008, IAUS, 245, 323
\bibitem[\protect\citeauthoryear{Molaeinezhad et al.}{2016}]{mola2016} Molaeinezhad, A., Falc\'on-Barroso, J., Mart\'inez-Valpuesta, I., Khosroshahi, H.G., Balcells, M., Peletier, R.F. 2016, MNRAS, 456, 692
\bibitem[\protect\citeauthoryear{Naab et al.}{2007}]{naab2007} Naab, T., Johansson, P. H., Ostriker, J. P., Efstathiou, G. 2007, ApJ, 658, 710
\bibitem[\protect\citeauthoryear{Naab et al.}{2014}]{naab2014} Naab, T., Oser, L., Emsellem, E. et al. 2014, MNRAS, 444, 3357
\bibitem[\protect\citeauthoryear{Nataf et al.}{2010}]{nataf2010} Nataf, D. M., Udalski, A., Gould, A., Fouqu\'e, P., Stanek, K. Z. 2010, ApJ, 721, 28
\bibitem[\protect\citeauthoryear{Negroponte $\&$ White}{1983}]{negroponte1983} Negroponte, J., White, S. D. M. 1983, MNRAS, 205, 1009 
\bibitem[\protect\citeauthoryear{Ness et al.}{2013}]{ness2013} Ness, M., Freeman, K., Athanassoula, E., Wylie-de-Boer, E. 2013, MNRAS, 432, 2092 
\bibitem[\protect\citeauthoryear{Ness $\&$ Lang}{2016}]{ness2016}  Ness, M. Lang, D. 2016, AJ, 152, 14
\bibitem[\protect\citeauthoryear{Neumann, Wisotzki $\&$ Choudhury }{2017}]{neumann2017} Neumann, J., Wisotzki, L., Choudhury, O.S. et al. 2017, AA, 604, 30
\bibitem[\protect\citeauthoryear{Noguchi}{1999}]{noguchi1999} Noguchi, M. 1999, ApJ, 514, 77
\bibitem[\protect\citeauthoryear{O'Neill $\&$ Dubinski}{2003}]{dubinski2003} O'Neill, J. K., Dubinski, J., 2003, MNRAS, 346, 251
\bibitem[\protect\citeauthoryear{Oppenheimer et al.}{2010}]{oppenheimer2010} Oppenheimer, B.D., Dav\'e, R., Kere\'s, D., Fardal, M., Katz, N., Kollmeier, J.A., Weinberg, D.H. 2010, MNRAS, 406, 2325
\bibitem[\protect\citeauthoryear{Oser et al.}{2010}]{oser2010} Oser, L., Ostriker, J.P., Naab, T., Johansson, P.H., Burkert, A. 2010, ApJ, 725, 2312
\bibitem[\protect\citeauthoryear{Patsis $\&$ Xilouris}{2006}]{patsis2006} Patsis, P. A., Xilouris, E. M., 2006, MNRAS, 366, 1121
\bibitem[\protect\citeauthoryear{Peng, Impey $\&$ Rix}{2010}]{peng2010} Peng, C., Ho, L., Impey, C., Rix, H-W. 2010, AJ, 139, 2097
\bibitem[\protect\citeauthoryear{P\'erez et al.}{2017}]{perez2017} P\'erez, I., Mart\'inez-Valpuesta, I., Ruiz-Lara, T., de Lorenzo-Caceres, A. et al. 2017, MNRAS, 471, 122
\bibitem[\protect\citeauthoryear{P\'erez-Villegas, Portail $\&$ Garhard}{2017}]{perezvillegas2017} P\'erez-Villegas, A., Portail, M., Gerhard, O. 2017, MNRAS, 464, 80
\bibitem[\protect\citeauthoryear{Pfenniger $\&$ Norman}{1990}]{pfenniger1990} Pfenniger, D., Norman, C. 1990, ApJ, 363, 391
\bibitem[\protect\citeauthoryear{Pietrukowicz et al.}{2015}]{pietrukowicz2015} Pietrukowicz, P., Kozlowski, S., Skowron, J. et al. 2015, ApJ, 811, 113
\bibitem[\protect\citeauthoryear{Portail, Wegg $\&$ Gerhard}{2015}]{portail2015} Portail, M., Wegg, C., Gerhard, O. 2015, MNRAS, 450, 66
\bibitem[\protect\citeauthoryear{Qu et al.}{2017}]{qu2017} Qu, Y., Helly, J. C., Bower, R. G., Theuns, T., Crain, R. A., Frenk, C. S., Furlong, M., McAlpine, S., Schaller, M., Schaye, J., White, S. D. M. 2017, MNRAS, 464, 1659
\bibitem[\protect\citeauthoryear{Quillen}{2002}]{quillen2002} Quillen, A. C., 2002, AJ, 124, 722 
\bibitem[\protect\citeauthoryear{Raha et al.}{1991}]{raha1991} Raha, N., Sellwood, J. A., James, R. A., Kahn, F. D., 1991, Nature, 352, 411
\bibitem[\protect\citeauthoryear{Ravikumar et al.}{2006}]{ravikumar2006} Ravikumar, C.D., Barway, S., Kembhavi, A., Mobasher, B., Kuriakose, V.C. 2006, AA, 446, 827
\bibitem[\protect\citeauthoryear{Rich et al.}{2008}]{rich2008} Rich, R. M., Howard, C., Reitzel, D. B., Zhao, H-S, de Propris, R. 2008, IAUS, 245, 333
\bibitem[\protect\citeauthoryear{Rowley}{1988}]{rowley1988} Rowley, G., 1988, ApJ, 331, 124
\bibitem[\protect\citeauthoryear{Saha, Graham $\&$ Rodr\'iquez-Herranz}{2018}]{saha2018} Saha, K., Graham, A.,  Rodr\'iquez-Herranz, I. 2018, ApJ, 852, 133
\bibitem[\protect\citeauthoryear{Salo et al.}{2015}]{salo2015} Salo, H., Laurikainen, E., Laine, J. et al. 2015, ApJS, 219, 4
\bibitem[\protect\citeauthoryear{Salo $\&$ Laurikainen}{2017}]{salo2017} Salo, H., Laurikainen, E. 2017, ApJ, 835, 252
\bibitem[\protect\citeauthoryear{S\'anchez et al.}{2012}]{sanchez2012} S\'anchez, S. F., Kennicutt, R. C., Gil de Paz, A. et al. 2012, AA, 538, A8
\bibitem[\protect\citeauthoryear{S\'anchez, Garc\'ia-Benito $\&$ Zibetti}{2016}]{sanchezGZ2016} S\'anchez, S. F., Garc\'ia-Benito, R., Zibetti, S. 2016, AA, 594, A36
\bibitem[\protect\citeauthoryear{S\'anchez et al.}{2016}]{sanchez2016b} S\'anchez, S. F., P\'erez, E., S\'anchez-Bl\'azquez, P. et al. 2016, Revista Mexicana de Astronomia y Astrofisica, 52, 171
\bibitem[\protect\citeauthoryear{S\'anchez-Bl\'azquez et al.}{2006}]{sanchez2006} S\'anchez-Bl\'azquez, P., Peletier, R. F., Jim\'enez-Vicente, J. et al. 2006, MNRAS, 371, 703
\bibitem[\protect\citeauthoryear{Salpeter}{1995}]{salpeter1995} Salpeter, E.E. 1995, ApJ, 161 1995
\bibitem[\protect\citeauthoryear{Scannapieco $\&$ Athanassoula}{2012}]{scannapieco2012} Scannapieco, C., Athanassoula, E. 2012, MNRAS, 425, L10
\bibitem[\protect\citeauthoryear{Shen $\&$ Sellwood}{2004}]{shen2004} Shen, J. $\&$ Sellwood, J. A. 2004, ApJ, 604, 614
\bibitem[\protect\citeauthoryear{Shen et al.}{2010}]{shen2010} Shen, J., Rich, R. M., Kormendy, J., Howard, C. D., De Propris, R., Kunder, A. 2010, ApJ, 720, 72
\bibitem[\protect\citeauthoryear{Springel}{2005}]{springel2005} Springel V. 2005, MNRAS, 364, 1105 
\bibitem[\protect\citeauthoryear{Stinson et al.}{2013}]{stinson2013} Stinson, G.S., Brook, C., Macci\'o, A.V., Wadsley, J., Quinn, T.R., Couchman, H.M.P. 2013, MNRAS, 428, 129
\bibitem[\protect\citeauthoryear{Tissera et al.}{2018}]{tissera2018} Tissera, P.B., Machado, R.E.G., Carollo, D., Minniti, D., Beers, T.C., Zoccali, M., Meza, A. 2018, MNRAS, 473, 1656
\bibitem[\protect\citeauthoryear{Toomre $\&$ Toomre}{1972}]{toomre1972} Toomre, A., Toomre, J. 1972, ApJ, 178, 623
\bibitem[\protect\citeauthoryear{Trujillo et al.}{2006}]{trujillo2006}  Trujillo, I., Feulner, G., Goranova, Y. et al. 2006, MNRAS, 373, 36
\bibitem[\protect\citeauthoryear{Ubler et al.}{2014}]{ubler2014} Ubler, H., Naab, T., Oser, L., Aumer, M., Sales, L.V., White, S.D.M. 2014, MNRAS, 443, 2092
\bibitem[\protect\citeauthoryear{van Dokkum et al.}{2010}]{dokkum2010} van Dokkum, P. G., Whitaker, K. E. et al. 2010, ApJ, 709, 1018 
\bibitem[\protect\citeauthoryear{van Dokkum et al.}{2013}]{dokkum2013} van Dokkum, P.G., Leja, J., Nelson, E.J., Patel, S., Skelton, R.E., Momcheva, I., Brammer, G., Whitaker, K.E., Lundgren, B., Fumagalli, M. et al.  2013, ApJ, 771, 35
\bibitem[\protect\citeauthoryear{van den Berg et al.}{1996}]{berg1996} van den Berg, S., Abraham, R., Ellis, R., Tanvir, N., Santiago, B., $\&$ Glazebrook, K. 1996, AL, 112, 359
\bibitem[\protect\citeauthoryear{Vazdekis et al.}{2010}]{vazdekis2010} Vazdekis, A., S\'anchez-Bl\'azques, P., Falc\'on-Barroso, J. et al. 2010, MNRAS, 404, 1639
\bibitem[\protect\citeauthoryear{Walcher, Wisotzki $\&$ Bekerait\'e}{2014}]{walcher2014} Walcher, C. J., Wisotzki, L., Bekerait\'e, S. 2014, AA, 569, A1
\bibitem[\protect\citeauthoryear{Wegg $\&$ Gerhard}{2013}]{wegg2013} Wegg, C. Gerhard, O. 2013, MNRAS, 345, 1874
\bibitem[\protect\citeauthoryear{Williams et al.}{2011}]{williams2011} Williams, M. J., Zamojski, M. A., Bureau, M., Kuntschner, H., Merrifield, M. R., de Zeeuw, P. T., Kuijken, K. 2011, MNRAD, 414, 2163
\bibitem[\protect\citeauthoryear{Williams et al.}{2012}]{williams2012} Williams, M. J., Bureau, M., Kuntschner, H. 2012, MNRAS, 427, 99
\bibitem[\protect\citeauthoryear{Woods et al.}{2014}]{woods2014} Woods, R.M., Wadsley, J., Couchman, H.M.P., Stinson, G., Shen, S. 2014, MNRAS, 442, 732
\bibitem[\protect\citeauthoryear{Yoshino $\&$ Yamauchi}{2015}]{yoshino2015}  Yoshino, A., Yamauchi, C. 2015, MNRAS, 446, 3749
\bibitem[\protect\citeauthoryear{Zoccali et al.}{2006}]{zoccali2006} Zoccali, M., Lecureur, A., Barbuy, B., Hill, V., Renzini, A., Minniti, D., Momany, Y., G\'omez, A., Ortolani, S. 2006, AA, 457, 1
\bibitem[\protect\citeauthoryear{Zoccali, Hill $\&$ Lecureur}{2008}]{zoccali2008} Zoccali, M., Hill, V., Lecureur, A. 2008, AA, 486, 177 
\bibitem[\protect\citeauthoryear{Zhu et al.}{2017}] {zhu2017} Zhu, L., van den Bosch, R.Rix H-W et al. 2017, astro-ph 1711.06728  
\bibitem[\protect\citeauthoryear{Zhu et al.}{2018}]{zhu2018} Zhu, L., van den Bosch, R., van de Ven, G., Lyubenova, M., Falc\'on-Barroso, J., Meidt, S. E., Martig, M., Shen, J., Li, Z-Y Yildirim, A., et al. 2018, MNRAS, 473, 4000

\end{thebibliography}

\begin{appendix} 

 \section{Preparing the images for decompositions}

\label{appendixA}
{\it Mosaic images and masks:}
\vskip 0.15cm
\noindent We use Montage software \citep{jacob2010} to create mosaics of SDSS
\emph{r'}-band images from SDSS DR12 \citep{alam2016}.  Montage uses
the calibrated frames from SDSS, and creates mosaics with user
defined image and pixel sizes, which are centered on the provided
coordinates. The images in SDSS are sky subtracted, but
Montage performs internal background matching and gradient removal to
the frames before combining them to a mosaic. We created mosaics with a
size of 1000 $\times$ 1000 pixels with the SDSS pixel size of 0.396
arcsec/pix. These images, covering 6.6 $\times$ 6.6 arcmin,
extend much further out than the optical radius of the CALIFA galaxies,
thus allowing reliable
estimation of possible remaining sky background.  We then
used SExtractor \citep{bertin1996} to create masks for the mosaic
images. The masks were checked and manually edited as in \citet{salo2015}.

We used a modified version of the pipeline developed for the S$^4$G
survey decompositions by \cite{salo2015} to measure the galaxy center
coordinates, sky background values, and the position angle and
ellipticity of the galaxy outer isophotes. 
The inclinations ($i$) and positions angles ($PA$) of the disk are
shown in Table \ref{appendixD}.
ls         
\vskip 0.25cm
\noindent {\it Sigma-images:}
\vskip 0.1cm
\noindent Sigma-images quantify the statistical uncertainties of the image
pixels.  We followed the description given in the SDSS web-pages to
calculate the pixel uncertainties in the corrected and calibrated
frames of the
SDSS\footnote{\url{https://data.sdss.org/datamodel/files/BOSS\_PHOTOOBJ/frames/RERUN/RUN/CAMCOL/frame.html}}.
The calibration frames were used to convert the SDSS images from
''nanomaggies'', the units used in SDSS, to raw "data numbers"
(``dn''), and vice-versa.  For each frame we also created sky, gain
(e$^-$/dn), and "dark variance'' (read-out noise + dark current)
images (CAMCOL header keyword and a table given in SDSS web-page were
used). 

Using Montage we then created sigma-image mosaics from the
individual sky, calibration, dark variance, and gain images, with the
same parameters as used to create the mosaic data images. Using the
calibration and sky mosaics, the image pixel values were converted to
"data numbers", and the previously subtracted sky background was added
back:

$$
\rm dn = \frac{\rm image}{\rm calibration} + \rm sky.
$$

\noindent The pixel value errors (dn$_{\rm err}$) were calculated using the gain and dark variance mosaic images:

$$
\rm dn_{\rm err} = \sqrt{\frac{\rm dn}{\rm gain} + \rm dark variance}.
$$

\noindent These errors were converted back to SDSS image ''nanomaggies'' using the calibration mosaic, 
which gives the initial sigma mosaic image ($\sigma_{\rm ini}$):
$$
\sigma_{\rm ini} = \rm dn_{\rm err} \times \rm calibration.
$$
The final sigma-image for the mosaic data image was calculated with:

$$
\sigma = \sigma_{\rm ini} \times \frac{1}{\sqrt{N}} \times \frac{2}{3},
$$

\noindent where $N$ is the number of frames used to construct each mosaic pixel.
The term $\frac{2}{3}$ is an empirical correction factor, used to match the 
sigma-image with the real noise measured from the mosaic images. 
The produced sigma-images match well the mean RMS noise calculated
from the sky measurement regions of the mosaic data images.

\vskip 0.25cm
\noindent {\it PSF-images:}
\vskip 0.1cm

\noindent Modeling the accurate Point Spread Function (PSF) has shown out to be critical to acquire the correct
bulge parameters in the structure decompositions (see
Salo et al. 2015). SDSS provides PSF information for all
the individual calibrated and corrected science frames (''psFields''
files), which were downloaded from SDSS DR12 for constructing
the mosaic images. PSF-image was re-constructed for each individual science mosaic frame. 
Mean SDSS stack PSF was created by normalizing and stacking the individual PSFs.
Then Gaussian + Moffat function was fitted to the mean SDSS PSF (= SDSS fit PSF).

The Gaussian + Moffat fit PSF was then used in our decompositions. 
We found a good agreement with the obtained ''fit PSF'' and the real
PSF extracted from the mosaic images, consisting also the low-flux
wings of the PSF.
We find that the PSF Full Width at Half Maximum (FWHM) values for the r'-band images vary between
0.8--1.4 arcseconds, with the mean FWHM being 1.15 arcsecs.  These values are in agreement with those used in
the decomposition study by \citet{mendez2017}, who used the same
SDSS data.

\clearpage
\onecolumn

\section{Barlens identifications}
\begin{longtable}{llllll}
\caption{\label{appendixB} Barlens (bl) identifications in the CALIFA
  survey, shown separately for the CALIFA ``mother'' and ``extended''
  samples, and for the galaxies in which barlenses and X-shape
  features appear at the same time.  Shown also are the redshifts (z),
  absolute r'-band magnitudes (M$_{\rm r'}$), and stellar masses
  (log M$_{\star}$/M$_{\odot}$), as given in the CALIFA database. The magnitudes are
  based on Petrosian aperture magnitudes, except for a few cases where
  they are based on curve-of-growth magnitudes (in
  italics). Calculation of the stellar masses was explained by Walcher
  et al. (2008).  Indicated is whether the CALIFA V1200 grating kinematics (K),
  and V500 grating stellar population SSP (S) data-cube is available. The galaxies
  previously decomposed by MA2017 are denoted with D. In comments
  uncertain bl-identifications are marked. In a few galaxies a nuclear
  bar (nbar), nuclear ring (nr), or nuclear lens (nl) was recognized.}
\\
\hline\hline
 \noalign{\smallskip}
Galaxy& z   & M$_{\rm r'}$ [mag] & log M$_{\star}$/M$_{\odot}$ & K S D & comment \\
 \noalign{\smallskip}
\hline
\endfirsthead
\caption{continued.}\\
\hline\hline
 \noalign{\smallskip}
Galaxy & z  & M$_{\rm r'}$ [mag] & log M$_{\star}$/M$_{\odot}$ & K S D &comment \\
 \noalign{\smallskip}
\hline
\endhead
\hline
\endfoot
 \noalign{\smallskip}
 \noalign{\smallskip}
bl(mother):  &    &   &   &  &  \\
 \noalign{\smallskip}
 \noalign{\smallskip}
ESO 540-G003         & 0.0108 & -19.99 & 9.980 &  K S D & uncertain  \\
IC 0674 & 0.0264     & -21.80 & 10.867 &  K S D &            \\     
IC 0994 & 0.0269     & -22.03 & 10.996 &      D &            \\
IC 1199 & 0.0179     & -21.26 & 10.432 &  K S D &            \\
IC 3376 & 0.0257     & -21.92 & 10.996 &        &            \\   
IC 3598 & 0.0274     & -21.66 & 10.900 &        &           \\       
IC 4534 & 0.0185     & -21.54 & 10.855 &      D &           \\
IC 4566 & 0.0210     & -21.72 & 10.820 &  K S D &           \\
KUG0210-078 & 0.0155 & -20.59 & 10.371 &      D &           \\               
NGC 0036 & 0.0197    & -21.94 & 10.899 &  K S D &           \\  
NGC 0165 & 0.0192    & -21.29 & 10.511 &      D &          \\ 
NGC 0171 & 0.0129    & -21.45 & 10.614 &  K S D &          \\  
NGC 0309 & 0.0184    & -21.89 & 11.143 &      D &          \\ 
NGC 0364 & 0.0166    & -21.41 & 10.874 &  K   D &          \\
NGC 0447 & 0.0183    & -21.51 & 10.980 &  K   D & nbar,nr  \\
NGC 0570 & 0.0178    & -21.59 & 10.803 &      D &          \\
NGC 0776 & 0.0159    & -21.30 & 10.489 &  K S D & nr       \\
NGC 1211 & 0.0103    & -20.88 & 10.809 &      D & nbar     \\
NGC 1645 & 0.0160    & -21.38 & 10.776 &  K S D &          \\
NGC 1666 & 0.0090    & -20.65 & 10.431 &      D &          \\
NGC 2253 & 0.0125    & -21.34 & 10.504 &  K S D &           \\
NGC 2449 & 0.0169    & -21.57 & 10.790 &  K S D & uncertain \\
NGC 2486 & 0.0161    & -21.09 & 10.655 &  K   D &          \\
NGC 2487 & 0.0168    & -21.63 & 10.998 &  K   D &          \\
NGC 2540 & 0.0216    & -21.49 & 10.214 &  K   D &          \\
NGC 2543 & 0.0091    & -20.50 & 10.265 &      D &          \\ 
NGC 2553 & 0.0165    & -21.06 & 10.627 &  K   D &          \\
NGC 2572 & 0.0272    & -21.73 & 10.855 &      D &          \\
NGC 2595 & 0.0153    & -21.02 & 10.305 &      D &          \\
NGC 2692 & 0.0144    & -21.08 & 10.768 &        &          \\ 
NGC 2860 & 0.0152    & -20.69 & 10.291 &        &          \\
NGC 2874 & 0.0138    & -20.63 & 10.519 &        &          \\ 
NGC 2880 & 0.0066    & -20.54 & 10.530 &  K S D & uncertain  \\
NGC 2927 & 0.0263    & -22.25 & 10.998 &        &          \\ 
NGC 2959 & 0.0157    & -21.89 & 10.863 &        &          \\ 
NGC 2968 & 0.0065    & -20.34 & 10.724 &        &          \\  
NGC 3230 & 0.0118    & -21.58 & 10.912 &        &         \\    
NGC 3237 & 0.0248    & -21.99 & 11.083 &        &         \\  
NGC 3300 & 0.0117    & -21.23 & 10.683 &  K   D &         \\
NGC 3304 & 0.0243    & -21.96 & 11.138 &        &         \\  
NGC 3540 & 0.0228    & -21.65 & 11.080 &        &          \\
NGC 3648 & 0.0082    & -20.34 & 10.360 &        & uncertain  \\    
NGC 3649 & 0.0162    & -20.74 & 10.444 &        &          \\  
NGC 3668 & 0.0130    & -21.28 & 10.539 &        &          \\ 
NGC 3674 & 0.0086    & -20.41 & 10.478 &        &          \\
NGC 3687 & 0.0100    & -20.64 & 10.509 &  K   D &          \\
NGC 3772 & 0.0136    & -20.80 & 10.485 &        &         \\
NGC 3825 & 0.0236    & -22.25 & 11.118 &        &         \\
NGC 3832 & 0.0248    & -21.97 & 10.814 &        &         \\
NGC 3947 & 0.0225    & -21.89 & 10.738 &        &         \\
NGC 3968 & 0.0232    & -22.63 & 11.078 &        &         \\
NGC 4003 & 0.0235    & -21.83 & 10.976 &  K S D &         \\
NGC 4210 & 0.0105    & -20.63 & 10.192 &  K S D &          \\
NGC 4227 & 0.0233    & -22.31 & 11.276 &        &         \\    
NGC 4233 & 0.0094    & -21.03 & 10.850 &        &         \\
NGC 4290 & 0.0116    & -21.30 & 10.592 &        &        \\
NGC 4475 & 0.0265    & -21.64 & 10.675 &        &        \\ 
NGC 4612 & 0.0033    & -19.15 &  9.877 &        &        \\        
NGC 4779 & 0.0111    & -20.78 & 10.289 &        &        \\    
NGC 4795 & 0.0109    & -21.12 & 10.684 &        &        \\   
NGC 5157 & 0.0263    & -22.27 & 11.011 &      D &        \\
NGC 5205 & 0.0075    & -19.81 &  9.887 &  K S D &        \\
NGC 5207 & 0.0276    & -22.13 & 11.206 &        &        \\     
NGC 5267 & 0.0216    & -22.02 & 10.986 &      D &        \\
NGC 5347 & 0.0097    & -20.52 & 10.171 &        &        \\
NGC 5350 & 0.0096    & -21.28 & 10.654 &        &        \\  
NGC 5378 & 0.0121    & -20.92 & 10.603 &  K S D &        \\
NGC 5406 & 0.0192    & -22.26 & 11.065 &  K S D &        \\
NGC 5443 & 0.0078    & -20.28 & 10.501 &      D &        \\
NGC 5473 & 0.0085    & -21.33 & 10.867 &      D &        \\
NGC 5657 & 0.0151    & -20.61 & 10.295 &  K   D &        \\
NGC 5720 & 0.0275    & -22.03 & 10.759 &  K S D &       \\
NGC 5876 & 0.0126    & -21.01 & 10.767 &  K   D &       \\
NGC 5947 & 0.0198    & -21.16 & 10.559 &  K S D &       \\
NGC 6004 & 0.0148    & -21.41 & 10.626 &  K S D &       \\
NGC 6154 & 0.0216    & {\it -22.24} & {\it 10.949} & S D &   \\
NGC 6186 & 0.0117    & -20.90 & 10.536 &  K   D &       \\
NGC 6278 & 0.0111    & -21.32 & 10.845 &  K   D &      \\ 
NGC 6427 & 0.0125    & -21.04 & 10.653 &  K S D &      \\ 
NGC 6497 & 0.0217    & -21.97 & 10.901 &  K S D &      \\
NGC 6945 & 0.0136    & -21.55 & 10.942 &  K   D &       \\
NGC 7321 & 0.0238    & -22.18 & 10.933 &  K S D &       \\
NGC 7563 & 0.0139    & -21.23 & 11.038 &  K S D &       \\
NGC 7611 & 0.0109    & -21.10 & 10.757 &  K   D &         \\
NGC 7623 & 0.0125    & -21.05 & 10.790 &  K   D & uncertain\\ 
NGC 7671 & 0.0137    & -21.54 & 10.946 &  K S D & nl, uncertain \\
NGC 7738 & 0.0222    & -21.96 & 11.096 &  K S D &         \\
NGC 7824 & 0.0201    & -22.10 & 11.086 &  K   D &         \\
UGC 01271 & 0.0164   & -21.21 & 10.709 &  K   D &         \\
UGC 02018 & 0.0199   & -21.51 & 10.731 &        &          \\ 
UGC 02222 & 0.0164   & -21.08 & 10.760 &  K S D & uncertain \\
UGC 02311 & 0.0230   & -21.85 & 10.433 &      D &         \\
UGC 03253 & 0.0145   & -21.00 & 10.328 &  K S D &        \\
UGC 03973 & 0.0226   & -21.61 & 10.216 &      D &         \\
UGC 04195 & 0.0170   & -21.11 & 10.420 &      D &         \\
UGC 04416 & 0.0192   & -21.44 & 10.747 &        &         \\ 
UGC 04515 & 0.0174   & -20.88 & 10.394 &        &         \\  
UGC 05859 & 0.0266   & -21.72 & 10.878 &        &        \\   
UGC 06062 & 0.0103   & -20.19 & 10.391 &        &        \\ 
UGC 06176 & 0.0104   & -19.92 &  9.980 &        &         \\   
UGC 07416 & 0.0246   & -21.89 & 10.515 &        &         \\  
UGC 08539 & 0.0264   & -22.00 & 10.849 &        &        \\   
UGC 09492 & 0.0299   & -22.09 & 11.000 &      D &        \\
UGC 10811 & 0.0303   & -21.70 & 10.818 &  K S D &        \\
UGC 11694 & 0.0175   & -21.59 & 11.318 &        &       \\  
UGC 12767 & 0.0173   & {\it -22.36} & {\it 10.992} &   &   \\ 
 \noalign{\smallskip}
bl(extended): &      &        &        &        &       \\
 \noalign{\smallskip}                                                    
UGC 04455     & 0.0310 & -21.70 & 10.878 &        &      \\   
NGC 0495      & 0.0135 & -20.80 & 10.902 &        &      \\                 
MCG+02-35-020 & 0.0247 & -20.13 & 10.076 &        &      \\                
NGC 5794      & 0.0139 & -20.72 & 10.627 &        &      \\                
IC 1078       & 0.0287 & -21.68 & 10.645 &        &       \\     
NGC 6977      & 0.0204 & -21.73 & 10.930 &        & uncertain \\
NGC 0515      & 0.0170 & -21.41 & 11.038 &        &          \\       
IC 0195       & 0.0122 & -20.48 & 10.455 &        &           \\      
NGC 5947      & 0.0198 & -21.16 & 10.559 &        &        \\
NGC 1281      & 0.0141 & -20.80 & 10.852 &        & uncertain \\
NGC 2767      & 0.0165 & -20.87 & 10.750 &        &          \\       
PGC 11179     & 0.0225 & -21.41 & 10.915 &        & uncertain \\
PGC 32873     & 0.0251 & -21.21 & 10.934 &        &        \\ 
 \noalign{\smallskip}
 \noalign{\smallskip}
bl+X:         &        &        &        &        &         \\
 \noalign{\smallskip}
IC 1755      & 0.0257  & -21.57 & 10.878 &     K     &   X/bl  \\
IC 2434      & 0.0248  & -21.85 & 10.748 &           &  X/bl  \\ 
NGC 0180     & 0.0172  & -21.45 & 10.960 &  K S D     &  X/bl      \\ 
NGC 1093     & 0.0172  & -21.29 & 10.431 &     K   D &  X,bl \\
NGC 5000     & 0.0207  & -21.52 & 10.552 &     K S D &  X/bl \\
NGC 5411     & 0.0216  & -21.65 & 11.001 &           &   X/bl \\
NGC 5735     & 0.0145  & -20.91 & 10.332 &         D &  X/bl \\
NGC 5957     & 0.0078  & {\it -20.83} & {\it 10.265} &   D &  X/bl \\
NGC 06941    & 0.0216  & -22.05 & 11.072 &     K S D &  X/bl \\
UGC 04280    & 0.0125  & -20.11 & 10.045 &     K     &  X/bl \\
UGC 06891    & 0.0245  & -20.63 & 10.308 &           &  X/bl \\
UGC 07145    & 0.0238  & -21.05 & 10.329 &     K   D &  X/bl \\
UGC 08781    & 0.0274  & -22.22 & 10.931 &     K S D &  X/bl \\
UGC 09842    & 0.0315  & -21.44 & 10.582 &         D &  X/bl \\
UGC 12185    & 0.0222  & -21.38 & 10.574 &     K S D &  X/bl \\
\end{longtable}
\clearpage

\section{X-shape galaxies}
\begin{longtable}{llllll}
\caption{\label{appendixC} Galaxies with X-shaped bars in the CALIFA
  survey, identified from the unsharp mask images. The columns are the
  same as in  Table \ref{appendixB}. In comments the bar identifications in the
  direct mosaic images are marked. In a few galaxies boxy rather than
  X-shape bar morphology appeared. } \\ 
\hline\hline \noalign{\smallskip}
Galaxy & z & M$_{\rm r'}$ [mag] & log M$_{\star}$/M$_{\odot}$ & K S D & comment \\ \noalign{\smallskip}
\hline
\endfirsthead
\caption{continued.}\\
\hline\hline
 \noalign{\smallskip}
Galaxy &  z  & M$_{\rm r'}$ [mag] & log M$_{\star}$/M$_{\odot}$ & K S D &comment \\
 \noalign{\smallskip}
\hline
\endhead
\hline
\endfoot
 \noalign{\smallskip}
X (mother):    &       &        &        &       &    \\
      \noalign{\smallskip}     
IC 0836 & 0.0107 & -19.52 & 10.409 &        &  bar\\
IC 1652 & 0.0169 & -20.83 & 10.599 &  K S   &  bar  \\
IC 2247 & 0.0150 & -20.35 & 10.496 &  K     &      \\ 
IC 2487 & 0.0156 & -20.89 & 10.511 &  K S   &       \\
IC 3203 & 0.0249 & -21.06 & 10.965 &        &       \\
IC 3704 & 0.0310 & -21.68 & 10.366 &        &      \\
MCG-01-01-012 & 0.0189 & -20.84 & 10.791 &    S   &      \\
MCG-01-10-015 & 0.0134 & -19.94 & 9.864 &      D &      \\
MCG-02-02-030 & 0.0115 & -20.75 & 10.301 &  K S D &      \\
MCG-02-08-014 & 0.0162 & -20.27 & 10.574 &        & boxy  \\              
NGC 0169 & 0.0151 & -21.12 & 11.233 &  K S   &      \\
NGC 0177 & 0.0126 & -20.49 & 10.366 &  K     &      \\       
NGC 0217 & 0.0130 & -21.56 & 10.994 &  K     &      \\
NGC 0833 & 0.0125 & -20.97 & 10.753 &        &       \\
NGC 0955 & 0.0048 & -19.52 & 10.261 &        &  bar \\
NGC 2481 & 0.0085 & -20.66 & 10.605 &  K     &  bar  \\
NGC 2530 & 0.0174 & -20.85 & 10.182 &      D &       \\
NGC 2558 & 0.0174 & -21.59 & 10.721 &      D &  bar  \\
NGC 2638 & 0.0136 & -20.91 & 10.715 &        &      \\
NGC 2735 & 0.0092 & -19.99 & 10.322 &        &      \\
NGC 2769 & 0.0170 & -21.62 & 11.066 &        &      \\
NGC 2826 & 0.0223 & -21.50 & 10.957 &        &      \\
NGC 2854 & 0.0104 & -20.30 & 10.053 &        & bar  \\
NGC 2906 & 0.0081 & -20.60 & 10.463 &  K S D &      \\
NGC 3160 & 0.0243 & -21.40 & 10.899 &  K S   &      \\
NGC 3303 & 0.0224 & -21.37 & 10.890 &  K S   &      \\
NGC 3697 & 0.0226 & -22.05 & 10.803 &        &      \\
NGC 3753 & 0.0308 & -22.13 & 11.264 &        &      \\
NGC 3762 & 0.0129 & -21.27 & 10.869 &        &      \\
NGC 3869 & 0.0117 & -21.18 & 10.930 &        &      \\
NGC 3958 & 0.0127 & -20.97 & 10.567 &        & bar  \\
NGC 3987 & 0.0168 & -20.48 & 10.709 &        &      \\
NGC 4012 & 0.0158 & -20.84 & 10.300 &        &      \\
NGC 4175 & 0.0152 & -19.90 & 10.525 &        & boxy\\
NGC 4180 & 0.0033 & -18.54 & 9.583 &        &     \\
NGC 4352 & 0.0034 & -18.16 & 9.400 &        &      \\
NGC 4405 & 0.0034 & -18.69 & 9.437 &        & bar  \\     
NGC 4474 & 0.0034 & -19.13 & 9.850 &        &     \\
NGC 4570 & 0.0033 & -19.91 & 10.390 &        &     \\
NGC 4675 & 0.0174 & -20.63 & 10.327 &        & bar \\
NGC 4676B & 0.0216 & -18.73 & 10.007 &  K S   &     \\
NGC 4686 & 0.0182 & -22.10 & 11.362 &        &     \\
NGC 4892 & 0.0216 & -21.33 & 10.720 &        & bar \\
NGC 4895 & 0.0302 & -22.32 & 11.316 &        &     \\
NGC 5081 & 0.0241 & -21.73 & 10.715 &        &  bar  \\
NGC 5166 & 0.0174 & -21.33 & 11.098 &        &       \\
NGC 5208 & 0.0247 & -22.32 & 11.355 &        &       \\
NGC 5305 & 0.0207 & -21.50 & 10.658 &        & bar  \\
NGC 5308 & 0.0084 & -21.30 & 10.919 &        &       \\
NGC 5349 & 0.0210 & -21.04 & 10.559 &        & bar \\
NGC 5353 & 0.0097 & -22.00 & 11.338 &        &     \\
NGC 5379 & 0.0071 & -19.23 & 9.779 &      D &     \\
NGC 5401 & 0.0145 & -20.69 & 10.494 &        &     \\
NGC 5439 & 0.0082 & -19.15 & 9.474 &        &     \\
NGC 5445 & 0.0150 & -21.19 & 10.800 &        & bar \\        
NGC 5448 & 0.0086 & -20.60 & 10.509 &        &    \\
NGC 5475 & 0.0075 & -20.21 & 10.165 &        &     \\
NGC 5587 & 0.0095 & -20.42 & 10.321 &      D &    \\
NGC 5610 & 0.0190 & -21.67 & 10.627 &      D &    \\
NGC 5659 & 0.0170 & -21.00 & 10.528 &      D &  bar \\
NGC 5689 & 0.0090 & -21.30 & 10.913 &        &     \\
NGC 5888 & 0.0308 & -22.55 & 11.212 &  K S D &  bar  \\
NGC 5908 & 0.0127 & -21.66 & 11.146 &  K S   &      \\
NGC 6032 & 0.0163 & -20.96 & 10.415 &  K S D & bar  \\
NGC 6081 & 0.0194 & -21.69 & 11.100 &  K S   &      \\
NGC 6150 & 0.0307 & -22.41 & 11.237 &  K S D &      \\
NGC 6361 & 0.0141 & -21.28 & 10.914 &        &     \\ 
NGC 6394 & 0.0294 & -21.79 & 10.862 &  K S D & bar \\ 
NGC 6762 & 0.0109 & -20.21 & 10.311 &  K S D &     \\ 
NGC 7631 & 0.0125 & -20.99 & 10.411 &  K S D &  bar \\
NGC 7783 & 0.0250 & -22.21 & 11.359 &  K S   &      \\
UGC 00987 & 0.0152 & -21.02 & 10.638 &  K   D &  bar \\
UGC 01062 & 0.0177 & -21.50 & 10.825 &        &  bar \\
UGC 01123 & 0.0159 & -20.84 & 10.581 &        &  bar   \\  
UGC 01274 & 0.0255 & -21.59 & 11.103 &        &  bar  \\   
UGC 01659 & 0.0267 & -21.63 & 10.497 &      D &  bar \\
UGC 01749 & 0.0261 & -21.35 & 10.762 &        &  bar \\
UGC 02134 & 0.0149 & -21.15 & 10.604 &     D  &  bar \\
UGC 02403 & 0.0133 & -20.66 & 10.530 &  K S D &     \\
UGC 02465 & 0.0165 & -21.40 & 10.742 &        &    \\
UGC 02628 & 0.0221 & -21.02 & 10.355 &        &    \\   
UGC 03151 & 0.0143 & -21.30 & 10.523 &  K     &    \\
UGC 04029 & 0.0153 & -20.41 & 10.425 &  K     &    \\
UGC 04136 & 0.0229 & -21.34 & 11.119 &        &    \\
UGC 04190 & 0.0172 & -21.20 & 10.948 &        &    \\      
UGC 04308 & 0.0127 & -20.48 & 10.911 &  K   D &  bar \\
UGC 04386 & 0.0163 & -21.27 & 10.904 &        &    \\
UGC 04461 & 0.0174 & -20.76 & 9.980 &      D &    \\
UGC 04546 & 0.0182 & -20.55 & 10.770 &        &    \\
UGC 04938 & 0.0310 & -21.67 & 11.025 &        &      \\
UGC 05113 & 0.0235 & -21.45 & 10.976 &  K     &  bar \\
UGC 05267 & 0.0200 & -21.21 & 10.815 &        &  bar \\
UGC 05481 & 0.0223 & -21.55 & 11.059 &        &     \\
UGC 05657 & 0.0242 & -20.93 & 10.675 &        &  bar \\   
UGC 05680 & 0.0243 & -21.60 & 11.001 &        &  bar  \\   
UGC 05713 & 0.0225 & -21.27 & 10.736 &        &  bar \\
UGC 05894 & 0.0233 & -21.59 & 10.648 &        &      \\
UGC 06106 & 0.0232 & -21.47 & 10.631 &        &     \\ 
UGC 06219 & 0.0224 & -21.47 & 10.864 &        &     \\
UGC 06273 & 0.0230 & -21.66 & 11.051 &        &    \\
UGC 06336 & 0.0276 & -21.11 & 10.801 &        &     \\         
UGC 06397 & 0.0226 & -21.08 & 10.842 &        &     \\    
UGC 06414 & 0.0272 & -20.66 & 10.680 &        &    \\
UGC 06545 & 0.0104 & -19.91 & 9.883 &        &    \\           
UGC 06653 & 0.0124 & -20.14 & 10.201 &        &    \\   
UGC 06677 & 0.0302 & -21.13 & 11.043 &        &    \\              
UGC 07141 & 0.0249 & -20.84 & 10.011 &        &    \\
UGC 07367 & 0.0153 & -21.51 & 10.928 &        & bar  \\    
UGC 08025 & 0.0230 & -21.36 & 11.008 &        &     \\             
UGC 08119 & 0.0305 & -22.04 & 11.159 &        &     \\
UGC 08498 & 0.0263 & -22.05 & 11.075 &        &     \\               
UGC 08778 & 0.0128 & -20.23 & 10.149 &  K S   &     \\
UGC 08902 & 0.0277 & -22.12 & 11.059 &        &     \\
UGC 08955 & 0.0143 & -19.82 & 10.018 &        &     \\  
UGC 09539 & 0.0233 & -21.29 & 10.719 &        & X/boxy \\   
UGC 09711 & 0.0302 & -21.27 & 10.963 &        &     \\
UGC 10337 & 0.0313 & -22.02 & 10.995 &  K   D & bar \\
UGC 10388 & 0.0175 & -20.86 & 10.547 &  K S D & bar \\
UGC 11740 & 0.0220 & -20.95 & 10.413 &      D &     \\
UGC 12274 & 0.0254 & -21.73 & 11.101 &  K S D &  bar \\
UGC 12348 & 0.0253 & -21.79 & 10.475 &      D &  bar \\
UGC 12810 & 0.0266 & -21.76 & 10.810 &  K   D &  bar \\
 \noalign{\smallskip}
 \noalign{\smallskip}
X (extended):  &       &         &       &       &      \\        
       \noalign{\smallskip}                     
NGC 3600 & 0.0046 & -18.54 & 9.132 &        &      \\
NGC 3990 & 0.0052 & -19.37 & 9.942 &        &  bar \\
NGC 0675 & 0.0175 & -17.07 & 9.719 &        &      \\
NGC 5358 & 0.0081 & -19.09 & 9.796 &        &      \\
PGC 32873 & 0.0251 & -21.21 & 10.934 &        &     \\
\end{longtable}

\clearpage

\section{Sizes of bars and barlengths}

\begin{longtable}{llrrrl}
\caption{\label{appendixD} Measured semi-major ($a$) and semi-minor ($b$) axis lengths of bars, and barlenses, given in arcseconds. Given also are the position angles ($PA$ in degrees) of these structures. For the disks positions angles and inclinations ($i$) are shown. The r'-band images have a pixel resolution  of 0.396 arcseconds.}\\
\hline\hline
 \noalign{\smallskip}
Galaxy& structure   & $a$ [``] & $b$ [``] & $PA$ [$\degr$] & $i$ [$\degr$] \\
 \noalign{\smallskip}
\hline
\endfirsthead
\caption{continued.}\\
\hline\hline
 \noalign{\smallskip}
Galaxy & structure  & $a$ [``]  & $b$ [``] & $PA$ [$\degr$] & $i$ \\
\hline
\endhead
\hline
\endfoot
 \noalign{\smallskip}
      \noalign{\smallskip}     
 \noalign{\smallskip}
IC 0674   &   bl   &     6.37  &   3.38  &  112.86  &    \\
          &   bar  &     5.29  &   1.32  &  175.91  &     \\
          &   disk &           &         &  117.75  &  69.08 \\
 \noalign{\smallskip}
IC 1199   &   bl   &    3.94   &  2.12   &  145.78  &      \\
          &   bar  &    4.28   &  1.07   &   49.40  &     \\
          &   disk &           &         &  158.28  & 68.67\\
 \noalign{\smallskip}
IC 1755   &   bl   &    4.66   &  1.66   &  127.30  & \\
          &   bar  &    3.36   &  0.84   &    9.46  & \\
          &   disk &           &         &  154.33  &75.97\\
 \noalign{\smallskip}
IC 4566   &   bl   &    5.97   &  3.94   &  141.86  & \\
          &   bar  &    7.77   &  1.94   &  132.84  & \\
          &   disk &           &         &  140.38  &54.87\\
 \noalign{\smallskip}
NGC 0036  &   bl   &    6.49   &  4.89   &   24.02  & \\
          &   bar  &    6.04   &  1.51   &  123.91  & \\
          &   disk &           &         &   16.84  &62.51\\
 \noalign{\smallskip}
NGC 0171  &   bl   &  10.39    & 8.52    &  106.27  &\\
          &   bar  &  15.08    & 3.77    &  124.32  &\\
          &   disk &           &         &   87.70  &13.79\\
 \noalign{\smallskip}
NGC 0177  &   bl   &   5.06    & 4.87    &   12.38  &\\
          &   bar  &   8.19    & 2.05    &    4.82  &\\
          &   disk &           &         &    8.44  &75.93\\
 \noalign{\smallskip}
NGC 0180  &   bl   &   6.46    & 6.13    &  157.37  & \\
          &   bar  &   10.49   & 2.62    &  141.82  & \\
          &   disk &           &         &  165.30  &48.44 \\
 \noalign{\smallskip}
NGC 0364  &   bl   &   6.03    & 4.80    &   35.45  & \\
          &   bar  &   6.10    & 1.52    &   94.87  & \\
          &   disk &           &         &31.05     & 46.25 \\
 \noalign{\smallskip}
NGC 0447  &   bl   &  11.66    & 10.27   &   99.53  &  \\
          &   bar  &  18.88    &  4.72   &   19.18  &  \\
          &   disk &           &         & 62.11    & 22.33 \\
 \noalign{\smallskip}
NGC 0776  &   bl   &   7.98    & 6.96    &   84.01  & \\
          &   bar  &   9.45    & 2.36    &  123.61  & \\
          &   disk &           &         & 56.46    &26.29 \\
 \noalign{\smallskip}
NGC 1093  &   bl   &   3.28    & 2.41    &   93.44  & \\
          &   bar  &   6.31    & 1.58    &  118.12  & \\
          &   disk &           &         & 97.44    &52.43 \\
 \noalign{\smallskip}
NGC 1645  &   bl   &   7.42    & 4.16    &   89.93  & \\
          &   bar  &   5.98    & 1.50    &   17.42  & \\
          &   disk &           &         &   82.67  &62.28 \\
 \noalign{\smallskip}
NGC 2253  &   bl   &   5.32    & 3.37    &  120.28  & \\
          &   bar  &   6.12    & 1.53    &  176.17  & \\
          &   disk &           &         &  124.64  & 41.18 \\
 \noalign{\smallskip}
NGC 2486  &   bl   &   6.27    & 4.64    &   95.78  & \\
          &   bar  &   6.28    & 1.57    &   45.95  & \\
          &   disk &           &         &   91.19  & 56.12 \\
 \noalign{\smallskip}
NGC 2487  &   bl   &  11.62    &10.07    &   85.89  &  \\
          &   bar  &  13.57    & 3.39    &   41.61  & \\
          &   disk &           &         &  123.13   &35.02 \\
 \noalign{\smallskip}
NGC 2540  &   bl   &   3.37    & 2.21    &  129.49  & \\
          &   bar  &   4.36    & 1.09    &   36.10  & \\
          &   disk &           &         &  129.58   &50.69 \\
 \noalign{\smallskip}
NGC 2553  &   bl   &   8.88    & 6.03    &  74.41   & \\
          &   bar  &   8.05    & 2.01    & 178.49   & \\
          &   disk &           &         &  61.29   &65.55 \\
 \noalign{\smallskip}
NGC 3300  &   bl   &   6.41    & 4.16    & 165.36   & \\
          &   bar  &   7.30    & 1.82    &  46.06   & \\
          &   disk &           &         & 173.19   &56.95 \\
 \noalign{\smallskip}
NGC 3687  &   bl   &   4.98    & 4.75    &  29.14   & \\
          &   bar  &   7.32    & 1.83    & 174.59   & \\
          &   disk &           &         & 144.82   &27.87 \\
 \noalign{\smallskip}
NGC 4003  &   bl   &   9.82    & 5.99    & 179.79   & \\
          &   bar  &  14.66    & 3.66    & 144.70   & \\
          &   disk &           &         & 173.73   &44.69 \\
 \noalign{\smallskip}
NGC 4210  &   bl   &   6.05    & 4.06    & 102.30   & \\
          &   bar  &   7.82    & 1.96    &  43.70   & \\
          &   disk &           &         &  98.42   &39.83 \\
 \noalign{\smallskip}
NGC 5000  &   bl   &   5.50    & 4.94    &  81.93   & \\
          &   bar  &  11.33    & 2.83    &  86.00   & \\
          &   disk &           &         &  16.85   &39.22 \\
 \noalign{\smallskip}
NGC 5205  &   bl   &   5.26    & 3.89    & 139.84   & \\
          &   bar  &   7.79    & 1.95    &  96.67   & \\
          &   disk &           &         & 168.13   &56.30 \\
 \noalign{\smallskip}
NGC 5378  &   bl   &  13.61    & 10.19   &  81.74   &  \\
          &   bar  &  16.68    &  4.17   &  39.44   & \\
          &   disk &           &         &  83.61    &41.86 \\
 \noalign{\smallskip}
NGC 5406  &   bl   &   7.46    & 6.23    &  93.73   & \\
          &   bar  &  10.76    & 2.69    &  50.09   & \\
          &   disk &           &         & 108.37   &36.61 \\
 \noalign{\smallskip}
NGC 5657  &   bl   &   4.96    & 1.73    & 149.22   & \\
          &   bar  &   6.99    & 1.75    &  30.37   & \\
          &   disk &           &         & 166.53   &67.42 \\
 \noalign{\smallskip}
NGC 5720  &   bl   &   4.36    & 3.33    & 123.71   & \\
          &   bar  &   4.36    & 1.09    &  51.25   & \\
          &   disk &           &         & 132.69   &43.72 \\
 \noalign{\smallskip}
NGC 5876  &   bl   &  11.23    & 7.00    &  54.18   & \\
          &   bar  &  11.26    & 2.81    & 177.18   & \\
          &   disk &           &         &  51.52   &65.32 \\
 \noalign{\smallskip}
NGC 5947  &   bl   &   4.07    & 4.04    &  25.21   & \\
          &   bar  &   6.42    & 1.60    &  25.29   &  \\
          &   disk &           &         &  69.49   &32.54 \\
 \noalign{\smallskip}
NGC 6004  &   bl   &   5.33    & 4.63    & 113.61   & \\
          &   bar  &   7.66    & 1.91    &  12.30   & \\
          &   disk &           &         &  93.41   &38.67 \\
 \noalign{\smallskip}
NGC 6186  &   bl   &  15.15    & 9.83    &  59.16   & \\
          &   bar  &  18.59    & 4.65    &  52.52   & \\
          &   disk &           &         &  49.16   &36.26 \\
 \noalign{\smallskip}
NGC 6278  &   bl   &   6.64    & 6.33    & 144.59   & \\
          &   bar  &  10.25    & 2.56    & 108.56   & \\
          &   disk &           &         & 126.40   &54.15 \\
 \noalign{\smallskip}
NGC 6497  &   bl   &   6.45    & 4.73    & 113.25   & \\
          &   bar  &   7.10    & 1.77    & 159.47   & \\
          &   disk &           &         & 111.97   & 61.73 \\
 \noalign{\smallskip}
NGC 6941  &   bl   &   6.41    & 5.02    & 122.27   & \\
          &   bar  &   9.31    & 2.33    & 113.09   & \\
          &   disk &           &         & 129.97   &43.49 \\
 \noalign{\smallskip}
NGC 6945  &   bl   &   6.58    & 4.14    & 123.94   & \\
          &   bar  &   6.94    & 1.73    &  74.90   & \\
          &   disk &           &         & 124.94   &50.28 \\
 \noalign{\smallskip}
NGC 7321  &   bl   &   4.26    & 2.66    & 178.34   & \\
          &   bar  &   7.26    & 1.81    &  65.60   & \\
          &   disk &           &         &  9.69    &44.49 \\
 \noalign{\smallskip}
NGC 7563  &   bl   &  10.60    & 7.73    & 148.14   & \\
          &   bar  &  11.39    & 2.85    &  84.67   & \\
          &   disk &           &         & 147.19   &53.01 \\
 \noalign{\smallskip}
NGC 7611  &   bl   &   4.25    & 3.85    & 125.66   & \\
          &   bar  &   4.13    & 1.03    &   0.43   & \\
          &   disk &           &         & 135.24   &63.37 \\
 \noalign{\smallskip}
NGC 7623  &   bl   &   9.45    & 8.52    &   5.84   & \\
          &   bar  &  11.28    & 2.82    & 157.40   & \\
          &   disk &           &         &   5.22   &40.91 \\
 \noalign{\smallskip}
NGC 7738  &   bl   &  15.71    &10.73    &  52.01   &  \\
          &   bar  &  27.55    & 6.89    &  37.05   & \\
          &   disk &           &         &  67.35   &47.23 \\
 \noalign{\smallskip}
NGC 7824  &   bl   &   4.02    & 2.56    & 131.16   & \\
          &   bar  &   4.75    & 1.19    &  17.24   & \\
          &   disk &           &         & 144.20   &50.52 \\
 \noalign{\smallskip}
UGC 01271 &   bl   &  7.08     &4.87     & 101.99   & \\
          &   bar  &  8.11     &2.03     &  48.75   & \\
          &   disk &           &         & 106.99   & 58.39 \\
 \noalign{\smallskip}
UGC 03253 &   bl   &  6.03     &3.42     &  92.45   & \\
          &   bar  &  8.01     &2.00     &  35.22   & \\
          &   disk &           &         &  84.74   & 59.30 \\
 \noalign{\smallskip}
UGC 08781 &   bl   &  6.11     &4.61     & 146.09   & \\
          &   bar  & 11.62     &2.90     & 167.68   & \\
          &   disk &           &         & 160.43   &55.33 \\
 \noalign{\smallskip}
UGC 10811 &   bl   &  3.83     &2.78     &  89.53   & \\
          &   bar  &  3.63     &0.91     & 147.67   & \\
          &   disk &           &         &  91.50   & 69.62 \\
\end{longtable}

\end{appendix}

\end{document}